\documentclass[]{JHEP3}
\usepackage{epsfig}

\def\simlt{\stackrel{<}{{}_\sim}}
\def\simgt{\stackrel{>}{{}_\sim}}
\def\be{\begin{equation}}
\def\ee{\end{equation}}
\def\beq{\begin{equation}}
\def\eeq{\end{equation}}
\def\bea{\begin{eqnarray}}
\def\eea{\end{eqnarray}}

\title{Neutrino Mass Models \footnote{Review article 
submitted for publication in Reports on Progress in Physics}}

\author{S. F. King \footnote{e-mail address: sfk@hep.phys.soton.ac.uk}\\
Department of Physics and Astronomy,
University of Southampton, Southampton, SO17 1BJ, U.K.}


\abstract{This is a review article about neutrino mass models,
particularly see-saw models involving three active neutrinos which
are capable of describing both the atmospheric neutrino oscillation
data, and the large mixing angle MSW solar solution, which is now
uniquely specified by recent data. We briefly review the current
experimental status, show how to parametrise and construct
the neutrino mixing matrix, and present the leading order neutrino
Majorana mass matrices. We then introduce the see-saw mechanism,
and discuss a natural application of it to current data
using the sequential dominance mechanism, which we compare to
an early proposal for obtaining large mixing angles.
We show how both the Standard Model and the Minimal
Supersymmetric Standard Model may be extended to incorporate
the see-saw mechanism, and show how the latter case leads to the
expectation of lepton flavour violation.
The see-saw mechanism motivates models with additional
symmetries such as unification and family symmetry models,
and we tabulate some possible models, before
focussing on two particular examples based on $SO(10)$ grand unification
and either $U(1)$ or $SU(3)$ family symmetry as specific examples.
The article contains extensive appendices which include
techniques for analytically diagonalising different types
of mass matrices involving two large mixing angles and one
small mixing angle, to leading order in the small mixing angle.}

\preprint{SHEP 03-15 \\ hep-ph/0310204}

\begin{document}

\section{Introduction}

\subsection{Overview for the non-specialist}

In 1930, the Austrian physicist Wolfgang Pauli proposed the
existence of particles called neutrinos as a ``desperate remedy'' to
account for the missing energy in a type of radioactivity called beta
decay. He deduced that some of the energy must have been taken away by
a new particle emitted in the decay process, the neutrino. Since then,
after decades of painstaking experimental and theoretical work,
neutrinos have become enshrined as an essential part of the accepted
quantum description of fundamental particles and forces, the Standard
Model of Particle Physics. This is a highly successful theory in which
elementary building blocks of matter are divided into three generations
of two kinds of particle - quarks and leptons. It also includes three
of the fundamental forces of Nature, but does not include gravity. The
leptons consist of the charged electron, muon and tau, together with
three electrically neutral particles - the electron neutrino, muon
neutrino and tau neutrino.

Unlike the case for quarks and charged leptons, however, the Standard
Model predicts that neutrinos have no mass! This might seem curious for
a matter particle, but the Standard Model
predicts that neutrinos always have a left handed
spin - rather like rifle bullets which spin counter clockwise
to the direction of travel. If right-handed neutrinos were to be added
to the Standard Model, then neutrinos could have the same sort of
masses as the quarks and charged leptons, and the theory would also
predict the existence of antineutrinos. However, even without
right-handed neutrinos, 
neutrinos with mass are possible, providing that the neutrino is
its own antiparticle. Such a mass is then called
a Majorana mass, named after the Sicilian physicist, Ettore Majorana.
But the current Standard Model
forbids such Majorana masses.
These subtle theoretical arguments about the nature of neutrinos
have now come to the fore, as the results from experiments detecting
neutrinos from the Sun, as well as atmospheric neutrinos produced by
cosmic rays, suggest that they do have mass after all.

The first clues came from an experiment deep underground, carried out
by an American scientist Raymond Davis Jr., detecting solar neutrinos. It
revealed only about one-third of the number predicted by theories of
how the Sun works. The result puzzled both solar and neutrino
physicists. However, some Russian researchers, Mikheyev and Smirnov,
developing ideas proposed previously by Wolfenstein in the U.S.,
suggested that the solar neutrinos might be changing into something
else.  Only electron neutrinos are emitted by the Sun and they could
be converting into muon and tau neutrinos which were not being
detected on Earth. This effect called neutrino oscillations as the
types of neutrino interconvert over time from one kind to another, was
first proposed some time earlier by Pontecorvo.  The precise mechanism
proposed by Mikheyev, Smirnov and Wolfenstein involved the resonant
enhancement of neutrino oscillations due to matter effects, and is
known as the MSW effect.

Most recently, the Sudbury Neutrino Observatory (SNO) in Canada
spectacularly showed this to be the case. The
experiment measured both the flux of the electron neutrinos and the
total flux of all three types of neutrinos. The data revealed that
physicists' theories of the Sun were correct after all. The idea of
neutrino oscillations had already gained support from the Japanese experiment
Super-Kamiokande which in 1998 showed that there was a deficit of muon
neutrinos reaching Earth when cosmic rays strike the upper atmosphere. The
results were interpreted as muon neutrinos oscillating into tau
neutrinos which could not be detected.

Such neutrino oscillations are analagous to coupled pendulums, where
oscillations in one pendulum induce oscillations in another pendulum.
The coupling strength is defined in terms of something
called the ``mixing angle''. Following the SNO results
several research groups showed that the electron
neutrino must have a mixing angle of about 30 degrees, and forms a mass
state of 0.007 electronvolts or greater
(by comparison the electron has a mass of about half a
megaelectronvolt). The muon and tau neutrinos
must have a (maximal) mixing angle of about 45 degrees,
and form a mass state of about 0.05 electronvolts or greater.

Experimental information on neutrino masses and mixings implies
new physics beyond the Standard Model, and there has been much
activity on the theoretical implications of these results.
An attractive mechanism for explaining small
neutrino masses is the so-called see-saw mechanism
proposed in 1979 by Murray Gell-Mann, Pierre Ramond
and Richard Slansky working in the U.S.,
and independently by Tsutomu Yanagida of Tokyo University.
The idea is to introduce right-handed neutrinos into the Standard Model
which are Majorana-type particles with very heavy masses, possibly
associated with large mass scale at which the three
forces of the Standard Model unify.
The Heisenberg Uncertainty Principle, which allows energy conservation
to be violated on small time intervals,
then allows a left-handed neutrino to convert spontaneously into a
heavy right handed neutrino for a brief moment before reverting back to
being a left-handed neutrino. This results in the very small
observed Majorana mass for the left-handed neutrino, its smallness
being associated with the heaviness of the right-handed neutrino,
rather like a flea and an elephant perched on either end of a see-saw.

An alternative explanation of small neutrino masses comes from the
concept of extra dimensions beyond the three that we
know of, motivated by theoretical attempts to extend the Standard Model
to include gravity. The extra dimensions are 'rolled up' on a very
small scale so that they are not normally observable. It has been
suggested that right-handed neutrinos (but not the rest of the Standard
Model particles) experience one or more of these extra dimensions. The
right handed neutrinos then only spend part of their time in our world,
leading to apparently small neutrino masses.

Cosmology today presents two major puzzles: why there is an
excess of matter over antimatter in the Universe; and what is the major
matter constituent of the Universe? Massive neutrinos may hold
important clues.

Matter and antimatter would have been created in equal amounts in the
Big Bang but all we see is a small amount of excess matter. The see-saw
mechanism allows for a novel resolution to this puzzle. The idea, due
to Masataka Fukugita and Tsutomu Yanagida of Tokyo University, is that
when the Universe was very hot, just after the Big Bang, the heavy
right-handed neutrinos would have been produced, and could have decayed
preferentially into leptons rather than antileptons.
The see-saw mechanism therefore opens up the possibility of generating the
baryon asymmetry of the universe via ``leptogenesis''. 
This process
requires CP violation for neutrinos which could be studied experimentally by
firing a very intense neutrino beam right through the Earth and
detecting it with a huge neutrino detector when it emerges.

Studies of the movements of galaxies and galaxy clusters suggest that
at least 90 per cent of the mass of the Universe is made of
unknown dark matter.
Cosmology is sensitive to the absolute values of neutrino masses,
in the form of relic hot dark matter.
Neutrinos could constitute anything
from 0.1 to 2 per cent of the mass of the Universe,
corresponding to the heaviest
neutrino being in the mass range 0.05 to about 0.23 electronvolt.
Neutrinos any heavier than this
would lead to galaxies being less clumped than actually
observed by the recent 2dF Galaxy Redshift Survey.
This illustrates the breathtaking rate at which neutrino
physics continues to advance.

\subsection{About this review}

There are many good reviews already in the literature, for example
\cite{King:2002gx},\cite{Altarelli:2002hx}, \cite{Mohapatra:2003qw},
\cite{bilenky}.
Three possible ways to extend the Standard Model
in order to account for the neutrino mass spectrum are
the see-saw mechanism \cite{seesaw},
extra dimensions \cite{Arkani-Hamed:1998vp},
and R-parity violating supersymmetry \cite{Drees:1997id}.
In this review we focus on theoretical
approaches to understanding neutrino masses and mixings
in the framework of the {\em see-saw mechanism}, 
assuming three active neutrinos.
The goal of such models is to account for two large mixing angles,
one small mixing angle, and a pattern of neutrino masses consistent
with observation. We are now in the unique position in the history of
neutrino physics of knowing not only that neutrino mass is real,
and hence the Standard Model at least in its minimal formulation
is incomplete, but also we have a unique solution to the solar
neutrino problem in the form of the large mixing angle
solution. In this sense a review of neutrino mass models
is very timely, since it has only been within the last year that
the solar solution has been uniquely specified.
That combined with the atmospheric oscillation data severely
constrains theoretical models, and in fact rules out many
possibilities which predicted other solar solutions.
Of course many possibilities remain, and we shall mention several
of them here. However this review is not supposed to be
an encyclopaedic review of all possible models, but instead
a review of useful approaches and techniques that may be applied to
constructing different classes of models.

We give a strong emphasis to classes of models where
the two large mixing angles can arise naturally and consistently
with a neutrino mass hierarchy, and although we classify all possible
neutrino mass structures, we do not spend much time on those
structures which apparently require a high degree of fine-tuning
to achieve. We show that if one of the right-handed neutrinos
contributes dominantly in the see-saw mechanism to the heaviest
neutrino mass, and a second right-handed neutrino contributes
dominantly to the second heaviest neutrino mass, then large
atmospheric and solar mixing angles may be interpreted
as simple ratios of Yukawa couplings. We refer to this
natural mechanism as sequential dominance. Although sequential
dominance looks very specialised it is not: either the
right-handed neutrinos contribute equally via the see-saw mechanism
to neutrino masses, or some of them contribute more than others.
The second possibility corresponds to sequential dominance, and
allows a very natural and intuitively appealing explanation of
the neutrino mass hierarchy with two large mixing angles.
Sequential dominance is not a model, it is a mechanism
in search of a model. The conditions for sequential
dominance, such as ratios of Yukawa couplings being of order unity
for large mixing angles, and the required pattern of right-handed
neutrino masses are put in by hand and require further
theoretical input. This motivates models with extra symmetry
such as unified models and models with family symmetry, which we
briefly review. There are a huge number of proposals in the literature,
but assuming sequential dominance, and the important clues
provided by quark masses and mixing angles, severely constrains the
possible successful models. We discuss one particularly
successful model as an example, but of course there may be others,
but maybe not so many as may be thought at first.

The layout of the remainder of the review article is as follows.
In section 2 we introduce and review the current status of
neutrino masses and mixing angles. We also parametrise the
neutrino mixing matrix in two different ways, whose equivalence
is discussed in an Appendix. We show how it may be constructed
theoretically from the underlying mass matrices,
and then show how the proceedure may be driven the other
way to derive the form of the neutrino mass matrix
whose leading order forms may be classified.
The properties of the matix corresponding to hierarchical
neutrino masses are explored. Section 3 introduces the see-saw
mechanism, which is central to this review, in both its
simplest version, and including more complicated versions.
In section 4 we show how the see-saw mechanism may be applied
to the hierarchical case in a very natural way using sequential
dominance, discuss different types of sequential dominance,
and a link with leptogenesis.
We also discuss an alternative early approach to obtaining
large mixing angles from the see-saw mechanism, and show
that it is quite different from sequential dominance.
Section 5 incorporates the see-saw mechanism into the Standard Model,
and its Supersymmetric version, where it leads to
lepton flavour violation. In section 6 we go beyond these
minimal extensions of the Standard Model, or its supersymmetric
version, and show how the see-saw mechanism motivates ideas
of Unification and Family Symmetry, and briefly review the
huge literature that has grown up around such approaches,
before focussing on two particular models based on
$SO(10)$ grand unification
and either $U(1)$ or $SU(3)$ family symmetry.
Section 7 concludes the review.

We also present extensive appendices which deal with more
technical issues, but which may provide useful
model building tools.
Appendix A
proves the equivalence between different parametrisations
of the neutrino mixing matrix and gives a useful dictionary.
Appendix B gives the full three family
neutrino oscillation formula (in vacuum).
Appendix C derives the formula given in the text for
charged lepton contributions to the neutrino mixing matrix.
Finally Appendix D discusses in detail how to diagonalise
different kinds of mass matrices involving two large mixing angles
analytically to leading order in the small mixing angle.

\section{Neutrino Masses and Mixing Angles}

The history of neutrino oscillations dates back to the work of
Pontecorvo who in 1957 \cite{Pontecorvo:1957qd}
proposed $\nu \rightarrow \bar{\nu}$ oscillations in 
analogy with $K \rightarrow \bar{K}$ oscillations,
described as the mixing of two Majorana neutrinos.
Pontecorvo was the first to
realise that what we call the ``electron neutrino'' for example is
really a linear combination of mass eigenstate neutrinos, and that
this feature could lead to neutrino oscillations
of the kind $\nu_e \rightarrow \nu_{\mu}$
\cite{Pontecorvo:fh}. Later on MSW proposed that such
neutrino oscillations could be resonantly enhanced in the Sun
\cite{MSW}. The present section introduces the basic formalism
of neutrino masses and mixing angles, gives an up-to-date
summary of the current experimental status of this fast moving field,
and discusses future experimental prospects.
Later in this section we also discuss some more theoretical aspects
such as charged lepton contributions to neutrino mixing angles,
and the neutrino mass matrix.

\subsection{Two state atmospheric neutrino mixing}

\begin{figure}[t]
\includegraphics[width=0.76\textwidth]{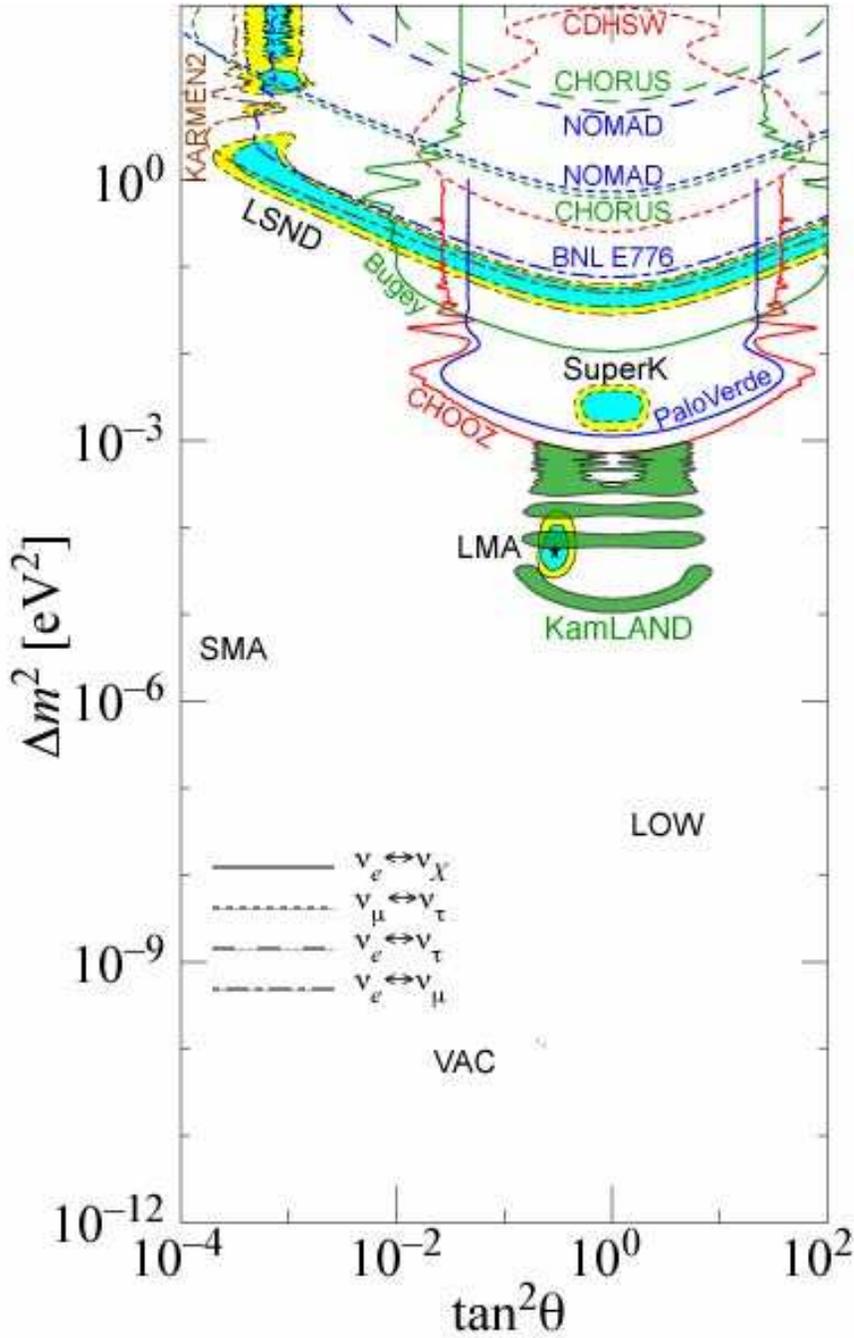}
\vspace*{-4mm}
    \caption{Summary of the currently allowed regions from a global analysis
of atmospheric and solar neutrino experiments including first results
from KamLAND (from H.Murayama's web site
http://hitoshi.berkeley.edu/neutrino/.)}
\label{fig0}
\vspace*{-2mm}
\end{figure}

In 1998 the Super-Kamiokande experiment published a
paper \cite{SK} 
which represents a watershed in the history of neutrino physics.
Super-Kamiokande measured the number of
electron and muon neutrinos that arrive at the Earth's surface as a
result of cosmic ray interactions in the upper atmosphere, which are
referred to as ``atmospheric neutrinos''. While the number and and
angular distribution of electron neutrinos is as expected,
Super-Kamiokande showed that the number of muon neutrinos is
significantly smaller than expected and that the flux of muon
neutrinos exhibits a strong dependence on the zenith angle. These
observations gave compelling evidence that muon neutrinos undergo
flavour oscillations and this in turn implies that at least one
neutrino flavour has a non-zero mass. The standard interpretation is
that muon neutrinos are oscillating into tau neutrinos.

Current atmospheric neutrino oscillation data are well described
by simple two-state mixing
\begin{equation}
\left(\begin{array}{c} \nu_\mu \\ \nu_\tau \end{array} \\ \right)=
\left(\begin{array}{cc}
 \cos \theta_{23} & \sin \theta_{23} \\
 -\sin \theta_{23} & \cos \theta_{23} \\
\end{array}\right)
\left(\begin{array}{c}
\nu_2 \\ \nu_3 \end{array} \\ \right)
\; ,
\end{equation}
and the two-state Probability oscillation formula
\be
P(\nu_{\mu }\rightarrow \nu_{\tau})=\sin^22\theta_{23}
\sin^2(1.27\Delta m_{32}^2{L}/{E})
\ee
where
\begin{equation}
\Delta m^2_{ij} \equiv m^2_i - m^2_j \;
\end{equation}
and $m_i$ are the physical neutrino mass eigenvalues associated
with the mass eigenstates $\nu_i$.
$\Delta m_{32}^2$ is in units of eV$^2$, the baseline $L$ is in km and
the beam energy $E$ is in GeV.

The atmospheric data is statistically dominated by the
Super-Kamiokande results and the latest reported data sample 
as of the time of writing leads
to:
\begin{equation}
\sin^22\theta_{23}>0.92, \ \
1.3\times 10^{-3} eV^2 < |\Delta m_{32}^2|
<3.0\times 10^{-3}eV^2 (90\% CL)
\end{equation}
The Super-Kamiokande  region is shown in Fig.\ref{fig0}.
The atmospheric neutrino data is thus consistent with
maximal $\nu_{\mu}- \nu_{\tau}$ neutrino mixing
$\theta_{23} \approx \pi/4$
with $|\Delta m_{32}^2|\approx 2.5\times 10^{-3}{\rm eV}^2$
and the sign of $\Delta m_{32}^2$ undetermined.
The maximal mixing angle means that
we identify the heavy atmospheric neutrino of mass $m_3$ as being approximately
\beq
\nu_3\approx \frac{\nu_{\mu}+\nu_{\tau}}{\sqrt{2}}
\eeq
and in addition there is a lighter orthogonal combination of mass $m_2$,
\be
\nu_2\approx \frac{\nu_{\mu}-\nu_{\tau}}{\sqrt{2}}.
\ee

\subsection{Three family solar neutrino mixing}

Super-Kamiokande is also sensitive to the electron neutrinos arriving
from the Sun, the ``solar neutrinos'', and has independently confirmed
the reported deficit of such solar neutrinos long reported by other
experiments.  For example Davis's Homestake Chlorine experiment
which began data taking in 1970 consists of 615 tons of
tetrachloroethylene, and uses radiochemical techniques to determine
the Ar37 production rate.  More recently the SAGE and Gallex
experiments contain large amounts of Ga71 which is converted to Ge71
by low energy electron neutrinos arising from the dominant pp reaction
in the Sun.  The combined data from these and other experiments
implies an energy dependent suppression of solar neutrinos which can
be interpreted as due to flavour oscillations. Taken together with the
atmospheric data, this requires that a second neutrino flavour has a
non-zero mass. The standard interpretation is that the electron
neutrinos $\nu_e$ oscillate into the light linear combination
$\nu_2\approx \frac{\nu_{\mu}-\nu_{\tau}}{\sqrt{2}}$.

SNO measurements of charged current (CC) reaction on deuterium is
sensitive exclusively to $\nu_e$'s, while the elastic scattering (ES)
off electrons also has a small sensitivity to $\nu_{\mu}$'s and
$\nu_{\tau}$'s.  The CC ratio is significantly smaller than the ES
ratio. This immediately disfavours oscillations of $\nu_e's$ to
sterile neutrinos which would lead to a diminished flux of electron
neutrinos, but equal CC and ES ratios.  On the other hand the
different ratios are consistent with oscillations of $\nu_e$'s to
active neutrinos $\nu_{\mu}$'s and $\nu_{\tau}$'s since this would
lead to a larger ES rate since this has a neutral current
component. The SNO analysis is nicely consistent with both the
hypothesis that electron neutrinos from the Sun oscillate into other
active flavours, and with the Standard Solar Model prediction. The
latest results from SNO including the data taken with salt inserted
into the detector to boost the efficiency of detecting the neutral
current events \cite{Ahmed:2003kj}, 
strongly favour the large mixing angle (LMA) MSW
solution.  In other words there is no longer any solar neutrino
problem: we have instead solar neutrino mass!

The minimal neutrino sector required to account for the
atmospheric and solar neutrino oscillation data thus consists of
three light physical neutrinos with left-handed flavour eigenstates,
$\nu_e$, $\nu_\mu$, and $\nu_\tau$, defined to be those states
that share the same electroweak doublet as the left-handed
charged lepton mass eigenstates.
Within the framework of three--neutrino oscillations,
the neutrino flavor eigenstates $\nu_e$, $\nu_\mu$, and $\nu_\tau$ are
related to the neutrino mass eigenstates $\nu_1$, $\nu_2$, and $\nu_3$
with mass $m_1$, $m_2$, and $m_3$, respectively, by a $3\times3$
unitary matrix called the lepton mixing matrix $U$
\cite{MNS}
\begin{equation}
\left(\begin{array}{c} \nu_e \\ \nu_\mu \\ \nu_\tau \end{array} \\ \right)=
\left(\begin{array}{ccc}
U_{e1} & U_{e2} & U_{e3} \\
U_{\mu1} & U_{\mu2} & U_{\mu3} \\
U_{\tau1} & U_{\tau2} & U_{\tau3} \\
\end{array}\right)
\left(\begin{array}{c} \nu_1 \\ \nu_2 \\ \nu_3 \end{array} \\ \right)
\; .
\label{MNS0}
\end{equation}

Assuming the light neutrinos are Majorana,
$U$ can be parameterized in terms of three mixing angles
$\theta_{ij}$ and three complex phases $\delta_{ij}$.
A unitary matrix has six phases but three of them are removed
by the phase symmetry of the charged lepton Dirac masses.
Since the neutrino masses are Majorana there is no additional
phase symmetry associated with them, unlike the case of quark
mixing where a further two phases may be removed.

If we suppose to begin with that the phases are zero, then
the lepton mixing matrix may be parametrised by
a product of three Euler rotations,
\begin{equation}
U=R_{23}R_{13}R_{12}
\label{euler}
\end{equation}
where
\begin{equation}
R_{23}=
\left(\begin{array}{ccc}
1 & 0 & 0 \\
0 & c_{23} & s_{23} \\
0 & -s_{23} & c_{23} \\
\end{array}\right)
\end{equation}
\begin{equation}
R_{13}=
\left(\begin{array}{ccc}
c_{13} & 0 & s_{13} \\
0 & 1 & 0 \\
-s_{13} & 0 & c_{13} \\
\end{array}\right)
\end{equation}
\begin{equation}
R_{12}=
\left(\begin{array}{ccc}
c_{12} & s_{12} & 0 \\
-s_{12} & c_{12} & 0\\
0 & 0 & 1 \\
\end{array}\right)
\end{equation}
where $c_{ij} = \cos\theta_{ij}$ and $s_{ij} = \sin\theta_{ij}$.
Note that the allowed range of the angles is
$0\leq \theta_{ij} \leq \pi/2$.

CHOOZ is a reactor experiment that falied to see any signal of
neutrino oscillations over the Super-Kamiokande mass range.  CHOOZ
data from $\bar{\nu}_{e}\rightarrow \bar{\nu}_{e}$ disappearance not
being observed provides a significant constraint on $\theta_{13}$ over
the Super-Kamiokande (SK) prefered range of $\Delta m_{32}^2$
\cite{Apollonio:1999ae}:
\begin{equation}
\sin^22\theta_{13}<0.1-0.3
\end{equation}
The CHOOZ experiment therefore limits $\theta_{13} \simlt 0.2$
over the favoured atmospheric range, as shown in Fig.\ref{fig0}.

KamLAND is a more powerful reactor experiment that measures
$\bar{\nu}_e$'s produced by surrounding nuclear
reactors. KamLAND has already seen
a signal of neutrino oscillations over the LMA MSW mass range,
and has recently confirmed the LMA MSW region ``in the laboratory''
\cite{Eguchi:2002dm}.
KamLAND and SNO results when combined
with other solar neutrino data especially that of Super-Kamiokande
uniquely specify the large mixing angle (LMA) MSW \cite{MSW} solar solution
with three active light neutrino states, a large solar angle
\beq
\tan^2 \theta_{12} \approx 0.4, \ \
\Delta m_{21}^2\approx 7\times 10^{-5}{\rm eV}^2.
\eeq
according to the most recent global fits \cite{GoPe} performed 
after the SNO salt data \cite{Ahmed:2003kj}.
KamLAND has thus not only confirmed solar neutrino oscillations,
but have also uniquely specified the large mixing angle (LMA) solar
solution, heralding a new era of precision neutrino physics.

The currently regions of atmospheric and solar parameter space
allowed by all experiments are depicted in
Figure \ref{fig0}.
\footnote{For more detailed most up to date plots of the LMA MSW region see
\cite{GoPe}.}
In Figure \ref{fig0} the atmospheric and
LMA MSW solar regions are clearly shown as elliptical regions,
with the SMA, LOW and VAC regions now having disappeared.
One of the KamLAND rate plus shape allowed regions shown in Figure \ref{fig0}
intersects the central part of the LMA ellipse near the best fit
LMA point as determined from the solar data alone, thereby confirming
the LMA MSM solution.

\subsection{Summary of neutrino mixing angles and mass patterns}

The current experimental situation is summarized by
$\theta_{23}\approx \pi/4$, $\theta_{13}\leq 0.2$, and
$\theta_{12}\approx \pi/6$.  Ignoring phases, the relation between the
neutrino flavor eigenstates $\nu_e$, $\nu_\mu$, and $\nu_\tau$ and the
neutrino mass eigenstates $\nu_1$, $\nu_2$, and $\nu_3$ is just given
as a product of three Euler rotations in Eq.\ref{euler} as depicted in
Fig.\ref{fig2}.
%
\FIGURE[h]{
\label{fig2}
 \unitlength=1in
\begin{picture}(4.5,2.7)
\put(-0.2,0){\epsfig{file=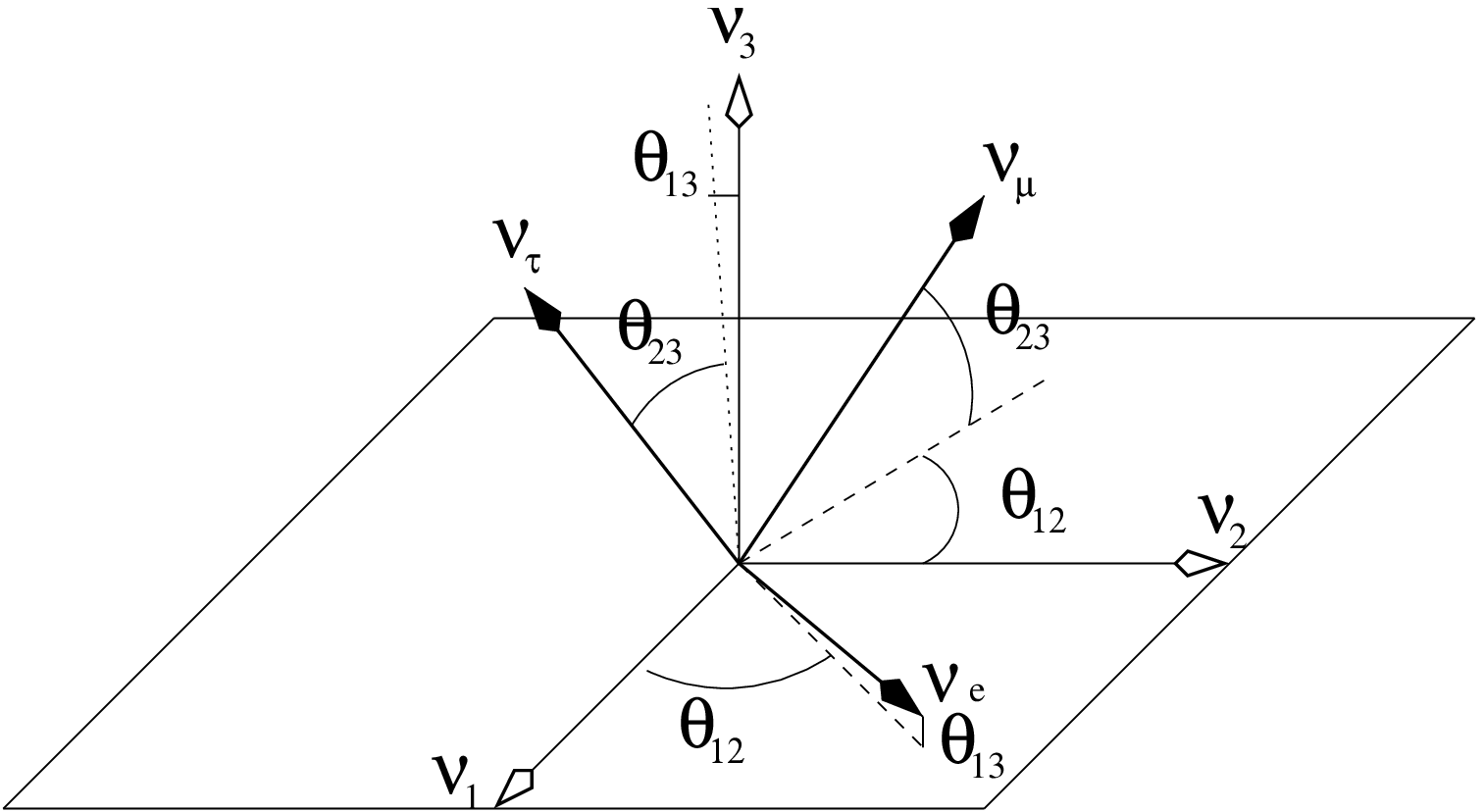, width=5in}}
\end{picture}
\caption
{The relation between
the neutrino flavor eigenstates $\nu_e$, $\nu_\mu$, and $\nu_\tau$ and
the neutrino mass eigenstates $\nu_1$, $\nu_2$, and $\nu_3$
in terms of the three mixing angles $\theta_{12}$,
$\theta_{13}$, $\theta_{23}$.
The atmospheric angle is $\theta_{23} \approx \pi/4$,
the CHOOZ angle is $\theta_{13} \simlt 0.2$, and the solar angle is
$\theta_{12} \approx \pi/6$. }
}
This corresponds to the approximate form of mixing matrix
\begin{equation}
U\approx
\left( \begin{array}{rrr}
c_{12} & s_{12} & \theta_{13}\\
-\frac{s_{12}}{\sqrt{2}} &
\frac{c_{12}}{\sqrt{2}} &
\frac{1}{\sqrt{2}}\\
\frac{s_{12}}{\sqrt{2}} &
-\frac{c_{12}}{\sqrt{2}}&
\frac{1}{\sqrt{2}}
\end{array}
\right)
\label{approxmix}
\end{equation}
where $\theta_{12}\approx \pi/6$ corresponds to  $s_{12}\approx \frac{1}{{2}}$,
$c_{12}\approx \frac{\sqrt{3}}{{2}}$.

\begin{figure}[t]
\includegraphics[width=0.96\textwidth]{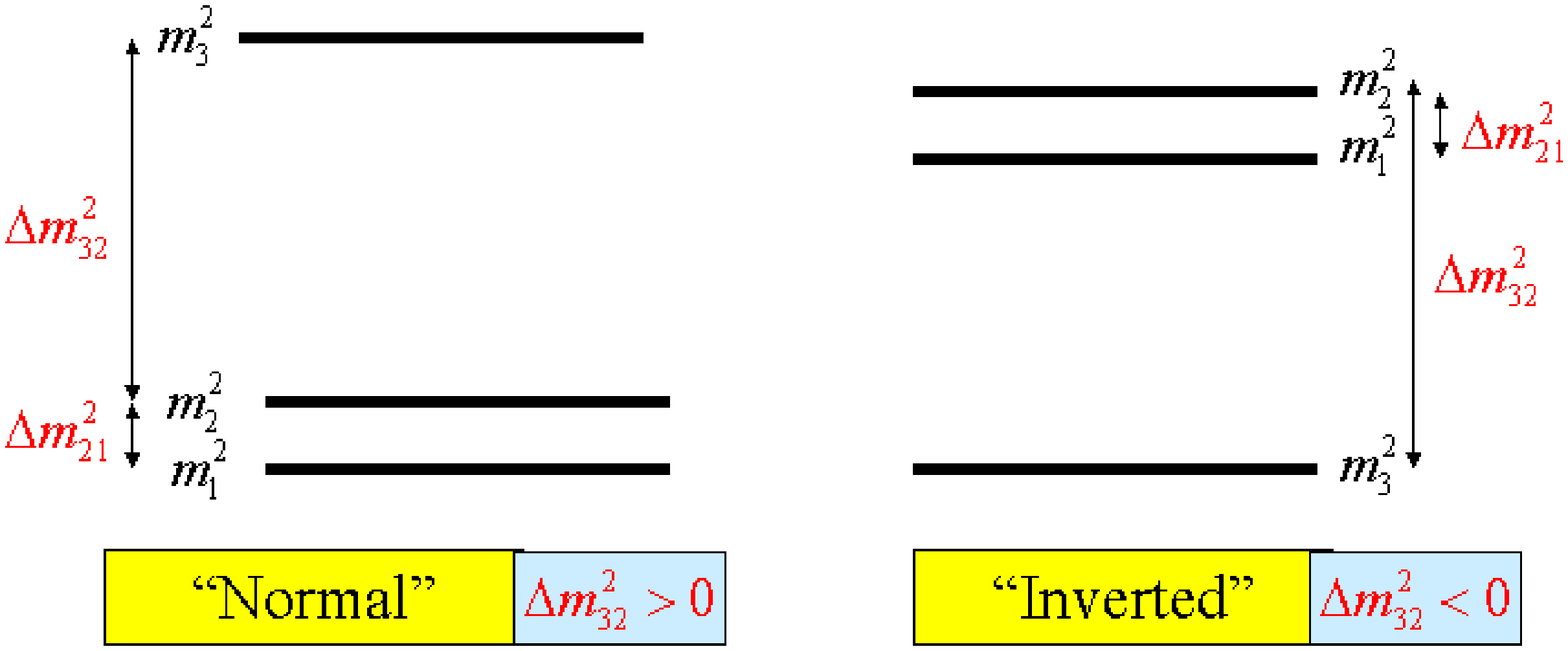}
\vspace*{-4mm}
    \caption{Alternative neutrino mass
patterns that are consistent with neutrino
oscillation explanations of the atmospheric and solar data.
The absolute scale of neutrino masses is not
fixed by oscillation data
and the lightest neutrino mass may vary from 0.0-0.23 eV.}
\label{fig1}
\vspace*{-2mm}
\end{figure}

It is clear that neutrino oscillations, which
only depend on $\Delta m_{ij}^2\equiv m_i^2-m_j^2$,
give no information about the absolute value of the neutrino mass squared
eigenvalues $m_i^2$,
and there are basically two
patterns of neutrino mass squared orderings
consistent with the atmospheric and solar data as shown in
Fig.\ref{fig1}.

\subsection{Three family neutrino mixing with phases}

Including the phases, assuming the light neutrinos are Majorana,
$U$ can be parameterized in terms of three mixing angles
$\theta_{ij}$, a Dirac phase $\delta$, together with
two Majorana phases $\beta_1,\beta_2$, as follows
\begin{equation}
U=R_{23}U_{13}R_{12}P_{12}
\end{equation}
where
\bea
U_{13} & = &
\left(\begin{array}{ccc}
c_{13} & 0 & s_{13}e^{-i\delta} \\
0 & 1 & 0 \\
-s_{13}e^{i\delta} & 0 & c_{13} \\
\end{array}\right), \\
P_{12}  & = &
\left(\begin{array}{ccc}
e^{i\beta_1} & 0 & 0 \\
0 & e^{i\beta_2} & 0\\
0 & 0 & 1 \\
\end{array}\right)
\label{MNS}
\eea
where $c_{ij} = \cos\theta_{ij}$ and $s_{ij} = \sin\theta_{ij}$,
and $R_{23},R_{12}$ were defined below Eq.\ref{euler}.

Alternatively the lepton mixing matrix
may be expressed as a product of three complex Euler rotations,
\begin{equation}
U=U_{23}U_{13}U_{12}
\end{equation}
where
\begin{equation}
U_{23}=
\left(\begin{array}{ccc}
1 & 0 & 0 \\
0 & c_{23} & s_{23}e^{-i\delta_{23}} \\
0 & -s_{23}e^{i\delta_{23}} & c_{23} \\
\end{array}\right)
\end{equation}

\begin{equation}
U_{13}=
\left(\begin{array}{ccc}
c_{13} & 0 & s_{13}e^{-i\delta_{13}} \\
0 & 1 & 0 \\
-s_{13}e^{i\delta_{13}} & 0 & c_{13} \\
\end{array}\right)
\end{equation}

\begin{equation}
U_{12}=
\left(\begin{array}{ccc}
c_{12} & s_{12}e^{-i\delta_{12}} & 0 \\
-s_{12}e^{i\delta_{12}} & c_{12} & 0\\
0 & 0 & 1 \\
\end{array}\right)
\end{equation}

The equivalence of different parametrisations of the 
lepton mixing matrix, and
the relation between them is discussed in Appendix A.

Three family
oscillation probabilities depend upon the time--of--flight (and hence the
baseline $L$), the $\Delta m^2_{ij}$, and $U$ (and hence
$\theta_{12}, \theta_{23}, \theta_{13}$, and $\delta$).
Three-state neutrino mixing is
discussed in Appendix B.
Since we have assumed that the neutrinos are Majorana,
there are two extra phases, but only one combination
$\delta = \delta_{13}-\delta_{23}-\delta_{12}$
affects oscillations.
If the neutrinos are Dirac, then the phases $\beta_1=\beta_2=0$,
but the phase $\delta$ remains.

\subsection{The LSND signal}

The signal of another independent mass
splitting from the LSND accelerator experiment \cite{Athanassopoulos:1997pv}
would either require a further light neutrino state with no weak
interactions (a so-called ``sterile neutrino'') or some other
non-standard physics.
This effect has not been confirmed by a similar
experiment KARMEN \cite{Eitel:2000by}, and currently a decisive
experiment MiniBooNE is underway to decide the issue.
In Figure \ref{fig0} the LSND signal region is indicated,
together with the KARMEN excluded region.

\subsection{Future experimental prospects}

Further experimental progress from SNO and
KamLAND will consist of pinning down LMA MSW parameters to high accuracy.
Neutrino physics has, now entered the precision era.
Future neutrino oscillation experiments,
will give accurate information about the mass squared splittings
$\Delta m_{ij}^2\equiv m_i^2-m_j^2$, mixing angles, and
CP violating phase.
In the near future much better solar neutrino measurements
will be available as KamLAND, SNO and Borexino furnish us with new and
better data.
The K2K long baseline (LBL) experiment from KEK to Super-Kamiokande
has recently reported results in its phases I and
II, which cover the atmospheric region and support the
Super-Kamiokande results.
In the longer term LBL experiments such as
MINOS and eventually the  CERN to Gran Sasso
experiments will given more accurate determinations
of the atmospheric parameters, eventually to 10\%.
J-PARC will be an  ``off-axis superbeam'' over a LBL of 295 km
to Super-Kamiokande due to start in 2008.
Its first goal is to measure $\theta_{13}$ or set a limit
on it of about 0.05 (as compared to the CHOOZ limit 
on $\theta_{13}$ of about 0.2).
Interestingly MINOS over a LBL of 735 km is more sensitive than
J-PARC to matter effects, so there should be some interesting
complementarity between these two experiments,
which could for example allow the sign of $\Delta m^2_{32}$ 
to be determined.
The ultimate goal of oscillation experiments however is to measure
the CP violating phase $\delta$. An upgraded J-PARC
with a 4MW proton driver and a 1 megaton Hyper-Kamiokande
detector, or some sort of Neutrino Factory based on
muon storage rings would seem to be required for this purpose
\cite{Autin:2003xk}.

Oscillation experiments are not capable of telling
us anything about the absolute scale of neutrino masses.
The Tritium beta decay experiment KATRIN will tell us about the
absolute scale of neutrino mass down to about 0.35 eV.
The neutrinoless double beta decay experiment GENIUS will
probe the Majorana nature of the electron neutrino down to about 0.01 eV
\cite{Klapdor-Kleingrothaus:1999hk}.
Recent results
from the 2dF galaxy redshift survey and WMAP, when combined with oscillation
data, give the strong limit on the absolute mass of each neutrino species of
about 0.23 eV \cite{Elgaroy:2002bi,Pierce:2003uh}.
Turning to astrophysics, a galactic supernova could give valuable
information about neutrino masses \cite{Dighe:2003be}.
In future detection of energetic
neutrinos from gamma ray bursts (GRBs), by neutrino telescopes
such as ANTARES or ICECUBE could also
provide important astrophysical information, and may provide another
means of probing neutrino mass, and even quantum gravity
\cite{Choubey:2002bh}.

\subsection{Charged lepton contributions to neutrino masses and mixing angles}

Although we refer to neutrino masses and mixing angles, it is worth
pointing out that in general they could originate, at least in part,
from the charged lepton sector.  The (low energy) Lagrangian involving
the charged lepton and neutrino mass matrices \bea {\cal
L}_{mass}&=-&(\bar{e}_{L1} \bar{e}_{L2} \bar{e}_{L3})
m^E_{LR}({e}_{R1} {e}_{R2} {e}_{R3})^T \nonumber \\
&-&\frac{1}{2}(\bar{\nu}_{L1} \bar{\nu}_{L2} \bar{\nu}_{L3}) m_{LL}
({\nu}_{L1}^c {\nu}_{L2}^c {\nu}_{L3}^c)^T +H.c.
\label{Lmass}
\eea
where $e_{Li}$ are the three left-handed charged lepton states,
$e_{Ri}$ are the right-handed charged lepton states,
$\nu_{Li}$ are the three left-handed neutrino states,
and $\nu_{Li}^c$ are their CP conjugates.
Note that the states $\nu_{Li}$ are not the mass eigenstate
neutrinos since $m_{LL}$ is not diagonal in general.
We shall refer to the mass eigenstate neutrinos as $\nu_i$
(without the L subscript), as in Eq.\ref{MNS0}.

In general the neutrino and charged lepton masses are given by the eigenvalues
of a complex charged lepton mass matrix
$m^E_{LR}$ and a complex symmetric neutrino Majorana matrix
$m_{LL}$, obtained by diagonalising these mass
matrices,
\beq
V^{E_L}m^E_{LR}{V^{E_R}}^{\dagger}=
\left( \begin{array}{ccc}
m_e & 0 & 0    \\
0 & m_{\mu} & 0 \\
0 & 0 & m_{\tau}
\end{array}
\right)
\label{diag1}
\eeq
\beq
V^{\nu_L}m_{LL}{V^{\nu_L}}^T=
\left( \begin{array}{ccc}
m_1 & 0 & 0    \\
0 & m_2 & 0 \\
0 & 0 & m_3
\end{array}
\right)
\label{diag2}
\eeq
where $V^{E_L}$, $V^{E_R}$, $V^{\nu_L}$ are unitary tranformations
on the left-handed charged lepton fields $E_L$, right-handed charged
lepton fields $E_R$, and left-handed neutrino fields $\nu_L$
which put the mass matrices into diagonal form with real
eigenvalues.

After having diagonalised the mass matrices,
the lepton mixing matrix is then constructed by
\beq
U=V^{E_L}{V^{\nu_L}}^{\dagger}
\label{MNS2}
\eeq
A unitary three dimensional matrix has six independent phases.
As discussed in Appendix B, the freedom in the charged lepton phase
enables three of the phases to be removed from $U$
to leave three phases. Since we have assumed that the neutrinos
are Majorana, there is no further phase freedom, and the three
remaining phases are physical (unlike the case of Dirac neutrinos
where a further two phases can be removed, analagous to the
case of the CKM matrix in the quark sector).
Having constructed the lepton mixing matrix as discussed above, it may then be
parametrised as discussed in section 2. Having done this one may
then ask how much of a contribution to a particular
mixing angle or phase, comes from the neutrino sector, and
how much comes from the charged lepton sector.
The lepton mixing matrix is constructed in Eq.\ref{MNS} as a product
of a unitary matrix from the charged lepton sector $V^{E_L}$
and a unitary matrix from the neutrino sector ${V^{\nu_L}}^{\dagger}$.
Each of these unitary matrices may be parametrised by its own
mixing angles and phases analagous to the 
lepton mixing matrix parameters.
As shown in Appendix C \cite{King:2002nf} the lepton mixing matrix 
can be expanded
in terms of neutrino and charged lepton mixing angles and phases
to leading order in the charged lepton mixing angles which
are assumed to be small,
\bea
s_{23}e^{-i\delta_{23}}
& \approx &
s_{23}^{\nu_L}e^{-i\delta_{23}^{\nu_L}}
-\theta_{23}^{E_L}
c_{23}^{\nu_L}e^{-i\delta_{23}^{E_L}}
\label{chlep23}
\\
\theta_{13}e^{-i\delta_{13}}
& \approx &
\theta_{13}^{\nu_L}e^{-i\delta_{13}^{\nu_L}}
-\theta_{13}^{E_L}c_{23}^{\nu_L}e^{-i\delta_{13}^{E_L}} \nonumber \\
& + &
\theta_{12}^{E_L}s_{23}^{\nu_L}e^{i(-\delta_{23}^{\nu_L}-\delta_{12}^{E_L})}
\label{chlep13}
\\
s_{12}e^{-i\delta_{12}}
& \approx &
s_{12}^{\nu_L}e^{-i\delta_{12}^{\nu_L}}
+\theta_{23}^{E_L}s_{12}^{\nu_L}e^{-i\delta_{12}^{\nu_L}}
\nonumber \\
& + & \theta_{13}^{E_L}
c_{12}^{\nu_L}s_{23}^{\nu_L}e^{i(\delta_{23}^{\nu_L}-\delta_{13}^{E_L})}
\nonumber \\
& - & \theta_{12}^{E_L}
c_{23}^{\nu_L}c_{12}^{\nu_L}e^{-i\delta_{12}^{E_L}}
\label{chlep12}
\eea
Clearly $\theta_{13}$
receives important contributions not just from $\theta_{13}^{\nu_L}$,
but also from the charged lepton angles
$\theta_{12}^{E_L}$, and $\theta_{13}^{E_L}$.
In models where $\theta_{13}^{\nu_L}$ is
extremely small, $\theta_{13}$ may originate almost entirely from
the charged lepton sector.
Charged lepton contributions could also be important in models
where $\theta_{12}^{\nu_L}=\pi /4$, since charged lepton mixing angles
may allow consistency with the LMA MSW solution.
Such effects are important for the inverted hierarchy model
\cite{King:2002nf}.

Note that it is useful and possible to be able to diagonalise the mass matrices
analytically, at least to first order in the small 13 mixing angles,
but allowing the 23 and 12 angles to be large, while retaining
all the phases.
The proceedure for doing this is discussed for a hierarchical
general mass matrix in Appendix D.1, for a
hierarchical neutrino mass matrix in Appendix D.2, and for
an inverted hierarchical neutrino mass matrix in Appendix D.3.
The analytic results in these Appendices enable the
separate mixing angles and phases associated with each of the
unitary transformations $V^{E_L}$ and ${V^{\nu_L}}^{\dagger}$
to be obtained in many useful cases of interest.

\subsection{The neutrino mass matrix}

For many (but not all) purposes it is convenient to forget about the
division between charged lepton and neutrino mixing angles and
work in a basis where the charged lepton mass matrix is diagonal.
Then the lepton mixing 
angles and phases simply correspond to the neutrino ones.
In this special basis the mass matrix is given from Eq.\ref{diag2} and
Eq.\ref{MNS} as
\beq
m_{LL}=
U
\left( \begin{array}{ccc}
m_1 & 0 & 0    \\
0 & m_2 & 0 \\
0 & 0 & m_3
\end{array}
\right)
U^T
\label{mLLnu}
\eeq
For a given assumed form of $U$ and set of neutrino masses
$m_i$ one may use Eq.\ref{mLLnu} to ``derive'' the form of the neutrino
mass matrix $m_{LL}$, and this results in the candidate
mass matrices in Table \ref{table1} \cite{Barbieri:1998mq}.
Only the leading order forms are displayed explicitly in Table
\ref{table1}, and more accurate structures may be obtained case by
case.

\begin{table*}[htb]
{\small
\caption{Leading order low energy neutrino Majorana mass matrices
$m_{LL}$ consistent with large atmospheric and solar mixing angles,
classified according to the rate of neutrinoless double beta decay
and the pattern of neutrino masses.}
\label{table1}
\newcommand{\m}{\hphantom{$-$}}
\newcommand{\cc}[1]{\multicolumn{1}{c}{#1}}
\renewcommand{\tabcolsep}{2pc} 
\renewcommand{\arraystretch}{1.2} 
\begin{tabular}{@{}|c|c|c|}
\hline
\hline
& Type I  & Type II \\
 & Small $\beta \beta_{0\nu}$ & Large $\beta \beta_{0\nu}$ \\
\hline
\hline
 & & \\
A & $\beta \beta_{0\nu}\simlt 0.0082$ eV & \\
Normal hierarchy  & & \\
$m_1^2,m_2^2\ll m_3^2$ &
$\left(
\begin{array}{ccc}
0 & 0 & 0 \\
0 & 1 & 1\\
0 & 1 & 1 \\
\end{array}
\right)\frac{m}{2}$ & -- \\
& & \\
\hline
 & & \\
B & $\beta \beta_{0\nu}\simlt
0.0082$ eV & $\beta \beta_{0\nu}\simgt 0.0085$ eV\\
Inverted hierarchy & & \\
$m_1^2\approx m_2^2\gg m_3^2$ &
$\left(
\begin{array}{ccc}
0 & 1 & 1 \\
1 & 0 & 0\\
1 & 0 & 0 \\
\end{array}
\right)\frac{m}{\sqrt{2}}$ &
$\left(
\begin{array}{ccc}
1 & 0 & 0 \\
0 & \frac{1}{2} & \frac{1}{2}\\
0 & \frac{1}{2} & \frac{1}{2} \\
\end{array}
\right)m$
\\
& & \\
\hline
 & & \\
C &  &
$\beta \beta_{0\nu}\simgt 0.035$ eV \\
Approximate degeneracy & & diag(1,1,1)m\\
$m_1^2\approx m_2^2\approx m_3^2$ &
$\left(
\begin{array}{ccc}
0 & \frac{1}{\sqrt{2}} & \frac{1}{\sqrt{2}} \\
\frac{1}{\sqrt{2}} & \frac{1}{2} & \frac{1}{2}\\
\frac{1}{\sqrt{2}} & \frac{1}{2} & \frac{1}{2} \\
\end{array}
\right)m$ &
$\left(
\begin{array}{ccc}
1 & 0 & 0 \\
0 & 0 & 1\\
0 & 1 & 0 \\
\end{array}
\right)m$\\
& & \\
\hline
\hline
\end{tabular}\\[2pt]
}
\end{table*}

In Table \ref{table1} the mass matrices are classified into two types:

Type I - small neutrinoless double beta decay

Type II - large neutrinoless double beta decay

They are also classified into the limiting cases consistent with the
mass squared orderings in Fig.\ref{fig1}:

A - Normal hierarchy $m_1^2,m_2^2\ll m_3^2$

B - Inverted hierarchy $m_1^2 \approx m_2^2 \gg m_3^2$

C - Approximate degeneracy $m_1^2\approx m_2^2\approx m_3^2$

Thus according to our classification there is only one neutrino
mass matrix consistent with the normal neutrino mass hierarchy which
we call Type IA, corresponding to the leading order neutrino masses
of the form $m_i=(0,0,m)$. For the inverted hierarchy there are two
cases, Type IB corresponding to $m_i=(m,-m,0)$ or Type IIB
corresponding to $m_i=(m,m,0)$. For the approximate degeneracy cases
there are three cases, Type IC correponding to $m_i=(m,-m,m)$
and two examples of Type IIC corresponding to either
$m_i=(m,m,m)$ or $m_i=(m,m,-m)$.

At present experiment allows any of the matrices in Table
\ref{table1}. In future it will be possible to uniquely specify the
neutrino matrix in the following way:

1. Neutrinoless double beta effectively measures the 11 element
of the mass matrix $m_{LL}$ corresponding to
\beq
\beta \beta_{0\nu}\equiv \sum_iU_{ei}^2m_i
\eeq
and is clearly capable of resolving Type I from Type II cases
according to the bounds given in Table \ref{table1} \cite{Pascoli:2002xq}.
There has been a recent claim of a signal in neutrinoless double
beta decay correponding to $\beta \beta_{0\nu}=0.11-0.56$ eV at 95\% C.L.
\cite{Klapdor-Kleingrothaus:2001ke}.
However this claim has been criticised by two groups
\cite{Feruglio:2002af}, \cite{Aalseth:2002dt} and in turn this
criticism has been refuted \cite{Klapdor-Kleingrothaus:2002kf}.
Since the Heidelberg-Moscow experiment has almost reached its full
sensitivity, we may have to wait for a next generation experiment
such as GENIUS \cite{Klapdor-Kleingrothaus:1999hk} which is capable
of pushing down the sensitivity to 0.01 eV to resolve this question.

2. A neutrino factory will measure the sign of $\Delta m_{32}^2$
and resolve A from B.

3. Tritium beta decay experiments are sensitive to C
since they
measure the ``electron neutrino mass'' defined by
\beq
|m_{\nu_e}|\equiv \sum_i|U_{ei}|^2|m_i|.
\eeq
For example the KATRIN \cite{Osipowicz:2001sq}
experiment has a proposed sensitivity of
0.35 eV. As already mentioned the galaxy power spectrum combined with
solar and atmospheric oscillation data already limits each
neutrino mass to be less than about 0.23 eV, and this limit is also
expected to improve in the future. Also it is worth mentioning that
in future it may be possible to measure neutrino masses from
gamma ray bursts using time of flight techniques in principle down to
0.001 eV \cite{Choubey:2002bh}.

Type IIB and C involve small fractional
mass splittings $|\Delta m_{ij}^2| \ll m^2$ which are unstable
under radiative corrections \cite{Ellis:1999my}, and even the most
natural Type IC case is difficult to implement
\cite{Barbieri:1999km},\cite{Chankowski:2001fp}.
Types IA and IB seem to be the most natural cases.

Consider the case of full neutrino mass hierarchy
$m_3\gg m_2\gg m_1 \approx 0$, which is a special case of Type IA,
where in this case $m_3 \sim \sqrt{|\Delta m_{32}^2|}\sim 5.10^{-2}$ eV and
$m_2 \sim \sqrt{|\Delta m_{21}^2|}\sim 7.10^{-3}$ eV.
From Eqs.\ref{approxmix},\ref{mLLnu} we find
the symmetric mass matrix,
\begin{equation}
m_{LL}\approx
\left( \begin{array}{ccc}
m_2s_{12}^2 &
\frac{1}{\sqrt{2}}(m_2s_{12}c_{12}+m_3\theta_{13}) &
-\frac{1}{\sqrt{2}}(m_2s_{12}c_{12}-m_3\theta_{13}) \\
. &
\frac{1}{{2}}(m_3+ m_2c^2_{12})&
\frac{1}{{2}}(m_3- m_2c^2_{12}) \\
. & . & \frac{1}{{2}}(m_3+ m_2c^2_{12})
\end{array}
\right)
\label{approxmLL}
\end{equation}
neglecting terms like $m_2\theta_{13}$.
Clearly this expression reduces to the leading Type IA form with $m=m_3$
in the approximation that $m_2$ and $\theta_{13}$ are
neglected. However the more exact
expression in Eq.\ref{approxmLL} shows that the required form of
$m_{LL}$ should have a very definite detailed structure,
which goes beyond the leading approximation in Table \ref{table1}.
For example the requirement $m_2 \ll m_3$ implies that
the sub-determinant of the mass matrix $m_{LL}$ is small:
\beq
det \left(
\begin{array}{cc}
m_{22} & m_{23}\\
m_{23} & m_{33} \\
\end{array}
\right)\ll m_3^2.
\label{det}
\eeq
This requirement in Eq.\ref{det}
is satisfied by Eq.\ref{approxmLL}, as may be readily
seen, and this condition must be reproduced in a natural way (without
fine-tuning) by any successful theory.

\section{The See-Saw Mechanism}
There are several different kinds of see-saw mechanism in the
literature. In this review we shall focus on the
simplest Type I see-saw mechanism, which we shall introduce below.
However for completeness we shall also discuss the type II see-saw
mechanism and the double see-saw mechanism.

\subsection{Type I See-Saw}
Before discussing the see-saw mechanism it is worth first reviewing
the different types of neutrino mass that are possible. So far we
have been assuming that neutrino masses are Majorana masses of the form
\beq
m_{LL}\overline{\nu_L}\nu_L^c
\label{mLL}
\eeq
where $\nu_L$ is a left-handed neutrino field and $\nu_L^c$ is
the CP conjugate of a left-handed neutrino field, in other words
a right-handed antineutrino field. Such Majorana masses are possible
to since both the neutrino and the antineutrino
are electrically neutral and so
Majorana masses are not forbidden by electric charge conservation.
For this reason a Majorana mass for the electron would
be strictly forbidden. However such Majorana neutrino masses
violate lepton number conservation, and in the standard model,
assuming only Higgs doublets are present, are
forbidden at the renormalisable level by gauge invariance.
The idea of the simplest version of the see-saw mechanism is to assume
that such terms are zero to begin with, but are generated effectively,
after right-handed neutrinos are introduced \cite{seesaw}.

If we introduce right-handed neutrino fields then there are two sorts
of additional neutrino mass terms that are possible. There are
additional Majorana masses of the form
\beq
M_{RR}\overline{\nu_R}\nu_R^c
\label{MRR}
\eeq
where $\nu_R$ is a right-handed neutrino field and $\nu_R^c$ is
the CP conjugate of a right-handed neutrino field, in other words
a left-handed antineutrino field. In addition there are
Dirac masses of the form
\beq
m_{LR}\overline{\nu_L}\nu_R.
\label{mLR}
\eeq
Such Dirac mass terms conserve lepton number, and are not forbidden
by electric charge conservation even for the charged leptons and
quarks.

Once this is done then the types of neutrino mass discussed
in Eqs.\ref{MRR},\ref{mLR} (but not Eq.\ref{mLL} since we
assume no Higgs triplets)
are permitted, and we have the mass matrix
\begin{equation}
\left(\begin{array}{cc} \overline{\nu_L} & \overline{\nu^c_R}
\end{array} \\ \right)
\left(\begin{array}{cc}
0 & m_{LR}\\
m_{LR}^T & M_{RR} \\
\end{array}\right)
\left(\begin{array}{c} \nu_L^c \\ \nu_R \end{array} \\ \right)
\label{matrix}
\end{equation}
Since the right-handed neutrinos are electroweak singlets
the Majorana masses of the right-handed neutrinos $M_{RR}$
may be orders of magnitude larger than the electroweak
scale. In the approximation that $M_{RR}\gg m_{LR}$
the matrix in Eq.\ref{matrix} may be diagonalised to
yield effective Majorana masses of the type in Eq.\ref{mLL},
\beq
m_{LL}=-m_{LR}M_{RR}^{-1}m_{LR}^T.
\label{seesaw}
\eeq
The effective left-handed Majorana masses $m_{LL}$ are naturally
suppressed by the heavy scale $M_{RR}$.
In a one family example if we take $m_{LR}=M_W$ and $M_{RR}=M_{GUT}$
then we find $m_{LL}\sim 10^{-3}$ eV which looks good for solar
neutrinos.
Atmospheric neutrino masses would require
a right-handed neutrino with a mass below the GUT scale.

The see-saw mechanism can be formally derived from the following Lagrangian
\be
{\cal L}=-\bar{\nu}_Lm_{LR}\nu_R-\frac{1}{2}\nu_R^TM_{RR}\nu_R+H.c.
\ee
where $\nu_L$ represents left-handed neutrino fields
(arising from electroweak doublets), $\nu_R$ represents right-handed
neutrino fields (arising from electroweak singlets), in a matrix
notation where the $m_{LR}$ matrix elements
are typically of order the charged lepton masses,
while the $M_{RR}$ matrix elements may be much larger than the
electroweak scale, and maybe up to the Planck scale.
The number of right-handed neutrinos is not fixed, but the number
of left-handed neutrinos is equal to three.
Below the mass scale of the right-handed neutrinos we can
integrate them out using the equations of motion
\be
\frac{d{\cal L}}{\nu_R}=0
\ee
which gives
\be
\nu_R^T=-\bar{\nu}_Lm_{LR}M_{RR}^{-1},\ \
\nu_R=-M_{RR}^{-1}m_{LR}^T\bar{\nu}_L^T
\ee
Substituting back into the original Lagrangian we find
\be
{\cal L}=-\frac{1}{2}\bar{\nu}_Lm_{LL}\nu_L^c+H.c
\ee
with $m_{LL}$ as in Eq.\ref{seesaw}.

\subsection{Type II See-Saw and Double See-Saw}

The version of the see-saw mechanism discussed so far
is sometimes called the Type I see-saw mechanism. It is the simplest
version of the see-saw mechanism, and can be thought of as resulting
from integrating out heavy right-handed neutrinos to produce
the effective dimension 5 neutrino mass
operator
\beq
-\frac{1}{2}H_uL^T\kappa H_uL
\label{dim5}
\eeq
where
\beq
\kappa = Y^{\nu}_{LR}M_{RR}^{-1}{Y^{\nu}_{LR}}^T.
\label{kappa}
\eeq
One might wonder if it is possible to simply write down an operator
by hand similar to Eq.\ref{dim5}, without worrying about its origin.
In fact, historically, such an operator was introduced suppressed by
the Planck scale (rather than the right-handed neutrino mass scales)
by Weinberg in order to account for small neutrino
masses \cite{Weinberg:sa}. The problem is that such a Planck scale
suppressed operator would lead to neutrino masses of order
$10^{-5}{\rm eV}$ which are too small to account for $m_2$ or $m_3$
(though they could account for $m_1$). To account for
$m_3$ requires dimension 5 operators suppressed by a mass scale
of order $3 \times 10^{14}$ GeV if the dimensionless coupling
of the operator is of order unity, and the Higgs vev is equal to that
of the Standard Model.

One might also wonder if the see-saw mechanism with right-handed
neutrinos is the only possibility? In fact it is possible to
generate the dimension 5 operator in Eq.\ref{dim5} by the exchange
of heavy Higgs triplets of $SU(2)_L$, referred to as the type II see=saw
mechanism.

Alternatively the see-saw can be implemented in a two-stage process
by introducing additional neutrino singlets beyond the three right-handed
neutrinos that we have considered so far. It is useful to distingush
between ``right-handed neutrinos'' $\nu_R$ which carry $B-L$ and perhaps
form $SU(2)_R$ doublets with right-handed charged leptons,
and ``neutrino singlets'' $S$ which have no Yukawa couplings
to the left-handed neutrinos, but which may couple to $\nu_R$.
If the singlets have Majorana masses $M_{SS}$, but the right-handed neutrinos
have a zero Majorana mass $M_{RR}=0$, the see-saw mechanism may
proceed via mass couplings of singlets to right-handed neutrinos
$M_{SR}$. In the basis $(\nu_L,\nu_R,S)$ the mass matrix is
\beq
\left( \begin{array}{ccc}
0& m_{LR} & 0    \\
m_{LR} & 0 & M_{RS} \\
0 & M_{RS}^T & M_{SS}
\end{array}
\right)
\eeq
Assuming $M_{SS}\ll M_{RS}$ the light physical left-handed
Majorana neutrino masses are then doubly suppressed,
\beq
m_{LL} = m_{LR}M_{RS}^{-1}M_{SS}{M_{RS}^T}^{-1}m_{LR}^T
\eeq
This is called the double see-saw mechanism.
It is often used in string inspired neutrino mass
models \cite{Mohapatra:bd}.

\section{Right-Handed Neutrino Dominance}

In this section we discuss an elegant and natural way of
accounting for a neutrino mass hierarchy and two large mixing angles,
by simply assuming that not all of the right-handed neutrinos
contribute equally to physical neutrino masses in the see-saw
mechanism. This mechanism, called sequential dominance, is
a technique rather than a model, and can be applied to
large classes of models. Indeed the conditions for
sequential dominance can only be understood within
particular models, and provide useful clues to the nature
of such models.

\subsection{Single Right-Handed Neutrino Dominance}

With three left-handed neutrinos and
three right-handed neutrinos the Dirac masses $m_{LR}$
are a $3\times 3$ (complex) matrix and the heavy Majorana masses $M_{RR}$
form a separate $3\times 3$ (complex symmetric) matrix.
The light effective Majorana
masses $m_{LL}$ are also a $3\times 3$ (complex symmetric) matrix and
continue to be given from Eq.\ref{seesaw} which
is now interpreted as a matrix product. From a model building
perspective the fundamental parameters which must be input
into the see-saw mechanism are the Dirac mass matrix $m_{LR}$ and
the heavy right-handed neutrino Majorana mass matrix $M_{RR}$.
The light effective left-handed Majorana mass
matrix $m_{LL}$ arises as an output according to the
see-saw formula in Eq.\ref{seesaw}. The goal of see-saw model
building is therefore to choose input see-saw matrices
$m_{LR}$ and $M_{RR}$ that will give rise to one of the successful
matrices $m_{LL}$ in Table \ref{table1}.

We now show how the input see-saw matrices can be simply chosen to
give the Type IA matrix, with the property of a
naturally small sub-determinant in Eq.\ref{det} using a mechanism
first suggested in \cite{King:1998jw}.
\footnote{See also \cite{Davidson:1998bi}}
The idea was developed in \cite{King:1999cm} where it was called
single right-handed neutrino dominance (SRHND) . SRHND was first successfully
applied to the LMA MSW solution in \cite{King:2000mb}.

To understand the basic idea of dominance, it is 
instructive to begin by discussing
a simple $2\times 2$ example, where we have in mind applying
this to the atmospheric mixing in the 23 sector:
\begin{equation}
M_{RR}=
\left( \begin{array}{cc}
Y & 0     \\
0 & X 
\end{array}
\right), \ \ \ \ 
m_{LR}=
\left( \begin{array}{cc}
e & b \\
f & c
\end{array}
\right)
\label{2by2dom}
\end{equation}
The see-saw formula in Eq.\ref{seesaw} 
$m_{LL}=m_{LR}M_{RR}^{-1}m_{LR}^T$ gives:
\beq
m_{LL}
=
\left( \begin{array}{cc}
\frac{e^2}{Y}+\frac{b^2}{X}
& \frac{ef}{Y}+\frac{bc}{X} \\
\frac{ef}{Y}+\frac{bc}{X}
& \frac{f^2}{Y}+\frac{c^2}{X}
\end{array}
\right)
\approx
\left( \begin{array}{cc}
\frac{e^2}{Y}
& \frac{ef}{Y} \\
\frac{ef}{Y}
& \frac{f^2}{Y}
\end{array}
\right)
\label{one}
\eeq
where the approximation in Eq.\ref{one} 
assumes that the right-handed neutrino of mass $Y$
is sufficiently light that it dominates in the see-saw mechanism:
\beq
\frac{e^2,f^2,ef}{Y}\gg
\frac{b^2,c^2,bc}{X}.
\label{srhndp}
\eeq
The neutrino mass
spectrum from Eq.\ref{one}
then consists of one neutrino with mass $m_3\approx (e^2+f^2)/Y$
and one naturally light neutrino $m_2\ll m_3$, 
since the determinant of Eq.\ref{one} is clearly
approximately vanishing, due to the dominance assumption
\cite{King:1998jw}. 
The atmospheric angle from Eq.\ref{one} is
$\tan \theta_{23} \approx e/f$ \cite{King:1998jw}
which can be large or maximal providing $e \approx f$,
even in the case $e,f,b\ll c$ that 
the neutrino Dirac mixing angles 
arising from Eq.\ref{2by2dom} are small.
Thus two crucial features, namely a neutrino mass hierarchy
$m_3^2\gg m_2^2$ and a 
large neutrino mixing angle $\tan \theta_{23} \approx 1$, 
can arise naturally from the 
see-saw mechanism assuming the dominance of a single right-handed neutrino.
It was also realised that small perturbations from
the sub-dominant right-handed neutrinos can then lead to a small
solar neutrino mass splitting \cite{King:1998jw}, as we now discuss.

\subsection{Sequential Right-Handed Neutrino Dominance}

In order to account for the solar and other mixing angles,
we must generalise the above discussion to the
$3\times 3$ case.
The SRHND mechanism is most simply described
assuming three right-handed neutrinos
in the basis where the right-handed neutrino mass matrix is diagonal
although it can also be developed in other bases
\cite{King:1999cm,King:2000mb}. In this basis we write the input
see-saw matrices as
\begin{equation}
M_{RR}=
\left( \begin{array}{ccc}
Y & 0 & 0    \\
0 & X & 0 \\
0 & 0 & X'
\end{array}
\right)
\label{seq1}
\end{equation}
\begin{equation}
m_{LR}=
\left( \begin{array}{ccc}
d & a & a'    \\
e & b & b'\\
f & c & c'
\end{array}
\right)
\label{dirac}
\end{equation}
In \cite{King:1998jw} it was suggested that one of the right-handed neutrinos
may dominante the contribution to $m_{LL}$ if it is lighter than
the other right-handed neutrinos.
The dominance condition was subsequently generalised to
include other cases where the right-handed neutrino may be
heavier than the other right-handed neutrinos but dominates due to its larger
Dirac mass couplings \cite{King:1999cm}.
In any case the dominant right-handed neutrino may be taken to be the one
with mass $Y$ without loss of generality.

It was subsequently shown how to
account for the LMA MSW solution with a large solar angle
\cite{King:2000mb} by careful consideration of the sub-dominant contributions.
One of the examples considered in \cite{King:2000mb} is when the
right-handed neutrinos dominate sequentially,
\beq
\frac{|e^2|,|f^2|,|ef|}{Y}\gg
\frac{|xy|}{X} \gg
\frac{|x'y'|}{X'}
\label{srhnd}
\eeq
which is the straightforward generalisation of Eq.\ref{srhndp}
where $x,y\in a,b,c$ and $x',y'\in a',b',c'$.
Assuming SRHND with sequential sub-dominance as in
Eq.\ref{srhnd}, then Eqs.\ref{seesaw}, \ref{seq1}, \ref{dirac} give
\beq
m_{LL}
\approx
\left( \begin{array}{ccc}
\frac{a^2}{X}+\frac{d^2}{Y}
& \frac{ab}{X}+ \frac{de}{Y}
& \frac{ac}{X}+\frac{df}{Y}    \\
.
& \frac{b^2}{X}+\frac{e^2}{Y}
& \frac{bc}{X}+\frac{ef}{Y}    \\
.
& .
& \frac{c^2}{X}+\frac{f^2}{Y}
\end{array}
\right)
\label{mLL2}
\eeq
where the contribution from the right-handed neutrino of mass $X'$ may be
neglected according to Eq.\ref{srhnd}.
If the couplings satisfy the sequential dominance
condition in Eq.\ref{srhnd}
then the matrix in Eq.\ref{mLL2} resembles the Type IA matrix,
and furthermore has a naturally small sub-determinant as in
Eq.\ref{det}. 
This leads to a full neutrino mass hierarchy
\beq
m_3^2\gg m_2^2\gg m_1^2
\eeq
and, ignoring phases, the solar angle only depends
on the sub-dominant couplings and is given by
$\tan \theta_{12} \approx a/(c_{23}b-s_{23}c)$ \cite{King:2000mb}.
The simple requirement for large solar angle is then $a\sim b-c$
\cite{King:2000mb}.

Including phases the neutrino masses
are given to leading order in $m_2/m_3$ by diagonalising
the mass matrix in Eq.\ref{mLL2} using the analytic proceedure
described in Appendix D \cite{King:2002nf}.
In the case that $d=0$, corresponding to a 11 texture zero
in Eq.\ref{dirac}, we have
\bea
m_1 & \sim & O(\frac{x'y'}{X'}) \label{m1} \\
m_2 & \approx &  \frac{|a|^2}{Xs_{12}^2} \label{m2} \\
m_3 & \approx & \frac{|e|^2+|f|^2}{Y}
\label{m3}
\eea
where $s_{12}=\sin \theta_{12}$ is given below.
Note that with SD each neutrino mass is generated
by a separate right-handed neutrino, and the sequential dominance condition
naturally results in a neutrino mass hierarchy $m_1\ll m_2\ll m_3$.
The neutrino mixing angles are given to leading order in $m_2/m_3$ by,
\bea
\tan \theta_{23} & \approx & \frac{|e|}{|f|} \label{23}\\
\tan \theta_{12} & \approx &
\frac{|a|}
{c_{23}|b|
\cos(\tilde{\phi}_b)-
s_{23}|c|
\cos(\tilde{\phi}_c)} \label{12} \\
\theta_{13} & \approx &
e^{i(\tilde{\phi}+\phi_a-\phi_e)}
\frac{|a|(e^*b+f^*c)}{[|e|^2+|f|^2]^{3/2}}
\frac{Y}{X}
\label{13}
\eea
where we have written some (but not all) complex Yukawa couplings as
$x=|x|e^{i\phi_x}$. The phase $\delta$
is fixed to give a real angle
$\theta_{12}$ by,
\beq
c_{23}|b|
\sin(\tilde{\phi}_b)
\approx
s_{23}|c|
\sin(\tilde{\phi}_c)
\label{chi1}
\eeq
where
\bea
\tilde{\phi}_b &\equiv &
\phi_b-\phi_a-\tilde{\phi}+\delta, \nonumber \\
\tilde{\phi}_c &\equiv &
\phi_c-\phi_a+\phi_e-\phi_f-\tilde{\phi}+\delta
\label{bpcp}
\eea
The phase $\tilde{\phi}$
is fixed to give a real angle
$\theta_{13}$ by,
\beq
\tilde{\phi} \approx  \phi_e-\phi_a -\phi_{\rm COSMO}
\label{phi2dsmall}
\eeq
where
\beq
\phi_{\rm COSMO}=\arg(e^*b+f^*c).
\label{lepto0}
\eeq
is the leptogenesis phase \cite{Buchmuller:2003jr}
corresponding to the interference diagram involving the
lightest and next-to-lightest right-handed neutrinos \cite{King:2002nf}.

\subsection{Types of Sequential Dominance}

Assuming sequential dominance,
there is still an ambiguity regarding the mass ordering of the
heavy Majorana right-handed neutrinos.
So far we have assumed that the dominant right-handed neutrino
of mass $Y$ is dominant because it is the lightest one.
We emphasise that this need not be the case.
The neutrino of mass $Y$ could be dominant even if it is the heaviest
right-handed neutrino, providing its Yukawa couplings are strong
enough to overcome its heaviness and satisfy the condition in
Eq.\ref{srhnd}. In hierarchical mass matrix models, it is
natural to order the right-handed neutrinos so that the heaviest
right-handed neutrino is the third one, the intermediate right-handed
neutrino is the second one, and the lightest right-handed neutrino
is the first one. It is also natural to assume that the 33 Yukawa
coupling is of order unity, due to the large top quark mass.
It is therefore possible that the dominant right-handed
neutrino is the heaviest (called heavy sequential dominance or
HSD), the lightest (called light sequential dominance or LSD), or
the intermediate one (called intermediate sequential dominance or
ISD). This leads to the six possible types of sequential dominance
corresponding to the six possible mass orderings of the right-handed
neutrinos as shown in Table\ref{table1}.
In each case the dominant right-handed neutrino is the one with mass
$Y$, and the leading subdominant right-handed neutrino is the one
with mass $X$. The resulting see-saw matrix $m_{LL}$ is invariant
under re-orderings of the right-handed neutrino columns,
but the leading order form of the neutrino Yukawa matrix $Y_{\nu}$
is not.

\begin{table*}[htb]
{\small
\caption{Types of sequential dominance (SD), classified
according to the mass ordering of the right-handed neutrinos.
Light sequential dominance (LSD) corresponds to the dominant
right-handed neutrino of mass $Y$ being the lightest.
Intermediate sequential dominance (ISD) corresponds to the dominant
right-handed neutrino of mass $Y$ being the intermediate one.
Heavy sequential dominance (HSD) corresponds to the dominant
right-handed neutrino of mass $Y$ being the heaviest.}
\label{table2}
\newcommand{\m}{\hphantom{$-$}}
\newcommand{\cc}[1]{\multicolumn{1}{c}{#1}}
\renewcommand{\tabcolsep}{2pc} 
\renewcommand{\arraystretch}{1.2} 
\begin{tabular}{|c|c|c|c|}
\hline
\hline
Type of SD & $M_{RR}$  & $Y_{\nu}$ & Leading $Y_{\nu}$\\
\hline
\hline
$\stackrel{\rm LSDa}{Y<X<X'}$  &
$\left(
\begin{array}{ccc}
Y & 0 & 0 \\
0 & X & 0\\
0 & 0 & X' \\
\end{array}
\right)$
&
$\left(
\begin{array}{ccc}
d & a & a'\\
e & b & b'\\
f & c & c'\\
\end{array}
\right)$
&
$\left(
\begin{array}{ccc}
0 & 0 & 0 \\
0 & 0 & 0\\
0 & 0 & 1 \\
\end{array}
\right)$
\\
\hline
\hline
$\stackrel{\rm LSDb}{Y<X'<X}$  &
$\left(
\begin{array}{ccc}
Y & 0 & 0 \\
0 & X' & 0\\
0 & 0 & X \\
\end{array}
\right)$
&
$\left(
\begin{array}{ccc}
d & a' & a \\
e & b' & b\\
f & c' & c \\
\end{array}
\right)$
&
$\left(
\begin{array}{ccc}
0 & 0 & 1\\
0 & 0 & 1\\
0 & 0 & 1 \\
\end{array}
\right)$
\\
\hline
\hline
$\stackrel{\rm ISDa}{X<Y<X'}$  &
$\left(
\begin{array}{ccc}
X & 0 & 0 \\
0 & Y & 0\\
0 & 0 & X' \\
\end{array}
\right)$
&
$\left(
\begin{array}{ccc}
a & d & a' \\
b & e & b'\\
c & f & c' \\
\end{array}
\right)$
&
$\left(
\begin{array}{ccc}
0 & 0 & 0 \\
0 & 0 & 0\\
0 & 0 & 1 \\
\end{array}
\right)$
\\
\hline
\hline
$\stackrel{\rm ISDb}{X'<Y<X}$  &
$\left(
\begin{array}{ccc}
X' & 0 & 0 \\
0 & Y & 0\\
0 & 0 & X \\
\end{array}
\right)$
&
$\left(
\begin{array}{ccc}
a' & d & a \\
b' & e & b\\
c' & f & c \\
\end{array}
\right)$
&
$\left(
\begin{array}{ccc}
0 & 0 & 1 \\
0 & 0 & 1\\
0 & 0 & 1 \\
\end{array}
\right)$
\\
\hline
\hline
$\stackrel{\rm HSDa}{X'<X<Y}$  &
$\left(
\begin{array}{ccc}
X' & 0 & 0 \\
0 & X & 0\\
0 & 0 & Y \\
\end{array}
\right)$
&
$\left(
\begin{array}{ccc}
a' & a & d \\
b' & b & e\\
c' & c & f \\
\end{array}
\right)$
&
$\left(
\begin{array}{ccc}
0 & 0 & 0 \\
0 & 0 & 1\\
0 & 0 & 1 \\
\end{array}
\right)$
\\
\hline
\hline
$\stackrel{\rm HSDb}{X<X'<Y}$  &
$\left(
\begin{array}{ccc}
X & 0 & 0 \\
0 & X' & 0\\
0 & 0 & Y \\
\end{array}
\right)$
&
$\left(
\begin{array}{ccc}
a & a' & d \\
b & b' & e\\
c & c' & f \\
\end{array}
\right)$
&
$\left(
\begin{array}{ccc}
0 & 0 & 0 \\
0 & 0 & 1\\
0 & 0 & 1 \\
\end{array}
\right)$\\
\hline
\hline
\end{tabular}\\[2pt]
}
\end{table*}

It is worth emphasising that since all the forms above give the
same light effective see-saw neutrino matrix $m_{LL}$ in
Eq.\ref{mLL2}, under the sequential dominance assumption
in Eq.\ref{srhnd}, this implies that the analytic results for neutrino masses
and mixing angles applies to all of these forms.
They are distinguished theoretically by different preferred leading order
forms of the neutrino Yukawa matrix $Y_{\nu}$ shown in the table.
These leading order forms follow from the
the large mixing angle requirements $e\sim f$ and
$a \sim b-c$. \footnote{Note that the leading order $Y_{\nu}$ in the Table
only gives the independent order unity entries in the matrix,
so that for example in LSDb we would expect $b-c\sim 1$ in general,
and not zero.}
Thus we see that LSDa, and ISDa are
consistent with a form of Yukawa matrix with small Dirac mixing
angles, while HSDa and HSDb correspond to the so called ``lop-sided''
forms. LSDb and ISDb correspond to the D-brane inspired
``single right-handed democracy'' form studied in \cite{Everett:2000up}.
They are also distinguished by
leptogenesis and lepton flavour violation as we shall see.

For example, suppose that we impose the theoretical requirement that
the neutrino Yukawa matrix resembles hierarchical quark matrices,
and have a large 33 element of order unity, but no other large off-diagonal
entries. Then the large mixing angle requirements $e\sim f$ and
$a \sim b-c$ immediately excludes HSDa, HSDb, LSDb and ISDb.
We are left with LSDa and ISDa as the remaining possibilities.
If we further impose the requirement of a 11 texture zero,
as motivated by the GST relation \cite{Gatto:ss}, then
$a \sim b-c$ excludes ISDa, and we are left uniquely with LSDa.
We shall later discuss an example of a realistic model of all
quark and lepton masses and mixing angles based on LSDa.
For now we note that that for LSDa in order to
satisfy the sequential dominance condition in Eq.\ref{srhnd}
the heavy Majorana masses must be necessarily strongly hierarchical,
\beq
Y\ll X \ll X'.
\eeq
The reason is that the heavy right-handed neutrino of mass $X'$
has order unity Yukawa couplings to left-handed neutrinos, which
implies that the lightest right-handed neutrino of mass $Y$ must be
significantly lighter in order to dominate.

\subsection{Leptogenesis Link}

It is interesting to note that in LSDa, assuming a 11 texture zero,
there is a link between the CP violation required for leptogenesis,
and the phase $\delta$ measurable in accurate neutrino oscillation
experiments. This can be seen from Eq.\ref{lepto0} which may be expressed as
\beq
\tan \phi_{\rm COSMO} \approx
\frac{|b|s_{23}s_2+|c|c_{23}s_3}{|b|s_{23}c_2+|c|c_{23}c_3}.
\label{phi121}
\eeq
From Eqs.\ref{phi2dsmall},\ref{chi1},\ref{bpcp},
\beq
\tan (\phi_{\rm COSMO}+\delta) \approx
\frac{|b|c_{23}s_2-|c|s_{23}s_3}{-|b|c_{23}c_2+|c|s_{23}c_3}
\label{phi12del}
\eeq
where we have written $s_i=\sin \eta_i, c_i=\cos \eta_i$
where
\beq
\eta_2\equiv \phi_b-\phi_e, \ \ \eta_3\equiv \phi_c-\phi_f
\label{eta}
\eeq
are invariant under a charged lepton phase transformation.
The reason that the see-saw parameters only
involve two invariant phases $\eta_2, \eta_3$ rather than the usual six
is due to the LSD
assumption which has the effect of decoupling the
heaviest right-handed neutrino, which removes three phases,
together with the assumption
of a 11 texture zero, which removes another phase.

Eq.\ref{phi12del} shows that
$\delta$ is a function of
the two see-saw phases
$\eta_2 , \eta_3$ that also determine $\phi_{\rm COSMO}$ in Eq.\ref{phi121}.
If both the phases $\eta_2 , \eta_3$ are zero,
then both $\phi_{\rm COSMO}$ and $\delta$ are necessarily zero.
This feature is absolutely crucial. It means that,
barring cancellations, measurement of a non-zero value for
the phase $\delta$ at a neutrino factory will be a signal of a
non-zero value of the leptogenesis phase $\phi_{\rm COSMO}$.
We also find the remarkable result
\beq
|\phi_{\rm COSMO}|=|\phi_{\beta \beta 0\nu}|.
\label{remarkable2}
\eeq
where $\phi_{\beta \beta 0\nu}$ is the phase which enters the rate
for neutrinoless double beta decay \cite{King:2002qh}.

\subsection{Comparison to the Smirnov Approach}

An early approach to obtaining large mixing angles from
the see-saw mechanism was proposed by \cite{Smirnov:af},
which is sometimes confused with sequential dominance.
The purpose of this subsection is to briefly review the
Smirnov approach, and explain how it differs from sequential dominance.
The Smirnov approach
for obtaining large mixing angles from the see-saw
mechanism, is based on the theoretical assumption of having no large
mixing angles in the Yukawa sector \cite{Smirnov:af}.

We shall briefly discuss the two family case considered in
\cite{Smirnov:af}. For this case the physical lepton mixing angle
is written as
\beq
\theta \equiv \theta_L^{e_{LR}}-\theta_L^{\nu_{LR}}+\theta_{SS}
\eeq
where $\theta_L^{e_{LR}}$ is the left-handed mixing angle which
diagonalises the charge lepton Yukawa matrix,
$\theta_L^{\nu_{LR}}$ is the left-handed mixing angle which
diagonalises the neutrino Yukawa matrix,
and $\theta_{SS}$ is defined to be the additional angle which
results from the presence of the see-saw mechanism.
The basic idea \cite{Smirnov:af}
was that a large mixing angle $\theta$ could originate
from the see-saw mechanism via $\theta_{SS}$ with
$\theta_L^{e_{LR}}$ and $\theta_L^{\nu_{LR}}$ being small.
\footnote{Note that this is not a requirement of sequential dominance,
although it may be satisfied by LSDa or ISDa, as discussed previously.}
Smirnov obtains an approximate analytic expression for $\theta_{SS}$ in the
two family case,
\beq
\tan \theta_{SS}\approx
-2\epsilon^D\frac{\tan(\theta_R^{\nu_{LR}}-\theta_R^{\nu_{RR}})}
{\tan^2(\theta_R^{\nu_{LR}}-\theta_R^{\nu_{RR}})+\epsilon^M}
\eeq
where $\theta_R^{\nu_{LR}}$ is the mixing angle which
diagonalises the neutrino Yukawa $Y^{\nu}_{LR}$ matrix on the right,
$\theta_R^{\nu_{RR}}$ is the mixing angle which
diagonalises the heavy Majorana neutrino matrix $M_{RR}$,
$\epsilon^D$ is the ratio of neutrino Yukawa (Dirac) matrix eigenvalues,
and $\epsilon^M$ is the ratio of heavy Majorana matrix eigenvalues.
The conditions that $\theta_{SS}$ is large are
\bea
\tan(\theta_R^{\nu_{LR}}-\theta_R^{\nu_{RR}}) & \leq & \epsilon^D
\label{smirnov1} \\
\epsilon^M & \leq &  (\epsilon^D)^2
\label{smirnov2}
\eea
which, for a typical quark-like hierarchy $\epsilon^D\ll 1$,
implies both a very accurate equality of mixing angles
$\theta_R^{\nu_{LR}}=\theta_R^{\nu_{RR}}$ and
very strongly hierarchical heavy Majorana masses
(much stronger than the Dirac mass hierarchy).

The conditions in Eqs.\ref{smirnov1},\ref{smirnov2}
are clearly nothing to do with sequential dominance in
general. For one thing since some versions of sequential dominance involve
large neutrino Yukawa mixing angles $\theta_L^{\nu_{LR}}$
and do not require $\theta_{SS}$ to be large, which is the basic
assumption of this approach. However there are classes of sequential
dominance model such as LSDa where $\theta_L^{\nu_{LR}}$ is small and
$\theta_{SS}$ is large. Furthermore in this class of model there is a
strong hierarchy of Majorana masses. One might be tempted to think
that LSDa is the same as the Smirnov approach, and this has led to
some confusion in the literature which we would like to clear up here.
The important point to emphasise is that \cite{Smirnov:af} never talks
about one of the right-handed neutrinos giving the dominant
contribution to the heaviest physical neutrino via the see-saw
mechanism, or indeed about the relative contribution of the
right-handed neutrinos to the see-saw mechanism in general.
Thus there is no natural mechanism present for generating a neutrino
mass hierarchy in \cite{Smirnov:af}, which is concerned only with
the condition for generating large mixing angles. The point about
sequential dominance is that it can naturally generate a neutrino mass
hierarchy and large mixing angles, as simple ratios of Yukawa
couplings of dominant and subdominant right-handed neutrinos.

A simple counter example will illustrate this point.
Condider the following matrices,
\begin{equation}
M_{RR}=
\left( \begin{array}{cc}
A_{11}\epsilon_D^2 & A_{12}\epsilon_D \\
A_{12}\epsilon_D & A_{22}
\end{array}
\right)M,\ \
{Y_{\nu}}_{LR}=
\left( \begin{array}{ccc}
a_{11}\epsilon_D & a_{12}\epsilon_D\\
a_{21}\epsilon_D & a_{22}
\end{array}
\right)
\label{counter}
\end{equation}
where $A_{ij},a_{ij}$ are order unity coefficients.
These matrices clearly satisfy the conditions
in Eqs.\ref{smirnov1},\ref{smirnov2}, since
$\theta_R^{\nu_{LR}}\sim \theta_R^{\nu_{RR}} \sim \epsilon_D$
and $\epsilon_M \sim \epsilon_D^2$.
However these matrices do not satisfy
the dominance conditions.
Both right-handed neutrinos will contribute equally at
$O(1/M)$ via the see-saw mechanism to the heaviest physical neutrino mass.
Without the dominance of a single right-handed neutrino
the neutrino mass hierarchy will require some tuning.
The tuning required for the atmospheric mixing angle involving
second and third families with be rather mild since $m_2/m_3$ is
not so small, however when this scheme is extended to all three
families, further tuning will be required to obtain a large solar
mixing angle in a natural way. In actual examples given in \cite{Smirnov:af}
even more tuning is likely to be required since the angles
$\theta_R^{\nu_{LR}},\theta_R^{\nu_{RR}}$ were both independently
supposed to be larger than $\epsilon^D$.

The conclusion is that Smirnov's approach did not recognise
right-handed neutrino dominance, contrary to some recent claims in the
literature, but it does provide a complementary approach to a large mixing
angles from the see-saw mechanism. At first sight it appears to have
some similarities to LSDa, however without the missing
ingredient of sequential dominance, to achieve two large mixing angles
together with a neutrino mass hierarchy will require some degree of
fine-tuning. The conditions proposed by Smirnov are therefore
neither necessary nor sufficient for right-handed neutrino dominance.

\section{See-Saw Standard Models}
In this section we show how the see-saw mechanism can be accomodated
in the Standard Model and its Supersymmetric Extension, where
it leads to lepton flavour violation.

\subsection{Minimal See-Saw Standard Model}

We now briefly discuss what the standard model looks like,
assuming a minimal see-saw extension.
In the standard model Dirac mass terms for charged leptons and quarks
are generated from Yukawa couplings
to Higgs doublets whose vacuum expectation value gives the Dirac
mass term.
Neutrino masses are zero in the Standard Model because
right-handed neutrinos are not present, and also
because the Majorana mass terms in Eq.\ref{mLL}
require Higgs triplets in order to be generated at the renormalisable
level. The simplest way to generate neutrino masses
from a renormalisable theory is to introduce right-handed neutrinos,
as in the Type I see-saw mechanism, which we assume here.
The Lagrangian for the lepton sector of the standard model
containing three right-handed neutrinos with heavy Majorana masses is
\footnote{We introduce two higgs doublets to pave the way for
the supersymmetric standard model.
For the same reason we express the standard model Lagrangian
in terms of left-handed fields, replacing right-handed fields
by their CP conjugates.
In the case of the minimal standard see-saw model with one Higgs doublet
one of the two Higgs doublets by the charge conjugate of the other,
$H_d\equiv H_u^c$.}
\be
{\cal L}_{mass}=-\epsilon_{ab}\left[\tilde{Y}^e_{ij}H_d^aL_i^be^c_j
-\tilde{Y}^{\nu}_{ij}H_u^aL_i^b\nu^c_j +
\frac{1}{2} \nu^c_i\tilde{M}_{RR}^{ij}\nu^c_j
\right] +H.c.
\label{SM}
\ee
where $\epsilon_{ab}=-\epsilon_{ba}$, $\epsilon_{12}=1$,
and the remaining notation is standard except that
the $3$ right-handed neutrinos $\nu_R^p$ have been replaced by their
CP conjugates $\nu^c_i$, and $\tilde{M}_{RR}^{ij}$ is a complex symmetric
Majorana matrix.
When the two Higgs doublets get their
VEVS $<H_u^2>=v_2$, $<H_d^1>=v_1$, where the ratio of VEVs
is defined to be $\tan \beta \equiv v_2/v_1$,
we find the terms
\be
{\cal L}_{mass}= -v_1\tilde{Y}^e_{ij}e_ie^c_j
-v_2\tilde{Y}^{\nu}_{ij}\nu_i\nu^c_j -
\frac{1}{2}\tilde{M}_{RR}^{ij}\nu^c_i\nu^c_j +H.c.
\ee
Replacing CP conjugate fields we can write in a matrix notation
\be
{\cal L}_{mass}=-\bar{e}_Lv_1\tilde{Y^e}^\ast e_R
-\bar{\nu}_Lv_2\tilde{Y^\nu}^\ast \nu_R -
\frac{1}{2}\nu^T_R\tilde{M}_{RR}^\ast \nu_R +H.c.
\ee
It is convenient to work in the diagonal charged lepton basis
\be
{\rm diag(m_e,m_{\mu},m_{\tau})}
= V_{E_L}v_1\tilde{Y^e}^\ast V_{E_R}^{\dag}
\ee
and the diagonal right-handed neutrino basis
\be
{\rm diag(M_{1},M_{2},M_{3})} =
V_{\nu_R}\tilde{M^\ast}_{RR}V_{\nu_R}^{T}
\ee
where $V_{eL},V_{eR},V_{\nu_R}$ are unitary transformations.
In this basis the neutrino Yukawa couplings are given by
\be
Y^{\nu}=V_{E_L} \tilde{Y}^{\nu^\ast} V_{\nu_R}^{T}
\ee
and the Lagrangian in this basis is
\bea
{\cal L}_{mass}&=-&(\bar{e}_L \bar{\mu}_L \bar{\tau}_L)
{\rm diag(m_e,m_{\mu},m_{\tau})} ({e}_R {\mu}_R {\tau}_R)^T
\nonumber \\
&-&(\bar{\nu_e}_L \bar{\nu_\mu}_L \bar{\nu_\tau}_L)Y^{\nu}v_2
(\nu_{R1} \nu_{R2} \nu_{R3} )^T
\nonumber \\
&-& (\nu_{R1} \nu_{R2} \nu_{R3} ){\rm diag(M_{1},M_{2},M_{3})}
(\nu_{R1} \nu_{R2} \nu_{R3} )^T
+H.c.
\eea
After integrating out the right-handed neutrinos (the see-saw mechanism)
we find
\bea
{\cal L}_{mass}&=-&(\bar{e}_L \bar{\mu}_L \bar{\tau}_L)
{\rm diag(m_e,m_{\mu},m_{\tau})} ({e}_R {\mu}_R {\tau}_R)^T
\nonumber \\
&-&\frac{1}{2}(\bar{\nu_e}_L \bar{\nu_\mu}_L \bar{\nu_\tau}_L)
m_{LL} ({\nu_e}_L^c {\nu_\mu}_L^c {\nu_\tau}_L^c)^T
+H.c.
\label{Lmass2}
\eea
where the light effective left-handed Majorana
neutrino mass matrix in the above basis is given by the
following see-saw formula which is equivalent to Eq.\ref{seesaw},
\be
m_{LL}=-v_2^2Y^{\nu}
{\rm diag(M_{1}^{-1},M_{2}^{-1},M_{3}^{-1})}  {Y^{\nu}}^T
\label{seesaw1}
\ee
Eq.\ref{Lmass2} is equivalent to Eq.\ref{Lmass} when expressed
in the charged lepton mass basis, which we have derived
starting from the standard model Lagrangian using the see-saw mechanism.

\subsection{Minimal Supersymmetric See-Saw Standard Model}

It is well known that large mass scales such as are required in the
see-saw mechanism can be stabilised by
assuming a TeV scale N=1 supersymmetry
which cancels the quadratic divergences
of the Higgs mass.
Thus it is natural to generalise the see-saw standard model to
include supersymmety. When this is done the leptonic part of the
superpotential with three right-handed neutrinos is given by
\be
\label{superpot}
W=\epsilon_{ab}[\hat{H}_d^a\hat{L}_i^b\tilde{Y}_e^{ij}\hat{e}^c_j
-\hat{H}_u^a\hat{L}^b_i\tilde{Y}_{\nu}^{ij}\hat{\nu}^c_j
+\frac{1}{2}\hat{\nu}^c_i\tilde{M}_{RR}^{ij}\hat{\nu}^c_j],
\ee
where $\epsilon_{ab}=-\epsilon_{ba}$ and $\epsilon_{12}=1$.
The $SU(2)$ representations of the lepton superfield doublets can
be expressed as follows (suppressing family indices for simplicity):
\bea
\hat{L}_i=\left(\begin{array}{c c}
\hat{\nu}_{i}\\ \hat{e}_{i}\end{array}\right).
\eea
The superfields are defined in the standard way as follows
(suppressing gauge indices):
\bea
\hat{\nu}_i&=&(\tilde{\nu}_{L_i}, \nu_{L_i})\nonumber\\
\hat{e}_i&=&(\tilde{e}_{L_i}, e_{L_i})\nonumber\\
\hat{e}^c_i&=&(\tilde{e}^c_{L_i}, e^c_{L_i})\nonumber\\
\hat{\nu}^c_i&=&(\tilde{\nu}^c_{L_i}, \nu^c_{L_i})\nonumber\\
\hat{H_u}&=&( H_u, \tilde{H}_u)\nonumber\\
\hat{H_d}&=&( H_d, \tilde{H}_d),
\eea
with $i,j=1\ldots 3$ labeling family indices.
The soft breaking Lagrangian ${\cal L}_{soft}$ in the lepton sector
takes the form (dropping
``helicity" indices):
\bea
-{\cal L}_{soft} &=&\epsilon_{ab}[
H^a_d\tilde{L}^b_i\tilde{A}^e_{ij}\tilde{e}^c_j
+{H}_u^a\tilde{L}^b_i\tilde{A}^{\nu}_{ij}\tilde{\nu}^c_j
+\frac{1}{2}\tilde{\nu}^c_i b^\nu_i\tilde{\nu}^c_i
+h.c.]\nonumber\\
&+& \tilde{L}^a_i{m^2_L}_{ij}\tilde{L}_j^{a*}
+\tilde{e}^{c*}_i{m^2_E}_{ij}\tilde{e}^c_j
+\tilde{\nu}^{c\ast }_im_{Nij}^2\tilde{\nu}^{c}_j.
\eea

The Yukawa terms in the Lagrangian
are given from the superpotential by replacing two of the superfields
by their fermion components, and one of the superfields by its scalar
component, and including an overall minus sign.
Then the leptonic part of the superpotential in Eq.\ref{superpot}
reduces to the standard model lagrangian in Eq.\ref{SM},
and the discussion then follows that of the previous section.
For the charged leptons, we have as before
\bea
{\rm diag}(m_e,m_{\mu},m_{\tau})&=&V_{E_L}v^*_1\tilde{Y}_e^*V_{E_R}^{\dagger};
\eea
in which
\bea
\label{leptonE}
\left ( \begin{array}{c} e_R \\\mu_R\\\tau_R \end{array} \right
)=V_{E_R}\left (\begin{array}{c}  e_{R_1}\\ e_{R_2}\\ e_{R_3} \end{array}
\right),
\; \; \left (\begin{array}{c} e_L \\\mu_L\\\tau_L \end{array} \right )
=V_{E_L}
\left (\begin{array}{c}  e_{L_1}\\ e_{L_2}\\ e_{L_3} \end{array}
\right).
\eea

The important new feature provided by SUSY is the existence of
scalar partners to the leptons (sleptons) which can give
lepton flavour violating (LFV) effects, which arise as discussed
in the following. To discuss these effects
we first need to express the sleptons in terms of their mass
eigenstates. It is usually convenient however to begin by
rotating the sleptons in exactly the same way as the
lepton.  In this basis, which we call the MNS basis, the photino
interactions conserve
flavor, while the wino (and higgsino) interactions violate flavor by
$U$, in analogy to the gauge boson interactions in the SM.
Therefore, the diagonalization of the scalar mass matrices proceeds in
two steps. First, the sleptons are rotated ``parallel'' to
their fermionic superpartners; {\it i.e.}, we do unto sleptons as we do
unto leptons:
\bea
\label{leptonMNS}
\left ( \begin{array}{c} \tilde{e}_{R} \\
\tilde{\mu}_{R} \\ \tilde{\tau}_{R} \end{array}
\right
)=V_{E_R}\left (\begin{array}{c} \tilde{e}_{R_1}\\ \tilde{e}_{R_2}\\
\tilde{e}_{R_3} \end{array}
\right),
\; \; \left (\begin{array}{c} \tilde{e}_{L} \\
\tilde{\mu}_{L} \\ \tilde{\tau}_{L} \end{array}
\right
) =V_{E_L}
\left (\begin{array}{c}  \tilde{e}_{L_1}\\ \tilde{e}_{L_2}\\
\tilde{e}_{L_3} \end{array} \right),
\eea
where in the MNS basis the slepton fields $(\tilde{e}_{L}, \tilde{\mu}_{L},
\tilde{\tau}_{L})$ are SUSY partners of the physical mass eigenstate
quarks $(e_L, \mu_L, \tau_L)$, respectively,
(i.e. $(\tilde{e}_{L}, {e}_{L})$ share the same superfield
where both components of the superfield have been subject to the
same rotation, thereby preserving the superfield structure),
and similarly for the other terms.

The slepton
fields expressed in the MNS basis are often more convenient to work
with, even though they are not mass eigenstates. Their $6 \times 6$ mass
matrices are obtained by adding the electroweak
symmetry breaking contributions and then rotating to the MNS basis.
They have the following form:
\bea
\label{pok1}
m^{2^{{\rm MNS}}}_{\tilde{E}} &=&
\left( \begin{array}{cc}
(m^2_{\tilde{E}})_{LL} + m_e^2
- \frac{\cos 2\beta}{6}(M_Z^2 + 2M_W^2)\hat{\mbox{\large 1}}  &
(m^2_{\tilde{E}})_{LR}
- \tan\beta \mu m_e\\
(m^2_{\tilde{E}})_{LR}^{\dagger}
- \tan\beta \mu^{\star}  m_e&
(m^2_{\tilde{E}})_{RR} + m_e^2
-\frac{\cos 2\beta}{3} M_Z^2\sin^2\theta_W\hat{\mbox{\large 1}} \\
\end{array}\right)\nonumber \\
\label{eq:sfmass}
\eea
in which  $\theta_W$ is the electroweak mixing angle, $\hat{\mbox{\large
1}}$ stands for the $3 \times 3$ unit matrix, and we have written
$m_e={\rm diag}(m_e,m_\mu,m_\tau)$.
The flavor-changing entries responsible for lepton flavour violation
are contained in the off-diagonal entries of the soft slepton mass matrices
above, which are given by
\bea
\label{pok1a}
\begin{array}{ccc}
(m^2_{\tilde{E}})_{LL} = V_{E_L} m^{2^*}_L V_{E_L}^{\dagger}
\hspace{0.5cm}&
(m^2_{\tilde{E}})_{RR} = V_{E_R} m^{2^*}_E V_{E_R}^{\dagger}
\hspace{0.5cm}&
(m^2_{\tilde{E}})_{LR} =
v_1^*V_{E_L} \tilde{A}^{e^*} V_{E_R}^{\dagger}.
\end{array}
\eea

\subsection{Lepton Flavour Violation}

The renormalisation group equations
(RGEs) contain additional
terms relative to the MSSM. The additional
terms imply that even if the soft slepton masses are diagonal
at the GUT scale, then in this case we would find that
three separate lepton numbers $L_e, L_{\mu}, L_{\tau}$
are not conserved at low energies, since the new RGE terms do not preserve
these symmetries in general if there are right-handed neutrinos below the
GUT scale. Below the mass scale of the right-handed
neutrinos we must decouple the heavy right-handed neutrinos from the
RGEs, and then the RGEs return to those of the MSSM.
Thus the lepton number violating additional terms
are only effective in the region between the GUT scale and the
mass scale of the lightest right-handed neutrino, and all the
effects of lepton number violation are generated by RGE effects
over this range. The effect of RGE running over this range
will lead to off-diagonal slepton masses at high energy,
which result in off-diagonal slepton masses at low energy,
and hence observable lepton flavor violation in experiments.

Assuming universal soft parameters at $M_{GUT}$,
$m_L^2(0)=m_N^2(0)=m_0^2I$, where $I$ is the
unit matrix, and $\tilde{A}_{\nu}(0)=A Y_{\nu}$,
the renormalisation group equation
(RGE) for the soft slepton doublet mass may be written as
\be
\frac{d m_L^2}{dt}  =  \left(\frac{d m_L^2}{dt}\right)_{Y_{\nu}=0}
 -  \frac{(3m_0^2+A^2)}{16\pi^2}\left[ Y_{\nu}Y_{\nu}^\dagger \right]
\nonumber
\ee
where in the basis in which the charged lepton Yukawa couplings
are diagonal, the first term on the right-hand side
is diagonal. In running the RGEs between $M_{GUT}$ and a right-handed
neutrino mass $M_{i}$ the neutrino
Yukawa couplings lead to an approximate contribution to the
slepton mass squared matrix of
\be
\delta m_L^2 \approx -\frac{1}{16\pi^2}
\ln \left(\frac{M_{GUT}^2}{M_{i}^2}\right)
(3m_0^2+A^2)
\left[ Y_{\nu}Y_{\nu}^\dagger \right]
\nonumber
\ee
This shows that, to leading log approximation,
off-diagonal slepton masses may be generated depending on the
form of the neutrino Yukawa matrix. The off-diagonal
slepton masses give rise to LFV processes such as
$\mu\rightarrow e \gamma$,
$\tau\rightarrow \mu \gamma$, $\tau\rightarrow e \gamma$.
From a future observation of these processes one may infer
information about the slepton mass matrix, and then use this
information to make inferences about the neutrino Yukawa matrix,
and heavy right-handed neutrino masses.
This proceedure would be impossible in the SM,
and is an example of how supersymmetry may in the future provide
a window into the Yukawa matrices which would not otherwise be
possible. This was originally discussed in
\cite{Borzumati:1986qx,Gabbiani:1996hi,Hisano:1995cp,King:1998nv}
and has been discussed recently in
\cite{Casas:2001sr,Blazek:2001zm,Lavignac:2001vp,Davidson:2001zk,Ellis:2002xt,Blazek:2002wq,Lavignac:2002gf}.

At leading order in a mass insertion approximation
the branching fractions of LFV processes are given by
\beq
{\rm BR}(l_i \rightarrow l_j \gamma)\approx
        \frac{\alpha^3}{G_F^2}
        f(M_2,\mu,m_{\tilde{\nu}})
        |m_{\tilde{L}_{ij}}^2|^2  \tan ^2 \beta
    \label{eq:BR(li_to_lj)}
\eeq
where $l_1=e, l_2=\mu , l_3=\tau$,
and where the off-diagonal slepton doublet mass squared is given
in the leading log approximation (LLA) by
\beq
m_{\tilde{L}_{ij}}^{2(LLA)}
\approx -\frac{(3m_0^2+A_0^2)}{8\pi ^2}C_{ij}
\label{lla}
\eeq
where in sequential dominance, in the notation of Eqs.\ref{seq1},\ref{dirac}
the leading log coefficients relevant for
$\mu \rightarrow e\gamma$ and $\tau \rightarrow \mu \gamma$
are given approximately as
\bea
C_{21} & = & ab\ln \frac{M_U}{X} +de\ln \frac{M_U}{Y} \nonumber \\
C_{32} & = & bc\ln \frac{M_U}{X} +ef\ln \frac{M_U}{Y}
\label{C2131}
\eea

A global analysis of LFV has been performed in the constrained
minimal supersymmetric standard model (CMSSM) for the case
of sequential dominance, focussing on the two cases
of HSD and LSD \cite{Blazek:2002wq}.
The results for HSD show a large rate for
$\tau \rightarrow \mu \gamma$ which is the characteristic expectation
of lop-sided models in general \cite{Blazek:2001zm} and HSD in
particular. The results are based on an exact calculation, and
the error incurred if the LLA were used
can be as much as 100\% \cite{Blazek:2002wq}.
For LSD $\tau \rightarrow \mu \gamma$
is well below observable values. Therefore $\tau \rightarrow \mu
\gamma$ provides a good discriminator between the HSD and LSD types
of dominance. The rate for $\mu \rightarrow e \gamma$ can be large
or small in each case.

\section{GUTs and Family Symmetry}

We have seen that atmospheric neutrino masses would seem to require
a right-handed neutrino with a mass below the GUT scale.
Such a mass scale demands an explanation,
and in fact one must then explain why the right-handed
neutrinos are so light compared to the Planck scale. In order
to explain this, one clearly needs a theory of right-handed neutrino
masses capable of protecting the right-handed neutrino masses
by some symmetry which is subsequently broken at some scale.
Suitable symmetries can correspond to either unification
or family symmetries, as we now discuss.

\subsection{Models Based on GUTs and Family Symmetry}

\begin{figure}[t]
\includegraphics[width=0.96\textwidth]{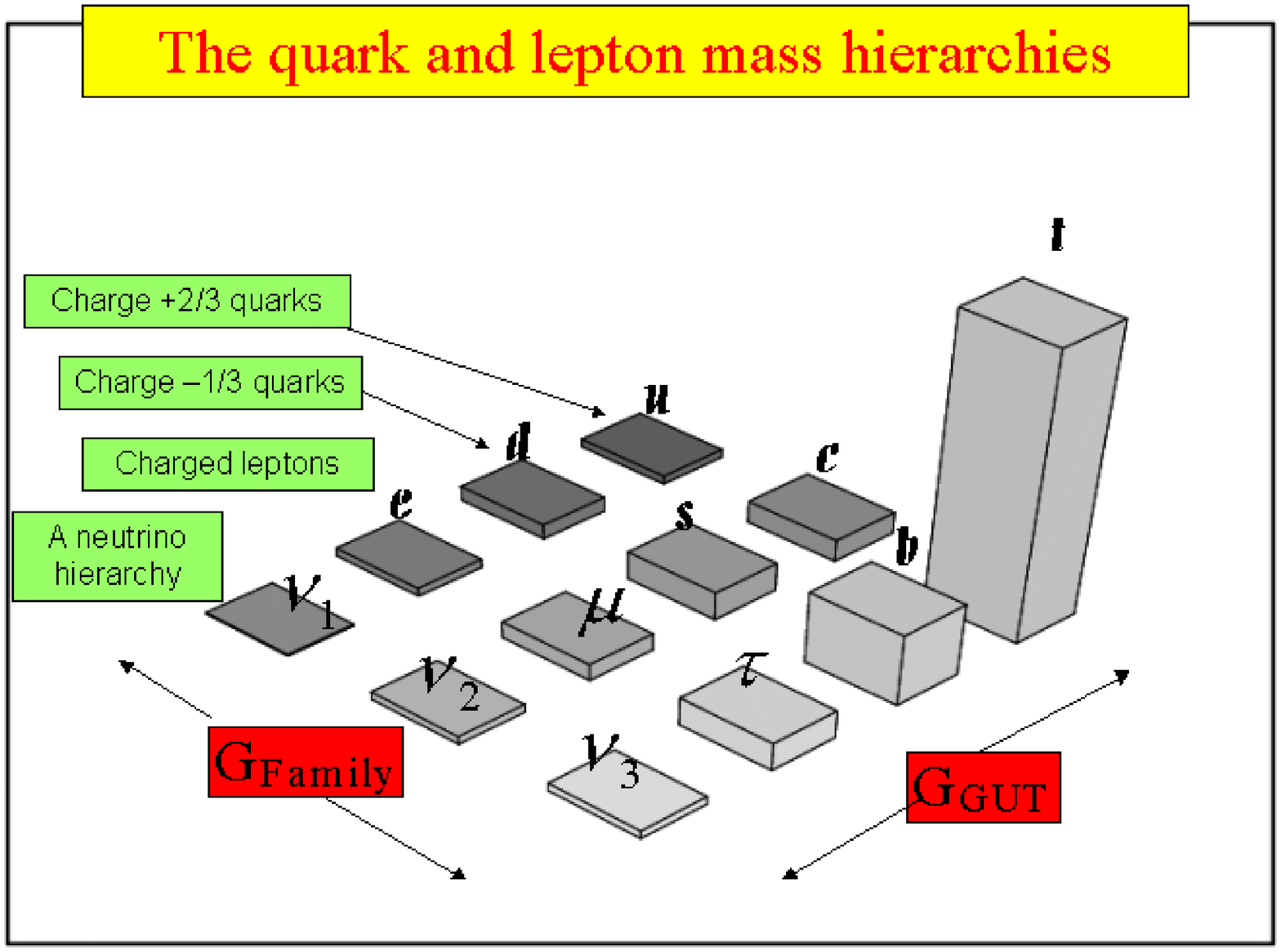}
\vspace*{-4mm}
    \caption{{\small The fermion masses are here represented by a lego plot.
We have multiplied the masses of the bottom, charm and tau by $10$,
the strange and muon by $10^2$, the up and down
by $10^3$, the electron by $10^4$ to make the lego blocks visible.
It is natural to assume a normal neutrino
hierarchy. We have multiplied the third neutrino mass by
$10^{11}$ and the second neutrino mass by $10^{12}$ to make the
lego blocks visible. This underlines how incredibly light the neutrinos are.
The symmetry groups $G_{{\rm GUT}}$ and $G_{{\rm Family}}$ act
in the directions indicated.}}
\label{fig3}
\vspace*{-2mm}
\end{figure}

One of the exciting things about the discovery of neutrino masses
and mixing angles is that this provides additional information
about the flavour problem - the problem of understanding the origin
of three families of quarks and leptons and their masses and mixing
angles (Fig.\ref{fig3}). Early approaches to the problem of quark
masses and mixing angles included the postulate that some
entries in the Yukawa matrices were equal to zero
(the so-called ``texture zeroes'') thereby
reducing the number of free parameters \cite{Fritzsch:1977za}.
In this approach the quark and lepton Yukawa matrices are assumed
to be hierarchical in nature with an order unity entry in the 33 entry.
Another complementary approach is to assume that the Yukawa
matrices are democratic with order unity entry everywhere
\cite{Fritzsch:1989qm}, and both approaches have been followed
for neutrino masses and mixings
\cite{Fritzsch:1999ee,Jezabek:1999ta,Branco:1999yf,Xing:1999wz,Roberts:2001zy}.
A specific model of the neutrino mass matrix with
texture zeroes, but with a texture zero in the 33 position
was proposed by Zee \cite{Zee:1980ai}, and this has been developed
recently by a number of authors
\cite{Kitabayashi:2001ex,Koide:2000jm,Balaji:2001ex,Jarlskog:1998uf}.
Unfortunately the simplest Zee texture is now excluded by experiment,
although a non-minimal Zee type model remains viable \cite{He}.

To understand the origin of the postulated forms
of Yukawa matrices, one must appeal to some sort of
Family symmetry $G_{{\rm Family}}$,
which acts in the direction shown in Fig.\ref{fig3}.
In the framework of the see-saw mechanism, new physics beyond the
standard model is required to violate lepton number and generate
right-handed neutrino masses which are typically around the
GUT scale. This is also exciting since it implies that
the origin of neutrino masses is also related to some
GUT symmetry group $G_{{\rm GUT}}$, which unifies the
fermions within each family as shown in Fig.\ref{fig3}.

Putting these ideas together we are suggestively led to a framework of
new physics beyond the standard model based on N=1 SUSY
\footnote{Supersymmetry enables the gauge couplings to
meet at the GUT scale to give a self-consistent unification picture.}
with commuting
GUT and Family symmetry groups,
\beq
G_{{\rm GUT}}\times G_{{\rm FAM}}
\label{symmetry}
\eeq
There are many possible candidate GUT and Family symmetry groups
some of which are listed in Table \ref{table3}. Unfortunately the
model dependence does not end there, since the details of the symmetry
breaking vacuum plays a crucial role in specifying the model and
determining the masses and mixing angles, resulting in many models
as given in \cite{Achiman:1993ce} - \cite{Wu:1999yz}
(listed alphabetically).
These models may be classified according to the particular
GUT and Family symmetry they assume as shown in Table \ref{table3}.

\begin{table}[htb]
\caption{Some candidate GUT and Family symmetry groups,
and the papers that use these symmetries to successfully describe the
LMA MSW solar solution and the atmospheric neutrino data.}
\label{table3}
{
\begin{tabular}{|l|c|c|c|c|c|c|c|}
\hline
\hline
$\begin{array}{cc}
 & G_{{\rm FAM}} \\
G_{{\rm GUT}} &
\end{array}$
&$SU(3)$ &$SU(2)$ &$U(1)$ &$Z_N$ &$SO(3)$ & $S(3)$  & {\rm None} \\
\hline
\hline
$E_6$ & & \cite{Bando:2001bj}&\cite{Ling:2003kr} & & & &\cite{Achiman:1992ke} \\
\hline
$SO(10)$ & \cite{Ross:2002fb,King:2003rf} &
\cite{Chen:2002pa,Blazek:1999hz}
&\cite{Achiman:1993ce} & & &
&\cite{Babu:1995hr} \\
         &  &
\cite{Raby:2003ay}
&\cite{Albright:2001xq,Shafi:1999au} & & &
&\cite{Nomura:1998gm,Pati:2003qi} \\
\hline
$SU(5)$ & &\cite{Aranda:2001rd} 
&\cite{Altarelli:1999dg,Grimus:2002zh} & & & &\cite{Barr:1996ar} \\
\hline
51 & & &\cite{Ellis:1998nk} &  & & & \\
\hline
422 & \cite{King:2003rf} & &\cite{King:2000ge,Blazek:2003wz} & & & &\cite{Pati:2003qi} \\
\hline
(321)$^3$ & & &\cite{Ling:2003kr} & & & &\cite{Froggatt:2002tb} \\
\hline
3221 & & & \cite{Mohapatra:1998bp}& & & & \\
\hline
321 & \cite{King:2001uz} &\cite{Kuchimanchi:2002yu}
&\cite{Choi:2001rm} &\cite{Ghosal:2000wg}
&\cite{Frampton:1999hk,Ghosal:1999jb} &
\cite{Mohapatra:1999zr,Babu:2002ex} &
\cite{Vissani:1998xg} \\
 & & &\cite{Grossman:1998jj,Ma:2001md} & & &\cite{Harrison:2002kp,Wu:1999yz} & \\
\hline
\hline
\end{tabular}\\[2pt]
}
\end{table}

We have used the notation that
\bea
51 & \equiv & SU(5)\times U(1) \\
422 & \equiv & SU(4)_{PS}\times SU(2)_L\times SU(2)_R \\
3221& \equiv & SU(3)_C\times SU(2)_L\times SU(2)_R \times U(1)_{B-L}\\
321& \equiv & SU(3)_C\times SU(2)_L\times U(1)_Y
\eea
where 422 is the Pati-Salam gauge group, 3221 is the left-right
symmetric gauge extension, 321 is the Standard Model gauge group.

Another complication is that the masses and mixing angles determined
in some high energy theory must be run down to low energies using
the renormalisation group equations (RGEs)
\cite{Babu:qv,Casas:1999tg,Dutta:2002nq,Antusch:2002ek,Mohapatra:2003tw}.
Large radiative corrections are seen
when the see-saw parameters \cite{Nomura:1998gm} are tuned,
since the spectrum is sensitive to small changes in the parameters,
and this effect is sometimes used to magnify small mixing angles
into large ones \cite{Babu:qv,Ellis:1999my,Balaji:2000ma,Antusch:2002fr}.
This idea has however been criticised in \cite{Casas:2003kh}.
In natural models based on SRHND the parameters are not tuned,
since the hierarchy and large atmospheric and
solar angles arise naturally as discussed in the previous section.
Therefore in SRHND models the radiative
corrections to neutrino masses and mixing angles are only expected
to be a few per cent, and this has been verified numerically
\cite{King:2000hk}.

\subsection{$SO(10)\times U(1)$}
As an example we shall here consider a model based on
a GUT group $SO(10)$ and a family symmetry $U(1)$.
We shall suppose that the GUT symmetry is broken
via a Pati-Salam group and define the model in terms
of the subgroup $422$ \cite{King:2000ge}.
This model provides an example of the use of both the $U(1)$ family
symmetry to generate inter-family hierarchies, and the use
of Clebsch-Gordon coefficients from the GUT group to generate
intra-family structure.

The left-handed quarks and leptons are accommodated in the following
422 representations,
\begin{equation}
{\psi^i}=(4,2,1)=
\left(\begin{array}{cccc}
u^R & u^B & u^G & \nu \\ d^R & d^B & d^G & e^-
\end{array} \right)^i
\label{psi}
\end{equation}
\begin{equation}
{\bar{\psi}}^i=(\bar{4},1,\bar{2})=
\left(\begin{array}{cccc}
\bar{d}^R & \bar{d}^B & \bar{d}^G & e^+  \\
\bar{u}^R & \bar{u}^B & \bar{u}^G & \bar{\nu}
\end{array} \right)^i
\label{psic}
\end{equation}
where $i=1\ldots 3$ is a family index.  The Higgs
fields are contained in the following representations,
\begin{equation}
h=(1,\bar{2},2)=
\left(\begin{array}{cc}
  {h_2}^+ & {h_1}^0 \\ {h_2}^0 & {h_1}^- \\
\end{array} \right) \label{h}
\end{equation}
(where $h_1$ and $h_2$ are the low energy Higgs superfields associated
with the MSSM.)

The two heavy Higgs representations are
\begin{equation}
{H}^{\alpha b}=(4,1,2)=
\left(\begin{array}{cccc}
u_H^R & u_H^B & u_H^G & \nu_H \\ d_H^R & d_H^B & d_H^G & e_H^-
\end{array} \right) \label{H}
\end{equation}
and
\begin{equation}
{\bar{H}}_{\alpha x}=(\bar{4},1,\bar{2})=
\left(\begin{array}{cccc}
\bar{d}_H^R & \bar{d}_H^B & \bar{d}_H^G & e_H^+ \\
\bar{u}_H^R & \bar{u}_H^B & \bar{u}_H^G & \bar{\nu}_H
\end{array} \right). \label{barH}
\end{equation}

The Higgs fields are assumed to develop VEVs,
\begin{equation}
<H>\equiv<\nu_H>\sim M_{GUT}, \ \ <\bar{H}>\equiv<\bar{\nu}_H>\sim M_{GUT}
\label{HVEV}
\end{equation}
leading to the symmetry breaking at $M_{GUT}$
\begin{equation}
\mbox{SU(4)}\otimes \mbox{SU(2)}_L \otimes \mbox{SU(2)}_R
\longrightarrow
\mbox{SU(3)}_C \otimes \mbox{SU(2)}_L \otimes \mbox{U(1)}_Y
\label{422to321}
\end{equation}
in the usual notation.  Under the symmetry breaking in
Eq.\ref{422to321}, the Higgs field $h$ in Eq.\ref{h} splits into two
Higgs doublets $h_1$, $h_2$ whose neutral components subsequently
develop weak scale VEVs,
\begin{equation}
<h_1^0>=v_1, \ \ <h_2^0>=v_2 \label{vevs}
\end{equation}
with $\tan \beta \equiv v_2/v_1$.

To construct the quark and lepton mass matrices
we make use of non-renormalisable operators \cite{Ops} of the form:
\begin{eqnarray}
&\left.i\right)& \hspace{1cm} (\psi^i \bar{\psi}^j )h\left(\frac{H\bar{H}}{M^2}\right)^n
\left(\frac{\theta}{M}\right)^{p_{ij}}\label{Yuk_op}\\
&\left.ii\right)& \hspace{1cm} (\bar{\psi}^i\bar{\psi}^j )\left(\frac{HH}{M^2}\right)
\left(\frac{H\bar{H}}{M^2}\right) ^m
\left(\frac{\theta}{M}\right) ^{q_{ij}}.\label{Maj_op}
\end{eqnarray}
The $\theta$ fields are Pati-Salam singlets
which carry $U(1)$ family charge and develop VEVs which break the
$U(1)$ family symmetry.
They are required to be present in the operators above to balance the
charge of the invariant operators.
After the $H$ and $\theta$ fields acquire VEVs, they generate a hierarchy in
$\left.i\right)$ effective Yukawa couplings and $\left.ii\right)$ Majorana masses.  These
operators are assumed to originate from additional interactions at the scale $M>M_{GUT}$.  The
value of the powers $p_{ij}$ and $q_{ij}$ are determined by the assignment of $U(1)$ charges,
with $X_{\theta}=-1$ then $p_{ij}=(X_{\psi^{i}}+X_{\bar{\psi}^{j}}+X_{h})$ and
$q_{ij}=(X_{\bar{\psi}^{i}}+X_{\bar{\psi}^{j}}+X_{h})$.

The contribution to the third family Yukawa coupling is assumed to be only from the renormalisable operator with $n=p=0$ leading to Yukawa unification.
The contribution of an operator, with a given power $n$, to the matrices
$Y_{f=u,d,\nu,e},\,M_{RR}$ is determined by the
relevant Clebsch factors coming from the gauge contractions within
that operator.  A list of Clebsch factors for all $n=1$ operators can
be found in the appendix of \cite{King:2000ge}.  These Clebsch factors give
zeros for some matrices and not for others, hence a choice of
operators can be made such that a large 23 entry can be given to
$Y_{\nu}$ and not $Y_{u,d,e}$.  We shall write,
\beq
\delta = \frac{<H><\bar{H}>}{M^2}=0.22, \ \ \epsilon =
\frac{<\theta>}{M^2}=0.22,
\label{eq:de}
\eeq
then we can identify $\delta$ with mass
splitting within generations and $\epsilon$ with splitting between
generations.

The choice of $U(1)$ charges are as in \cite{King:2000ge}
and can be summarised as
$X_{\psi^{i}}~=~\left(1,0,0\right)$,
$X_{\bar{\psi}^{i}}~=~\left(4,2,0\right)$,
$X_{h}=0$, $X_{H}=0$ and $X_{\bar{H}}=0$.
This fixes the powers of $\epsilon$ in each entry of the Yukawa
matrix, but does not specify the complete operator.
The Yukawa couplings are specified by a particular choice of operators,
\cite{Blazek:2003wz,King:2000ge} with the property
\beq
{\cal O}\sim (H\bar{H})\sim \delta, \ \
{\cal O}^{\prime}\sim (H\bar{H})^2\sim \delta^2, \ \
{\cal O}^{\prime\prime}\sim (H\bar{H})^3\sim \delta^3.\label{OO'O''}
\eeq
The Clebsch factors play an important part in determing the
form of the Yukawa matrices.
The particular operator choice in \cite{King:2000ge} leads to the
quark and lepton mass matrices below. For example the Clebsch
coeficients from the leading operator in the 22 positon
gives the ratio $0:1:3$ in the $Y_{U,D,E}$ matrices.
This ratio along
with subleading corrections provides the correct $m_{c}:m_{s}:m_{\mu}$
ratio \cite{Georgi:1979df}.

The final form of the Yukawa matrices is \cite{Blazek:2003wz},
\begin{eqnarray}
Y^{u} &\approx &\left(
\begin{array}{lcr}
\sqrt{2}\delta^3\epsilon^5 & \sqrt{2}\delta^2\epsilon^3 &
\frac{2}{\sqrt{5}}\delta^2\epsilon \\
0 & \frac{8}{5\sqrt{5}}\delta^2\epsilon^2 & 0\\
0 & \frac{8}{5}\delta^2\epsilon^2 & r_t
\end{array}%
\right),  \label{Yu1} \\
Y^{d} &\approx &\left(
\begin{array}{llr}
\frac{8}{5}\delta\epsilon^5 & -\sqrt{2}\delta^2\epsilon^3 &
\frac{4}{\sqrt{5}}\delta^2\epsilon \\
\frac{2}{\sqrt{5}}\delta\epsilon^4  &
\left[\sqrt{\frac{2}{5}}\delta\epsilon^2+
\frac{16}{5\sqrt{5}}\delta^2\epsilon^2 \right]& \sqrt{\frac{2}{5}}\delta^2\\
\frac{8}{5}\delta\epsilon^5 & \sqrt{2}\delta\epsilon^2 & r_b
\end{array}%
\right),  \label{Yd1} \\
Y^{e} &\approx &\left(
\begin{array}{llr}
\frac{6}{5}\delta\epsilon^5 & 0 & 0 \\
\frac{4}{\sqrt{5}}\delta \epsilon^4  &
\left[-3\sqrt{\frac{2}{5}}\delta \epsilon^2+
\frac{12}{5\sqrt{5}}\delta^2\epsilon^2 \right]& -3\sqrt{\frac{2}{5}}\delta^2\\
\frac{6}{5}\delta \epsilon^5 & \sqrt{2}\delta \epsilon^2 & 1
\end{array}%
\right),  \label{Ye1} \\
Y^{\nu} &\approx &\left(
\begin{array}{llr}
\sqrt{2}\delta^3\epsilon^5 & {2}\delta \epsilon^3 & 0 \\
0 & \frac{6}{5\sqrt{5}}\delta^2\epsilon^2 & 2\delta\\
0 & \frac{6}{5}\delta^2\epsilon^2 & r_{\nu}
\end{array}%
\right),  \label{Ynu1}
\end{eqnarray}
where the numerical Clebsch factors are displayed explicitly, and
$r_t,r_b,r_\nu$ are order unity parameters which quantify the
deviations from exact Yukawa unification \cite{Blazek:2003wz},
but all other order unity coefficients have been dropped.

The Majorana operators are assumed to arise from an $m=0$ operator
in the 33 position and $m=1$ operators elsewhere, resulting in
\begin{equation}
M_{RR}\approx \left(
\begin{array}{ccr}
\delta \epsilon^8 & \delta \epsilon^6 & \delta \epsilon^4 \\
\delta \epsilon^6 & \delta \epsilon^4 & \delta \epsilon^2 \\
\delta \epsilon^4 & \delta \epsilon^2 & 1%
\end{array}%
\right) M_{3}.  \label{MRR1}
\end{equation}

In the neutrino sector the matrices above satisfy the
condition of sequential dominance in which a
neutrino mass hierarchy naturally results with the
heaviest (third) right-handed neutrino being mainly responsible
for the atmospheric neutrino mass, and the second heaviest
right-handed neutrino being mainly responsible for the solar neutrino
mass. Thus this model corresponds to HSDa in Table \ref{table2}.
Using the HSDa ordering in Table \ref{table2} with the
matrices in Eqs.\ref{Ynu1},\ref{MRR1} we can use the analytic results
in Eqs.\ref{m1}-\ref{13} to give estimates of neutrino masses
\begin{eqnarray}
m_{1} &\sim & {\delta^5}{\epsilon^2} \frac{v_{2}^{2}}{M_{3}}\\
m_{2} &\approx & \frac{4\delta \epsilon^2}{s_{12}^2}\frac{v_{2}^{2}}{M_{3}}\\
m_{3} &\approx & (4\delta^2+r_{\nu}^2)\frac{v_{2}^{2}}{M_{3}}
\end{eqnarray}%
and neutrino mixing angles:
\begin{eqnarray}
\tan \theta _{23}^{\nu } &\approx & \frac{2\delta}{r_{\nu}}\\
\tan \theta _{12}^{\nu } &\approx &
\frac{2}{\left( c_{23}\frac{6}{5\sqrt{2}}-s_{23}\frac{6}{5}  \right)}
\frac{\epsilon}{\delta}
\\
\theta _{13}^{\nu } &\approx &
\frac{\frac{12}{5}\delta^2\epsilon\left(\frac{2\delta}{\sqrt{5}}+r_{\nu}
\right)}
{\left((2\delta)^2+r_{\nu}^2\right)^{3/2}}
\end{eqnarray}%
which are a good fit to the LMA MSW solution for
$\epsilon$ and $\delta$ as in Eq.\ref{eq:de}.

\subsection{$SO(10)\times SU(3)$}
As an example of a model based on a non-Abelian family symmetry,
we briefly review the model proposed in \cite{King:2003rf}.
The model uses the largest family symmetry $SU(3)$ consistent
with $SO(10)$ GUTs. An important further
motivation for $SU(3)$ family symmetry is, in the framework of
sequential dominance,
to relate the second and third entries of the Yukawa matrix, as required to
obtain an almost maximal 23 mixing in the atmospheric neutrino
sector \cite{King:2001uz}.
In this framework we already saw that the theoretical requirements that
the neutrino Yukawa matrix resembles the quark Yukawa matrices,
and therefore has a large 33 element with no large off-diagonal
elements and a texture zero in the 11 position \cite{Gatto:ss},
leads uniquely to
LSDa in Table \ref{table2}, where the dominant right-handed neutrino
is the first (lightest) one. Assuming this then
the atmospheric neutrino mixing angle
is given by
$\tan \theta_{23}^{\nu}\approx Y^{\nu}_{21}/Y^{\nu}_{31}\approx 1$.
The sequential dominance conditions which were assumed in
Eq.\ref{srhnd} will here be derived from the symmetries of the model.
Thus this model provides an example of the application of sequential
dominance to realistic models of flavour, and shows how the conditions
of sequential dominance which were simply assumed earlier can motivate
models based on GUTs and family symmetry which are capable of
explaining these conditions. In other words, the conditions for
sequential dominance can provide clues to the nature of the underlying
flavour theory.

The starting point of the model is the observation that an excellent
fit to all quark data is given by the approximately symmetric form
of quark Yukawa matrices \cite{Roberts:2001zy}
\begin{equation}
Y^{u}\propto \left(
\begin{array}{ccc}
0 & \epsilon ^{3} & O(\epsilon ^{3}) \\
. & \epsilon ^{2} & O(\epsilon ^{2}) \\
. & . & 1%
\end{array}%
\right) ,\ \ \ \ Y^{d}\propto \left(
\begin{array}{crc}
0 & 1.5\bar{\epsilon}^{3} & 0.4\bar{\epsilon}^{3} \\
. & \bar{\epsilon}^{2} & 1.3\bar{\epsilon}^{2} \\
. & . & 1%
\end{array}%
\right)  \label{yuk}
\end{equation}%
where the expansion parameters $\epsilon $ and $\bar{\epsilon}$ are given by
\begin{equation}
\epsilon \approx 0.05,\ \ \bar{\epsilon}\approx 0.15.  \label{exp}
\end{equation}

This motivates a particular model in which
the three families are unified as triplets under
an $SU(3)$ family symmetry, and $16's$ under an $SO(10)$ GUT
\cite{King:2001uz,Ross:2002fb,King:2003rf},
\begin{equation}
\psi_i= (3,16),
\end{equation}
where as before the $SO(10)$ is broken via the Pati-Salam group giving
the equivalent 422 reps in Eqs.\ref{psi},\ref{psic},
\begin{equation}
{\psi_i}=(3,4,2,1),\ \ {\bar{\psi}}_i=(3,\bar{4},1,\bar{2}).
\end{equation}
Further symmetries $R\times Z_2\times U(1)$
are assumed to ensure that
the vacuum alignment leads to a universal form of Dirac mass matrices
for the neutrinos, charged leptons and quarks \cite{King:2003rf}.
To build a viable model we also need spontaneous breaking of the family
symmetry
\begin{equation}
SU(3)\longrightarrow SU(2)\longrightarrow {\rm Nothing}  \label{fsb}
\end{equation}%
To achieve this symmetry breaking additional Higgs fields $\phi
_{3},$ $\overline{\phi }_{3},$ $\phi _{23}$ and $\overline{\phi }_{23}$
are required. The largeness of the third
family fermion masses implies that $SU(3)$ must be strongly broken by new
Higgs antitriplet fields $\phi _{3}$ which develop a vev in the third $SU(3)$
component $<\phi _{3}>^{T}=(0,0,a_{3})$ as in \cite{King:2001uz}.
$\phi _{3}^{i}$ transforms under
$SU(2)_{R}$ as ${\bf 3\oplus 1}$ rather than being $SU(2)_{R}$ singlets as
assumed in \cite{King:2001uz}, and develops vevs in the $SU(3)\times
SU(2)_{R}$ directions
\begin{equation}
<\phi _{3}>=<\overline{\phi _{3}}>=\left(
\begin{array}{c}
0 \\
0 \\
1%
\end{array}%
\right) \otimes \left(
\begin{array}{cc}
a_{3}^{u} & 0 \\
0 & a_{3}^{d}%
\end{array}%
\right) .  \label{phi3vev}
\end{equation}%
The symmetry breaking also involves the $SU(3)$ antitriplets $\phi _{23}$
which develop vevs \cite{King:2001uz}
\begin{equation}
<\phi _{23}>=\left(
\begin{array}{c}
0 \\
1 \\
1%
\end{array}%
\right) b,  \label{phi23vevs}
\end{equation}%
where, as in \cite{King:2001uz}, vacuum alignment ensures that the vevs are
aligned in the 23 direction. Due to D-flatness there must also be
accompanying Higgs triplets such as $\overline{\phi _{23}}$ which develop
vevs \cite{King:2001uz}
\begin{equation}
<\overline{\phi _{23}}>=\left(
\begin{array}{c}
0 \\
1 \\
1%
\end{array}%
\right) b.  \label{phibar23vevs}
\end{equation}%
We also introduce an
adjoint $\Sigma$ field which develops vevs in the $SU(4)_{PS}\times SU(2)_R$
direction which preserves the hypercharge generator $Y=T_{3R}+(B-L)/2$, and
implies that any coupling of the $\Sigma$ to a fermion and a messenger such
as $\Sigma^{a \alpha}_{b \beta}\psi^c_{a\alpha}\chi^{b\beta}$, where the $%
SU(2)_R$ and $SU(4)_{PS}$ indices have been displayed explicitly, is
proportional to the hypercharge $Y$ of the particular fermion component of $%
\psi^c$ times the vev $\sigma$. In addition a $\theta$ field is
required for the construction of Majorana neutrino masses.

The leading operators allowed by the symmetries are
\begin{eqnarray}
P_{{\rm Yuk}} &\sim &\frac{1}{M^{2}}\psi _{i}\phi _{3}^{i}\bar{\psi} _{j}\phi
_{3}^{j}h  \label{op1} \\
&+&\frac{\Sigma }{M^{3}}\psi _{i}\phi _{23}^{i}\bar{\psi} _{j}\phi _{23}^{j}h
\label{op2} \\
P_{{\rm Maj}} &\sim &\frac{1}{M}\bar{\psi} _{i}\theta ^{i}\theta ^{j}
\bar{\psi}_{j}  \label{mop1}
\end{eqnarray}%
where the operator mass scales, generically denoted by $%
M$ may differ and we have suppressed couplings of $O(1).$

The final form of the Yukawa matrices and heavy Majorana matrix
after inserting a particular choice of order unity coefficients is
\cite{King:2003rf}
\begin{eqnarray}
Y^{u} &\approx &\left(
\begin{array}{llr}
0 & 1.2\epsilon ^{3}& 0.9\epsilon ^{3} \\
-1.2\epsilon ^{3} & -\frac{2}{3}\epsilon ^{2} & -\frac{2}{3}\epsilon
^{2} \\
-0.9\epsilon ^{3} & -\frac{2}{3}\epsilon ^{2} & 1
\end{array}%
\right) \bar{\epsilon},  \label{Yu} \\
Y^{d} &\approx &\left(
\begin{array}{llr}
0 & 1.6\bar{\epsilon}^{3} & 0.7\bar{\epsilon}^{3}
\\
-1.6\bar{\epsilon}^{3} & \bar{\epsilon}^{2} & \bar{\epsilon}^{2}+
\bar{\epsilon}^{\frac{5}{2}} \\
-0.7\bar{\epsilon}^{3} & \bar{\epsilon}^{2} & 1
\end{array}%
\right) \bar{\epsilon},  \label{Yd} \\
Y^{e} &\approx &\left(
\begin{array}{llr}
0 & 1.6\bar{\epsilon}^{3} & 0.7\bar{\epsilon}^{3}
\\
-1.6\bar{\epsilon}^{3} & 3\bar{\epsilon}^{2} & 3\bar{\epsilon}^{2}\\
-0.7\bar{\epsilon}^{3} & 3\bar{\epsilon}^{2} & 1
\end{array}%
\right) \bar{\epsilon},  \label{Ye} \\
Y^{\nu } &\approx &\left(
\begin{array}{llr}
0 & 1.2\epsilon ^{3} & 0.9\epsilon ^{3} \\
-1.2\epsilon ^{3} & -\alpha \epsilon ^{2} & -\alpha \epsilon ^{2}\\
-0.9\epsilon ^{3} & -\alpha \epsilon ^{2}-\epsilon ^{3} & 1
\end{array}%
\right) \bar{\epsilon}.  \label{Ynu}
\end{eqnarray}
\begin{equation}
M_{RR}\approx \left(
\begin{array}{ccr}
\epsilon ^{6}\bar{\epsilon}^{3} & 0 & 0 \\
0 & \epsilon ^{6}\bar{\epsilon}^{2} & 0 \\
0 & 0 & 1%
\end{array}%
\right) M_{3}.  \label{MRRA}
\end{equation}

The model gives excellent agreement with the quark and
lepton masses and mixing angles. For the up and down quarks the form of $%
Y^{u}$ and $Y^{d}$ given in Eq.\ref{Yu}, \ref{Yd} is consistent with the
phenomenological fit in Eq.\ref{yuk}.
The charged lepton mass matrix is of the Georgi-Jarslkog \cite{Georgi:1979df}
form which, after
including radiative corrections, gives an excellent description of the
charged lepton masses. In the neutrino sector the parameters satisfy the
conditions of sequential dominance \ref{srhnd},
with the lightest right-handed neutrino
giving the dominant contribution to the heaviest physical neutrino mass, and
the second right-handed neutrino giving the leading subdominant
contribution, providing that $\alpha \sim \epsilon $.
It thus falls into the category of LSDa in Table \ref{table2}.

Analytic estimates of neutrino masses and mixing angles for sequential
dominance were derived in \cite{King:2002nf}, and for the special case here
of LSDa, with the 11 neutrino Yukawa coupling equal to
zero, they are given in
Eqs.\ref{m1}-\ref{13} from which
the analytic estimates below for the neutrino masses are obtained,
\begin{eqnarray}
m_{1} &\sim &\bar{\epsilon}^{2}\frac{v_{2}^{2}}{M_{3}} \\
m_{2} &\approx & 5.8\frac{v_{2}^{2}}{M_{3}} \\
m_{3} &\approx &15\frac{v_{2}^{2}}{M_{3}}
\end{eqnarray}%
and neutrino mixing angles:
\begin{eqnarray}
\tan \theta _{23}^{\nu } &\approx &  1.3 \\
\tan \theta _{12}^{\nu } &\approx &  0.66 \\
\theta _{13}^{\nu } &\approx &1.6\bar{\epsilon}
\end{eqnarray}%
Note that the physical lepton mixing angle $\theta_{13}$ receives
a large contribution from the neutrino sector $\theta_{13}^{\nu}\sim 0.3$
at the high energy scale, for this choice of parameters,
compared to the current CHOOZ limit
$\theta_{13}\leq 0.2$ \cite{Apollonio:1999ae}. However the physical
mixing angles will receive charged lepton contributions
\cite{King:2002nf} and all
the parameters are subject to radiative corrections in running from the high
energy scale to low energies, although in sequential dominance
models these corrections are only a few per cent
\cite{King:2000hk}. Thus the
model predicts that $\theta_{13}$ is close
to the current CHOOZ limit, and could be observed by
the next generation of long baseline experiments such as MINOS or OPERA.

\section{Conclusions}

This focus of the review has been on ``mainstream'' neutrino mass
models, defined as see-saw models involving three active neutrinos which
are capable of describing both the atmospheric neutrino oscillation
data, and the LMA MSW solar solution, which is now
uniquely specified by recent data. We have briefly reviewed the current
experimental status, showed how to parametrise and construct
the neutrino mixing matrix, and presented the leading order neutrino
Majorana mass matrices. We then introduced the see-saw mechanism,
and discussed a natural application of it to current data
using the sequential dominance mechanism, which we compared to
an early proposal for obtaining large mixing angles.
We showed how both the Standard Model and the Minimal
Supersymmetric Standard Model may be extended to incorporate
the see-saw mechanism, and showed how the latter case leads to the
expectation of lepton flavour violation.
The see-saw mechanism motivates models with additional
symmetries such as unification and family symmetry models,
and we tabulated some possible models, before
focussing on two particular models based on $SO(10)$ grand unification
and either $U(1)$ or $SU(3)$ family symmetry as specific examples.
We have provided extensive appendices which include
techniques for analytically diagonalising different types
of mass matrices involving two large mixing angles and one
small mixing angle, to leading order in the small mixing angle.

Neutrino physics has witnessed a renaissance
period with the watershed provided by Super-Kamiokande in 1998.
Before then we did not know whether atmospheric and solar neutrino
oscillations were fact or fancy. Now we know they are fact whose explanation
requires two large mixing angles. We have seen in this review that
there is no shortage of theoretical models which can account for
these data. Even discounting those theoretical
models which predicted a small solar angle, or vacuum oscillation
mass splittings, there are many many models that can describe
the current data. In this review we have tried to emphasise useful
approaches and techniques, rather than giving a detailed catalogue
of all possible models. We make no apology for emphasising the see-saw
mechanism, which is probably the most elegant way of accounting
for small neutrino masses. We have further shown that
the see-saw mechanism may be successfully applied to the atmospheric and
solar data to yield a neutrino mass hierarchy and two large mixing angles
in a technically natural and elegant way using the
idea of sequential dominance. Sequential dominance requires certain
mild conditions to apply, and we have seen that these conditions
may in turn arise from the symmetries of realistic models.

The problem of neutrino masses and mixings should be addressed
in the wider context of the problem of all quark and lepton masses
and mixing angles, and in this wider context we have emphasised
ideas such as unification and family symmetry which will surely play
a role in the ultimate solution to the problem of flavour.
In Table \ref{table3} we have classified successful models according to
the different unification and family symmetries upon which they are
based. It remains to be seen if any of these models will turn out to
provide the solution to the problem of flavour. If this turns out
to be not the case, then the effort will not have been in vain, since
it is quite likely that some of the ideas on which these models are
based will survive. Here we have emphasised particularly promising
ideas such as the see-saw mechanism, sequential dominance, supersymmetry,
unification and family symmetry, which when combined with the neutrino
data could help to unlock the whole mystery of flavour.

\begin{center}
{\bf Acknowledgements}
\end{center}
I would like to thank Stefan Antusch for reading the manuscript and
PPARC for the support of a Senior Fellowship.

\appendix

\section{Equivalence of different parametrisations}
\label{equivalence}

In this appendix we exhibit the equivalence of different
parametrisations of the lepton mixing matrix.
A $3\times 3$ unitary matrix may be parametrised by 3 angles and 6
phases. We shall find it convenient to parametrise a unitary
matrix $V^{\dagger}$ by
\footnote{It is convenient to define the
parametrisation of $V^{\dagger}$ rather than $V$
because the lepton mixing matrix involves ${V^{\nu_L}}^{\dagger}$
and the neutrino mixing angles will play a central r\^{o}le.}:
\beq
V^{\dagger}=P_2R_{23}R_{13}P_1R_{12}P_3
\label{V1}
\eeq
where $R_{ij}$ are a sequence of real rotations corresponding to the
Euler angles $\theta_{ij}$, and $P_i$ are diagonal phase matrices.
The Euler matrices are given by
\begin{equation}
R_{23}=
\left(\begin{array}{ccc}
1 & 0 & 0 \\
0 & c_{23} & s_{23} \\
0 & -s_{23} & c_{23} \\
\end{array}\right)
\label{R23}
\end{equation}
\begin{equation}
R_{13}=
\left(\begin{array}{ccc}
c_{13} & 0 & s_{13} \\
0 & 1 & 0 \\
-s_{13} & 0 & c_{13} \\
\end{array}\right)
\label{R13}
\end{equation}
\begin{equation}
R_{12}=
\left(\begin{array}{ccc}
c_{12} & s_{12} & 0 \\
-s_{12} & c_{12} & 0\\
0 & 0 & 1 \\
\end{array}\right)
\label{R12}
\end{equation}
where $c_{ij} = \cos\theta_{ij}$ and $s_{ij} = \sin\theta_{ij}$.
The phase matrices are given by
\beq
P_1=
\left( \begin{array}{ccc}
1 & 0 & 0    \\
0 & e^{i\chi} & 0 \\
0 & 0 & 1
\end{array}
\right)
\label{P1}
\eeq
\beq
P_2=
\left( \begin{array}{ccc}
1 & 0 & 0    \\
0 & e^{i\phi_2} & 0 \\
0 & 0 & e^{i\phi_3}
\end{array}
\right)
\label{P2}
\eeq
\beq
P_3=
\left( \begin{array}{ccc}
e^{i\omega_1} & 0 & 0    \\
0 & e^{i\omega_2} & 0 \\
0 & 0 & e^{i\omega_3}
\end{array}
\right)
\label{P3}
\eeq

By commuting the phase matrices to the left, it is not difficult to
show that the parametrisation in Eq.\ref{V1} is equivalent to
\beq
V^{\dagger}=PU_{23}U_{13}U_{12}
\label{V2}
\eeq
where $P=P_1P_2P_3$ and
\begin{equation}
U_{23}=
\left(\begin{array}{ccc}
1 & 0 & 0 \\
0 & c_{23} & s_{23}e^{-i\delta_{23}} \\
0 & -s_{23}e^{i\delta_{23}} & c_{23} \\
\end{array}\right)
\label{U23}
\end{equation}
\begin{equation}
U_{13}=
\left(\begin{array}{ccc}
c_{13} & 0 & s_{13}e^{-i\delta_{13}} \\
0 & 1 & 0 \\
-s_{13}e^{i\delta_{13}} & 0 & c_{13} \\
\end{array}\right)
\label{U13}
\end{equation}
\begin{equation}
U_{12}=
\left(\begin{array}{ccc}
c_{12} & s_{12}e^{-i\delta_{12}} & 0 \\
-s_{12}e^{i\delta_{12}} & c_{12} & 0\\
0 & 0 & 1 \\
\end{array}\right)
\label{U12}
\end{equation}
where
\beq
\delta_{23}=\chi+\omega_2-\omega_3
\eeq
\beq
\delta_{13}=\omega_1-\omega_3
\eeq
\beq
\delta_{12}=\omega_1-\omega_2
\eeq
The matrix $U$ is an example of a unitary matrix,
and as such
it may be parametrised by either of the equivalent forms
in Eqs.\ref{V1} or \ref{V2}.
If we use the form in Eq.\ref{V2} then the phase matrix $P$ on the
left may always be removed by an additional charged lepton phase rotation
$\Delta V^{E_L}=P^\dagger$,
which is always possible since
right-handed charged lepton phase rotations can always make the charged
lepton masses real. Therefore $U$ can always be parametrised by
\beq
U=U_{23}U_{13}U_{12}
\label{MNS2A}
\eeq
which involves just three irremoveable physical phases $\delta_{ij}$.
In this parametrisation the Dirac phase $\delta$
which enters the CP odd part of
neutrino oscillation probabilities is given by
\beq
\delta = \delta_{13}-\delta_{23}-\delta_{12}.
\label{DiracA}
\eeq

Another common parametrisation of the lepton mixing matrix is
\beq
U=R_{23}U_{13}R_{12}P_0
\label{MNS3}
\eeq
where
\beq
P_0=
\left( \begin{array}{ccc}
e^{i\beta_1} & 0 & 0    \\
0 & e^{i\beta_2} & 0 \\
0 & 0 & 1
\end{array}
\right)
\eeq
and in Eq.\ref{MNS3} $U_{13}$ is of the form in
Eq.\ref{U13} but with $\delta_{13}$ replaced by the Dirac phase $\delta$.
The parametrisation in Eq.\ref{MNS3} can be transformed into
the parametrisation in Eq.\ref{MNS2A} by commuting the phase matrix $P_0$
in Eq.\ref{MNS3} to the left, and then removing the phases
on the left-hand side by charged lepton phase rotations.
The two parametrisations are then related by the phase relations
\beq
\delta_{23}=\beta_2
\eeq
\beq
\delta_{13}=\delta + \beta_1
\eeq
\beq
\delta_{12}=\beta_1-\beta_2
\eeq
The use of the parametrisation in Eq.\ref{MNS3} is widespread in the
literature, however for the reasons discussed in the next sub-section
we prefer to use the parametrisation in Eq.\ref{MNS2A}
which is trivially related to Eq.\ref{MNS3} by the above phase
relations.

\section{Three Family Oscillation Formulae}

At a Neutrino Factory it is relatively
straightforward to measure the angle $\theta_{13}$
using the Golden Signature
of ``wrong sign'' muons. The effect relies on the full three family
oscillation formulae which we discuss in this Appendix.
For example suppose there are positive
muons circulating in the storage ring, then these decay as
$\mu^+ \rightarrow e^+\nu_e \bar{\nu}_\mu$ giving a mixed beam
of electron neutrinos and muon anti-neutrinos. The
muon anti-neutrinos will interact in the far detector to produce
positive muons. Any ``wrong sign'' negative muons which
may be observed can only arise from the neutrino oscillation
of electron neutrinos into muon neutrinos with probability given
by  a CP conserving part $P^{+}$ and a  a CP violating part
$P^{-}$. The exact formulae in vacuum are given by:
\be
P(\nu_{e}\rightarrow \nu_{\mu})= P^+(\nu_{e}\rightarrow \nu_{\mu})+
P^-(\nu_{e}\rightarrow \nu_{\mu})
\ee
\be
P(\bar{\nu}_{e}\rightarrow \bar{\nu}_{\mu})= P^+(\bar{\nu}_{e}\rightarrow \bar{\nu}_{\mu})+
P^-(\bar{\nu}_{e}\rightarrow \bar{\nu}_{\mu})
\ee
where the CP conserving parts are
\bea
P^+(\nu_{e}\rightarrow \nu_{\mu})  =
P^+(\bar{\nu}_{e}\rightarrow \bar{\nu}_{\mu})  =
& - & 4Re(U_{e1}U_{\mu 1}^*U_{e2}^*U_{\mu 2})
\sin^2(1.27\Delta m_{21}^2L/E) \nonumber \\
& - & 4Re(U_{e1}U_{\mu 1}^*U_{e3}^*U_{\mu 3})
 \sin^2(1.27\Delta m_{31}^2L/E) \nonumber \\
& - & 4Re(U_{e2}U_{\mu 2}^*U_{e3}^*U_{\mu 3})
\sin^2(1.27\Delta m_{32}^2L/E)
\label{P+}
\eea
and the CP violating parts are
\bea
& & P^-(\nu_{e}\rightarrow \nu_{\mu})
= -P^-(\bar{\nu}_{e}\rightarrow \bar{\nu}_{\mu})  =
 - c_{13}\sin2\theta_{13}\sin2\theta_{12}\sin2\theta_{23}\sin \delta
\nonumber \\
& \times & \sin (1.27\Delta m_{21}^2L/E) \sin (1.27\Delta m_{31}^2L/E)
\sin (1.27\Delta m_{32}^2L/E)
\label{P-}
\eea
Note that $P^-$ requires all three families to contribute, and it
vanishes if any mixing angle or mass splitting is zero.
The angle $\theta_{13}$ may easily be extracted from
$U_{e3}$ in the dominant CP conserving term $P^+$.

In order to determine the CP violating phase $\sin \delta$
it is necessary to measure the CP violating term $P^-$.
In order to do this one must compare the result for
$P(\nu_{e}\rightarrow \nu_{\mu})$ to the result to the case where
the positive muons in the storage ring are replaced by negative muons
and the analagous experiment is performed to measure
$P(\bar{\nu}_{e}\rightarrow \bar{\nu}_{\mu})$.
The CP violating asymmetry due to the CP violating phase $\delta$
is given by
\be
A^{\delta}= \frac{P(\nu_{e}\rightarrow \nu_{\mu})-P(\bar{\nu}_{e}\rightarrow \bar{\nu}_{\mu})}
{P(\nu_{e}\rightarrow \nu_{\mu})+P(\bar{\nu}_{e}\rightarrow \bar{\nu}_{\mu})}
\label{ACP}
\ee
from which we obtain
\be
A^{\delta}= \frac{P^-(\nu_{e}\rightarrow \nu_{\mu})}
{P^+(\nu_{e}\rightarrow \nu_{\mu})}
\approx \frac{\sin 2\theta_{12}\sin \delta}{\sin \theta_{13}}
\sin (1.27\Delta m_{21}^2L/E)
\ee
It is clear that in order to measure the CP asymmetry we require large
$\theta_{12}$ and large $\Delta m_{21}^2$ and this corresponds to
the LMA MSW solution. In addition we require large $\sin \delta$.
Also it would seem that having small $\theta_{13}$ enhances the CP
asymmetry, however it should be remembered that the CP asymmetric
rate $P^-$ in Eq.\ref{P-} is proportional to $\sin2\theta_{13}$,
and so $\theta_{13}$ should not be too small otherwise the number of
events will be too small.

Unfortunately life is not quite as simple as the above discussion
portrays. The Earth is made from matter and not anti-matter and so
CP will be violated by matter effects as the neutrino beam passes
through the Earth from the muon storage ring to the far detector.
For example the matter effects will modify the formulas for
$P(\nu_{e}\rightarrow \nu_{\mu})$ involving
$\theta_{13}$ and $\Delta m_{31}^2$ as follows:
\bea
\sin2\theta_{13} & \rightarrow &
\frac{\sin2\theta_{13}}
{\left( \frac{A}{\Delta
m_{31}^2}-\cos2\theta_{13}\right)^2+\sin^22\theta_{13}}
\nonumber \\
\Delta m_{31}^2 & \rightarrow &
\Delta m_{31}^2
\sqrt{\left( \frac{A}{\Delta
m_{31}^2}-\cos2\theta_{13}\right)^2+\sin^22\theta_{13}}
\label{matter}
\eea
where
\be
A=7.6\times 10^{-5}\rho E
\ee
where $\rho$ is the density of the Earth in gcm$^{-3}$ and $E$ is
the beam energy in GeV. The point is that
for $P(\bar{\nu}_{e}\rightarrow \bar{\nu}_{\mu})$
the sign of $A$ is reversed.
From one point of view
this is good news, since unlike the
vacuum oscillation formulae, $\Delta m_{31}^2$ enters linearly,
not quadratically, and so matter effects enable
the sign of the mass squared splitting to be determined in a
rather straightforward way.

However from the point of view of
measuring $\sin \delta$ it leads to complications since the
asymmetry in the rate in Eq.\ref{ACP} can get contributions from
both intrinsic CP violation and from matter induced CP violation,
and the measured asymmetry is a sum of the two effects
\be
A^{CP}=A^{\delta} + A^{matter}
\ee
Since both effects are by themselves rather small, it will be a
very difficult job to disentangle them, and the optimal
strategy continues to be studied \cite{Autin:2003xk}.
The optimal place to sit in order to observe CP violation
seems to be at the peak of $\sin (1.27\Delta m_{32}^2L/E)$ in
order to maximise $P^-$ according to Eq.\ref{P-}
(certainly we should avoid being at its node otherwise CP
violation vanishes).
In order to do this efficiently
it may be desirable to have energy-tunable beams, and it is certainly
necessary to have a good understanding of the density profile
of the Earth. Assuming the LMA solution, the prospects for
measuring CP violation at a Neutrino Factory are good.

\section{Charged lepton contributions to the lepton mixing matrix }
\label{charged}

In this appendix we discuss the contribution of the
charged lepton mixing angles to the lepton mixing matrix.
The lepton mixing matrix is constructed in Eq.\ref{MNS} as a product
of a unitary matrix from the charged lepton sector $V^{E_L}$
and a unitary matrix from the neutrino sector ${V^{\nu_L}}^{\dagger}$.
Each of these unitary matrices may be parametrised by the
parametrisation of $V^{\dagger}$ in Eq.\ref{V1}.
Thus we write
\beq
{V^{\nu_L}}^{\dagger}
=P_2^{\nu_L}R_{23}^{\nu_L}R_{13}^{\nu_L}P_1^{\nu_L}R_{12}^{\nu_L}P_3^{\nu_L}
\label{VnuLB}
\eeq
\beq
{V^{E_L}}^{\dagger}
=P_2^{E_L}R_{23}^{E_L}R_{13}^{E_L}P_1^{E_L}R_{12}^{E_L}P_3^{E_L}
\label{VELB}
\eeq
where the Euler angles and phases are defined as in
Eqs.\ref{R23}-\ref{P3}
but now there are independent angles
and phases for the left-handed neutrino and charged lepton sectors
distinguished by the superscripts $\nu_L$ and $E_L$.
The lepton mixing matrix from Eqs.\ref{MNS},\ref{VnuLB},\ref{VELB} is then
\beq
U=
{P_3^{E_L}}^{\dagger}{R_{12}^{E_L}}^{\dagger}{P_1^{E_L}}^{\dagger}
{R_{13}^{E_L}}^{\dagger}{R_{23}^{E_L}}^{\dagger}{P_2^{E_L}}^{\dagger}
P_2^{\nu_L}R_{23}^{\nu_L}R_{13}^{\nu_L}P_1^{\nu_L}R_{12}^{\nu_L}P_3^{\nu_L}
\label{MNS4}
\eeq
As before we commute all the phase matrices to the left, then choose
${P_3^{E_L}}^{\dagger}$ to cancel all the phases on the left-hand
side, to leave just
\beq
U=
{U_{12}^{E_L}}^{\dagger}
{U_{13}^{E_L}}^{\dagger}{U_{23}^{E_L}}^{\dagger}
U_{23}^{\nu_L}U_{13}^{\nu_L}U_{12}^{\nu_L}
\label{MNS5}
\eeq
with independent phases and angles
for the left-handed neutrino and charged lepton sectors,
in the convention of Eqs.\ref{U23},\ref{U13},\ref{U12}.
The phases in Eq.\ref{MNS5} are given in terms of the phases in
Eqs.\ref{VnuLB}, \ref{VELB} by
\bea
\delta_{12}^{\nu_L}&=&\omega_1^{\nu_L}-\omega_2^{\nu_L}
\label{B5}\\
\delta_{13}^{\nu_L}&=&\omega_1^{\nu_L}-\omega_3^{\nu_L}
\label{B6}\\
\delta_{23}^{\nu_L}&=&\chi^{\nu_L}+\omega_2^{\nu_L}-\omega_3^{\nu_L}
\label{B7}\\
\delta_{23}^{E_L}&=&
-\phi_2^{E_L}+\phi_3^{E_L}+\phi_2^{\nu_L}-\phi_3^{\nu_L}
+\chi^{\nu_L}+\omega_2^{\nu_L}-\omega_3^{\nu_L}
\\
\delta_{13}^{E_L}&=&\phi_3^{E_L}-\phi_3^{\nu_L}
+\omega_1^{\nu_L}-\omega_3^{\nu_L}
\\
\delta_{12}^{E_L}&=&\chi^{E_L}+\phi_2^{E_L}-\phi_2^{\nu_L}
-\chi^{\nu_L}+\omega_1^{\nu_L}-\omega_2^{\nu_L}
\eea
The form of $U$ in Eq.\ref{MNS5} is similar to the
parametrisation in Eq.\ref{MNS2}, which is the practical reason
why we prefer that form rather than that in Eq.\ref{MNS3}.

We now discuss the lepton mixing matrix to leading order
in $\theta_{13}$.
From Eqs.\ref{MNS2A},\ref{U23},\ref{U13},\ref{U12}, we find to leading
order in $\theta_{13}$ that $U$ may be expanded as:
\bea
&&U \approx   \nonumber \\
&& \left(\begin{array}{ccc}
c_{12} & s_{12}e^{-i\delta_{12}} & \theta_{13}e^{-i\delta_{13}} \\
-s_{12}c_{23}e^{i\delta_{12}}-c_{12}s_{23}\theta_{13}
e^{i(\delta_{13}-\delta_{23})}
& c_{12}c_{23}-s_{12}s_{23}\theta_{13}
e^{i(-\delta_{23}+\delta_{13}-\delta_{12})}
& s_{23}e^{-i\delta_{23}} \\
s_{12}s_{23}e^{i(\delta_{23}+\delta_{12})}
-c_{12}c_{23}\theta_{13}e^{i\delta_{13}}
& -c_{12}s_{23}e^{i\delta_{23}}
-s_{12}c_{23}\theta_{13}e^{i(\delta_{13}-\delta_{12})}
& c_{23} \\
\end{array}\right) \nonumber \\
\label{MNS6}
\eea
For $\theta_{13} = 0.1$, close to the CHOOZ limit,
the approximate form in Eq.\ref{MNS6} is accurate to 1\%.

We now wish to expand the MNS matrix in terms of neutrino
and charged lepton mixing angles and phases
to leading order in small angles,
using Eq.\ref{MNS5}.
In technically natural theories, based on right-handed
neutrino dominance, the contribution to $\theta_{23}$ comes mainly
from the neutrino sector, $\theta_{23}\approx \theta_{23}^{\nu_L}$.
Furthermore in natural theories we expect
that the contributions to $\theta_{13}$
are all separately small so that the smallness of this angle
does not rely on accidental cancellations.
Clearly this implies that $\theta_{13}^{\nu_L}$ and $\theta_{13}^{E_L}$
must both be $\simlt \theta_{13}$. Since the 13 element
of $U$ also receives a contribution from the charged lepton
sector proportional to $s_{12}^{E_L}s_{23}^{\nu_L}$, the
same argument also implies that
$\theta_{12}^{E_L}\simlt \theta_{13}$.
Therefore the natural expectation is that all the charged lepton
mixing angles are small! Expanding Eq.\ref{MNS5} to leading order
in small angles $\theta_{12}^{E_L}$,
$\theta_{23}^{E_L}$, $\theta_{13}^{E_L}$, $\theta_{13}^{\nu_L}$, we find
\bea
&&U \approx   \nonumber \\
&& \left(\begin{array}{ccc}
c_{12}^{\nu_L} & s_{12}^{\nu_L}e^{-i\delta_{12}^{\nu_L}} &
\theta_{13}^{\nu_L}e^{-i\delta_{13}^{\nu_L}} \\
-s_{12}^{\nu_L}c_{23}^{\nu_L}e^{i\delta_{12}^{\nu_L}}
-c_{12}^{\nu_L}s_{23}^{\nu_L}\theta_{13}^{\nu_L}
e^{i(\delta_{13}^{\nu_L}-\delta_{23}^{\nu_L})}
& c_{12}^{\nu_L}c_{23}^{\nu_L}-s_{12}^{\nu_L}s_{23}^{\nu_L}\theta_{13}^{\nu_L}
e^{i(-\delta_{23}^{\nu_L}+\delta_{13}^{\nu_L}-\delta_{12}^{\nu_L})}
& s_{23}^{\nu_L}e^{-i\delta_{23}^{\nu_L}} \\
s_{12}^{\nu_L}s_{23}^{\nu_L}e^{i(\delta_{23}^{\nu_L}+\delta_{12}^{\nu_L})}
-c_{12}^{\nu_L}c_{23}^{\nu_L}\theta_{13}^{\nu_L}e^{i\delta_{13}^{\nu_L}}
& -c_{12}^{\nu_L}s_{23}^{\nu_L}e^{i\delta_{23}^{\nu_L}}
-s_{12}^{\nu_L}c_{23}^{\nu_L}\theta_{13}^{\nu_L}
e^{i(\delta_{13}^{\nu_L}-\delta_{12}^{\nu_L})}
& c_{23}^{\nu_L} \nonumber \\
\end{array}\right) \nonumber \\
&&+\theta_{23}^{E_L}
\left(\begin{array}{ccc}
c_{12}^{\nu_L} &  s_{12}^{\nu_L}e^{-i\delta_{12}^{\nu_L}} & 0
\nonumber \\
-s_{23}^{\nu_L}s_{12}^{\nu_L}
e^{i(\delta_{23}^{\nu_L}-\delta_{23}^{E_L}
+\delta_{12}^{\nu_L})} &
s_{23}^{\nu_L}c_{12}^{\nu_L}
e^{i(\delta_{23}^{\nu_L}-\delta_{23}^{E_L})} &
-c_{23}^{\nu_L}e^{-i\delta_{23}^{E_L}}
\nonumber \\
-c_{23}^{\nu_L}s_{12}^{\nu_L}
e^{i(\delta_{23}^{E_L}+\delta_{12}^{\nu_L})} &
c_{23}^{\nu_L}c_{12}^{\nu_L}e^{i\delta_{23}^{E_L}} &
s_{23}^{\nu_L}e^{i(\delta_{23}^{E_L}-\delta_{23}^{\nu_L})}
\nonumber \\
\end{array}\right) \nonumber \\
&&+\theta_{13}^{E_L}
\left(\begin{array}{ccc}
-s_{12}^{\nu_L}s_{23}^{\nu_L}
e^{i(\delta_{12}^{\nu_L}+\delta_{23}^{\nu_L}-\delta_{13}^{E_L})} &
c_{12}^{\nu_L}s_{23}^{\nu_L}e^{i(\delta_{23}^{\nu_L}-\delta_{13}^{E_L})} &
-c_{23}^{\nu_L}e^{-i\delta_{13}^{E_L}} \nonumber \\
0 & 0 & 0 \nonumber \\
c_{12}^{\nu_L}e^{i\delta_{13}^{E_L}}
& s_{12}^{\nu_L}e^{i(-\delta_{12}^{\nu_L}+\delta_{13}^{E_L})} & 0
\nonumber \\
\end{array}\right) \nonumber \\
&&+\theta_{12}^{E_L}
\left(\begin{array}{ccc}
c_{23}^{\nu_L}s_{12}^{\nu_L}e^{i(\delta_{12}^{\nu_L}-\delta_{12}^{E_L})} &
-c_{23}^{\nu_L}c_{12}^{\nu_L}e^{-i\delta_{12}^{E_L}} &
s_{23}^{\nu_L}e^{i(-\delta_{23}^{\nu_L}-\delta_{12}^{E_L})} \nonumber \\
c_{12}^{\nu_L}e^{i\delta_{12}^{E_L}} &
s_{12}^{\nu_L}e^{i(-\delta_{12}^{\nu_L}+\delta_{12}^{E_L})} & 0 \nonumber \\
0 & 0 & 0  \\
\end{array}\right) \\
\label{MNS7}
\eea
where we have dropped terms of order $\theta_{23}^{E_L}\theta_{13}$.
The first matrix on the right hand side of
Eq.\ref{MNS7} gives the contribution to the
lepton mixing matrix 
from the neutrino mixing angles and phases, and is of
the same form as Eq.\ref{MNS6}.
The subsequent matrices give the corrections to the
lepton mixing matrix
from the charged lepton mixing angles
$\theta_{23}^{E_L}$, $\theta_{13}^{E_L}$, and $\theta_{12}^{E_L}$.

\section{Analytic Approach to Diagonalising Mass Matrices}

\subsection{Proceedure for diagonalising hierarchical mass matrices}
\label{proceedure}
In this appendix we discuss the diagonalisation of
a general complex hierarchical matrix $m$, assuming two large mixing
angles and one small mixing angle, to leading order in the small
mixing angle, where
\beq
m=
\left( \begin{array}{ccc}
m_{11} & m_{12} & m_{13} \\
m_{21} & m_{22} & m_{23} \\
m_{31} & m_{32} & m_{33}
\end{array}
\right)
\label{m11}
\eeq
The matrix $m$ is diagonalised by a sequence of tranformations:
\beq
{P_3^{L}}^{\ast}{R_{12}^{L}}^{T}{P_1^{L}}^{\ast}
{R_{13}^{L}}^{T}{R_{23}^{L}}^{T}{P_2^{L}}^{\ast}
m
P_2^{R}R_{23}^{R}R_{13}^{R}P_1^{R}R_{12}^{R}P_3^{R}=
\left( \begin{array}{ccc}
m_1 & 0 & 0    \\
0 & m_2 & 0 \\
0 & 0 & m_3
\end{array}
\right)
\label{diag5}
\eeq
In the case of the charged lepton mass matrix, all the rotation
angles are small, while in the case of the neutrino mass matrix it is
symmetric. The results for the general complex matrix
$m$ will be sufficiently general to allow us to
apply them to both of the physical cases of interest as limiting cases.

The proceedure for diagonalising a general hierarchical matrix $m$
involves the following steps.

1. The first step involves multiplying the mass matrix $m$ by the inner
phase matrices $P_2$ defined in Eq.\ref{P2}:
\beq
{P_2^{L}}^{\ast}
m
P_2^{R}=
\left( \begin{array}{ccc}
m_{11} & m_{12}e^{i\phi_2^R} & m_{13}e^{i\phi_3^R} \\
m_{21}e^{-i\phi_2^L} & m_{22}e^{i(\phi_2^R-\phi_2^L)}
& m_{23}e^{i(\phi_3^R-\phi_2^L)} \\
m_{31}e^{-i\phi_3^L}
& m_{32}e^{i(\phi_2^R-\phi_3^L)}
& m_{33}e^{i(\phi_3^R-\phi_3^L)}
\end{array}
\right)
\equiv
\left( \begin{array}{ccc}
m_{11} & m_{12}'& m_{13}' \\
m_{21}' & m_{22}' & m_{23}' \\
m_{31}' & m_{32}' & m_{33}'
\end{array}
\right)
\label{step1}
\eeq
The purpose of this re-phasing is to facilitate steps 2,3 using
real rotation angles $\theta_{23}$, $\theta_{13}$, as we shall see.

2. The second step is to perform the real rotations $R_{23}$
defined in Eq.\ref{R23} on the re-phased matrix from step1.
The purpose is to put zeroes in the 23,32 elements of the
resulting matrix:
\beq
{R_{23}^{L}}^{T}
\left( \begin{array}{ccc}
m_{11} & m_{12}'& m_{13}' \\
m_{21}' & m_{22}' & m_{23}' \\
m_{31}' & m_{32}' & m_{33}'
\end{array}
\right)
R_{23}^{R}
\equiv
\left( \begin{array}{ccc}
m_{11} & \tilde{m}_{12} & \tilde{m}_{13} \\
\tilde{m}_{21} &  \tilde{m}_{22} &  0 \\
\tilde{m}_{31} &  0 &  m_3'
\end{array}
\right)
\label{step2}
\eeq
The zeroes in the 23,32 positions are achieved by diagonalising the
lower 23 block, using the reduced matrix $R_{23}$ obtained by striking
out the row and column in which the unit element appears, to leave
a $2\times 2$ rotation,
\beq
{R_{23}^{L}}^{T}
\left( \begin{array}{cc}
m_{22}' & m_{23}' \\
m_{32}' & m_{33}'
\end{array}
\right)
R_{23}^{R}
\equiv
\left( \begin{array}{cc}
\tilde{m}_{22} &  0 \\
0 &  m_3'
\end{array}
\right)
\label{step21}
\eeq
which implies
\beq
\tan 2\theta_{23}^L=
\frac{2\left[m_{33}'m_{23}'+m_{22}'m_{32}'\right]}
{\left[{m_{33}'}^2-{m_{22}'}^2
+{m_{32}'}^2-{m_{23}'}^2\right]}
\label{23L}
\eeq
\beq
\tan 2\theta_{23}^R=
\frac{2\left[m_{33}'m_{32}'
+m_{22}'m_{23}'\right]}
{\left[{m_{33}'}^2-{m_{22}'}^2
+{m_{23}'}^2-{m_{32}'}^2\right]}
\label{23R}
\eeq
The requirement that the angles $\theta_{23}^L$ and $\theta_{23}^R$
are real means that the numerators and denominators must have equal
phases, and this is achieved by adjusting the relative phases
$\phi_i^R-\phi_j^L$ which appear in the lower block of Eq.\ref{step1}.
The remaining elements are then given by the reduced rotations
\beq
\left( \begin{array}{cc}
\tilde{m}_{12} & \tilde{m}_{13}
\end{array}
\right)
=
\left( \begin{array}{cc}
m_{12}'& m_{13}'
\end{array}
\right)
R_{23}^{R}
\label{step22}
\eeq
\beq
\left( \begin{array}{c}
\tilde{m}_{21} \\
\tilde{m}_{31}
\end{array}
\right)
=
{R_{23}^{L}}^{T}
\left( \begin{array}{c}
m_{21}' \\
m_{31}'
\end{array}
\right)
\label{step23}
\eeq

3. The third step is to perform the real small angle rotations $R_{13}$
defined in Eq.\ref{R13} on the matrix from step 2.
The purpose is to put zeroes in the 13,31 elements of the
resulting matrix:
\beq
{R_{13}^{L}}^{T}
\left( \begin{array}{ccc}
m_{11} & \tilde{m}_{12} & \tilde{m}_{13} \\
\tilde{m}_{21} &  \tilde{m}_{22} &  0 \\
\tilde{m}_{31} &  0 &  m_3'
\end{array}
\right)
R_{13}^{R}
\approx
\left( \begin{array}{ccc}
\tilde{m}_{11} & \tilde{m}_{12} & 0 \\
\tilde{m}_{21} &  \tilde{m}_{22} &  0 \\
0 &  0 &  m_3'
\end{array}
\right)
\label{step3}
\eeq
The zeroes in the 13,31 positions are achieved by diagonalising the
outer 13 block, using the reduced matrix $R_{13}$ obtained by striking
out the row and column in which the unit element appears, to leave
a $2\times 2$ rotation,
\beq
{R_{13}^{L}}^{T}
\left( \begin{array}{cc}
m_{11} & \tilde{m}_{13} \\
\tilde{m}_{31} & m_3'
\end{array}
\right)
R_{13}^{R}
\approx
\left( \begin{array}{cc}
\tilde{m}_{11} &  0 \\
0 &  m_3'
\end{array}
\right)
\label{step31}
\eeq
which implies
\beq
\theta_{13}^L\approx
\frac{\tilde{m}_{13}}{m_3'}+
\frac{\tilde{m}_{31}m_{11}}{(m_3')^2}
\label{13L}
\eeq
\beq
\theta_{13}^R\approx
\frac{\tilde{m}_{31}}{m_3'}+
\frac{\tilde{m}_{13}m_{11}}{(m_3')^2}
\label{13R}
\eeq
The requirement that the angles $\theta_{13}^L$ and $\theta_{31}^R$
are real fixes the absolute value of the phases
$\phi_i^R+\phi_j^L$, since only the relative phases were fixed
previously. This uses up all the phase freedom and thus
all the resulting mass matrix elements in Eq.\ref{step3} remain complex.
Note that Eq.\ref{step3} is written
to leading order in the small angles $\theta_{13}$,
and as discussed previously the 23,32 elements remain zero to this
order. The large complex element $m_3'$ is
approximately unchanged to this order.
Due to the zeroes in the 23,32 position of the matrix
the elements $\tilde{m}_{12}$ and $\tilde{m}_{21}$ are also unchanged
to leading order. The element $\tilde{m}_{22}$ is also unchanged
of course since it is not present in the reduced matrix.
The only new element is therefore
\beq
\tilde{m}_{11}\approx m_{11}-
\frac{\tilde{m}_{13}\tilde{m}_{31}}{m_3'}
\label{m111}
\eeq

4. The fourth step involves multiplying the mass matrix
resulting from Eq.\ref{step3} by the
phase matrices $P_1$ defined in Eq.\ref{P1}:
\beq
{P_1^{L}}^{\ast}
\left( \begin{array}{ccc}
\tilde{m}_{11} & \tilde{m}_{12} & 0 \\
\tilde{m}_{21} &  \tilde{m}_{22} &  0 \\
0 &  0 &  m_3'
\end{array}
\right)
P_1^{R}=
\left( \begin{array}{ccc}
\tilde{m}_{11} & \tilde{m}_{12}e^{i\chi^R} & 0 \\
\tilde{m}_{21}e^{-i\chi^L} &  \tilde{m}_{22}e^{i(\chi^R-\chi^L)} &  0 \\
0 &  0 &  m_3'
\end{array}
\right)
\equiv
\left( \begin{array}{ccc}
\tilde{m}_{11} & \tilde{m}_{12}'& 0 \\
\tilde{m}_{21}' & \tilde{m}_{22}' & 0 \\
0 & 0 & m_3'
\end{array}
\right)
\label{step4}
\eeq
The purpose of this re-phasing is to facilitate step 5 using
real rotation angle $\theta_{12}$.

5. The fifth step is to perform the real rotations $R_{12}$
defined in Eq.\ref{R12} on the re-phased matrix from step 4.
The purpose is to put zeroes in the 12,21 elements of the
resulting matrix:
\beq
{R_{12}^{L}}^{T}
\left( \begin{array}{ccc}
\tilde{m}_{11} & \tilde{m}_{12}'& 0 \\
\tilde{m}_{21}' & \tilde{m}_{22}' & 0 \\
0 & 0 & m_3'
\end{array}
\right)
R_{12}^{R}
\equiv
\left( \begin{array}{ccc}
m_1' & 0 & 0 \\
0 & m_2' &  0 \\
0 &  0 & m_3'
\end{array}
\right)
\label{step5}
\eeq
The zeroes in the 12,21 positions are achieved by diagonalising the
upper 12 block, using the reduced matrix $R_{12}$ obtained by striking
out the row and column in which the unit element appears, to leave
a $2\times 2$ rotation,
\beq
{R_{12}^{L}}^{T}
\left( \begin{array}{cc}
\tilde{m}_{11} & \tilde{m}_{12}'\\
\tilde{m}_{21}' & \tilde{m}_{22}'
\end{array}
\right)
R_{12}^{R}
\equiv
\left( \begin{array}{cc}
m_1' & 0 \\
0 & m_2'
\end{array}
\right)
\label{step51}
\eeq
which implies
\beq
\tan 2\theta_{12}^L=
\frac{2\left[\tilde{m}_{22}'\tilde{m}_{12}'
+\tilde{m}_{11}\tilde{m}_{21}'\right]}
{\left[(\tilde{m}_{22}')^2-(\tilde{m}_{11})^2
+(\tilde{m}_{21}')^2-(\tilde{m}_{12}')^2\right]}
\label{12L}
\eeq
\beq
\tan 2\theta_{12}^R=
\frac{2\left[\tilde{m}_{22}'\tilde{m}_{21}'
+\tilde{m}_{11}\tilde{m}_{12}'\right]}
{\left[(\tilde{m}_{22}')^2-(\tilde{m}_{11})^2
+(\tilde{m}_{12}')^2-(\tilde{m}_{21}')^2\right]}
\label{12R}
\eeq
The requirement that the angles $\theta_{12}^L$ and $\theta_{21}^R$
are real means that the numerators and denominators must have equal
phases, and this is achieved by adjusting the phases $\chi_L$, $\chi_R$.

6. The sixth step involves multiplying the complex diagonal mass matrix
resulting from Eq.\ref{step5} by the
phase matrices $P_3$ defined in Eq.\ref{P3}:
\beq
{P_3^{L}}^{\ast}
\left( \begin{array}{ccc}
m_1' & 0 & 0 \\
0 & m_2' &  0 \\
0 &  0 & m_3'
\end{array}
\right)
P_3^{R}=
\left( \begin{array}{ccc}
m_1 & 0 & 0 \\
0 & m_2 &  0 \\
0 &  0 & m_3
\end{array}
\right)
\label{step6}
\eeq
The result of this re-phasing is a diagonal matrix with real
eigenvalues. In the case of charged leptons this last step can be
achieved by a suitable $P_3^R$ for any choice of $P_3^L$.
This freedom in $P_3^L$ enables three phases to be removed from the
lepton mixing matrix.

\subsection{Diagonalising the hierarchical neutrino mass matrix}
\label{hierarchical}

In this appendix we shall apply the results of
appendix \ref{proceedure} to the case of the
complex symmetric hierarchical neutrino mass matrix of the
leading order form of Type IA as shown in Table 1, which will be
written in full generality as
\beq
m^{\nu}_{LL}=
\left( \begin{array}{ccc}
m^{\nu}_{11} & m^{\nu}_{12} & m^{\nu}_{13} \\
m^{\nu}_{12} & m^{\nu}_{22} & m^{\nu}_{23} \\
m^{\nu}_{13} & m^{\nu}_{23} & m^{\nu}_{33}
\end{array}
\right)
\equiv
\left( \begin{array}{ccc}
|m^{\nu}_{11}|e^{i\phi^{\nu}_{11}}
& |m^{\nu}_{12}|e^{i\phi^{\nu}_{12}}
& |m^{\nu}_{13}|e^{i\phi^{\nu}_{13}} \\
|m^{\nu}_{12}|e^{i\phi^{\nu}_{12}}
& |m^{\nu}_{22}|e^{i\phi^{\nu}_{22}}
& |m^{\nu}_{23}|e^{i\phi^{\nu}_{23}} \\
|m^{\nu}_{13}|e^{i\phi^{\nu}_{13}}
& |m^{\nu}_{23}|e^{i\phi^{\nu}_{23}}
& |m^{\nu}_{33}|e^{i\phi^{\nu}_{33}}
\end{array}
\right)
\label{mnu1}
\eeq
where it should be remembered that for a Type IA matrix the elements in the
lower 23 block are larger than the other elements.

The proceedure outlined in appendix \ref{hierarchical}
for diagonalising $m^{\nu}_{LL}$ is to
work our way from the inner transformations to the
outer transformations as follows.
\begin{enumerate}
\item Re-phase $m^{\nu}_{LL}$ using the $P^{\nu_L}_2$.
\item Put zeroes in the 23=32 positions using $R^{\nu_L}_{23}$.
\item Put zeroes in the 13=31 positions using $R^{\nu_L}_{13}$.
\item Re-phase the mass matrix using $P^{\nu_L}_1$.
\item Put zeroes in the 12=21 positions using $R^{\nu_L}_{12}$.
\item Make the diagonal elements real using the $P^{\nu_L}_3$.
\end{enumerate}
If $\theta^{\nu_L}_{13}$ is small, then for the
hierarchical case $m_3 \gg m_2$ this proceedure
will result in an approximately diagonal matrix to
leading order in $\theta^{\nu_L}_{13}$.
One might object that
after step 3 the $R^{\nu_L}_{13}$ rotations will
``fill-in'' the zeroes in the 23,32 positions with terms
of order $\theta^{\nu_L}_{13}$ multiplied by
$m^{\nu_L}_{12},m^{\nu_L}_{13}$. However in the hierarchical case
$m^{\nu_L}_{12},m^{\nu_L}_{13}$ are smaller than $m^{\nu_L}_{33}$
by a factor of $\theta^{\nu_L}_{13}$ which
means that the ``filled-in'' 23,32 entries
are suppressed by a total factor of $(\theta^{\nu_L}_{13})^2$ compared to the
33 element.
This means that after the 5 steps above a hierarchical
matrix will be diagonal
to leading order in $\theta^{\nu_L}_{13}$, as claimed.
For the inverted hierarchical neutrino case
a different proceedure must be followed, as discussed in the
next sub-section. Here we shall systematically diagonalise the
hierarchical neutrino mass matrix in Eq.\ref{mnu1} by following
the above proceedure as follows.

The first step is to re-phase the matrix in Eq.\ref{mnu1}
using ${P_2^{\nu_L}}^{\ast}$ so that
the neutrino mass matrix becomes,
\beq
\left( \begin{array}{ccc}
|m^{\nu}_{11}|e^{i\phi^{\nu}_{11}}
& |m^{\nu}_{12}|e^{i(\phi^{\nu}_{12}-\phi^{\nu_L}_2)}
& |m^{\nu}_{13}|e^{i(\phi^{\nu}_{13}-\phi^{\nu_L}_3)}  \\
|m^{\nu}_{12}|e^{i(\phi^{\nu}_{12}-\phi^{\nu_L}_2)}
& |m^{\nu}_{22}|e^{i(\phi^{\nu}_{22}-2\phi^{\nu_L}_2)}
& |m^{\nu}_{23}|e^{i(\phi^{\nu}_{23}-\phi^{\nu_L}_2-\phi^{\nu_L}_3)} \\
|m^{\nu}_{13}|e^{i(\phi^{\nu}_{13}-\phi^{\nu_L}_3)}
& |m^{\nu}_{23}|e^{i(\phi^{\nu}_{23}-\phi^{\nu_L}_2-\phi^{\nu_L}_3)}
& |m^{\nu}_{33}|e^{i(\phi^{\nu}_{33}-2\phi^{\nu_L}_3)}
\end{array}
\right)
\label{mnu2}
\eeq
To determine the 23 neutrino mixing angle
$\theta_{23}^{\nu_L}$ we perform a 23 rotation which
diagonalises the lower 23 block of Eq.\ref{mnu2}.
From Eq.\ref{23L} we find the 23 neutrino mixing angle
$\theta_{23}^{\nu_L}$ as
\beq
\tan 2\theta_{23}^{\nu_L}=
\frac{2\left[|m^{\nu}_{23}|
e^{i(\phi^{\nu}_{23}-\phi^{\nu_L}_2-\phi^{\nu_L}_3)}
\right]}
{\left[|m^{\nu}_{33}|e^{i(\phi^{\nu}_{33}-2\phi^{\nu_L}_3)}
-|m^{\nu}_{22}|e^{i(\phi^{\nu}_{22}-2\phi^{\nu_L}_2)}
\right]}
\label{nu23L}
\eeq
The relative phase $\phi^{\nu_L}_2-\phi^{\nu_L}_3$ is fixed by the
requirement that the angle $\theta_{23}^{\nu_L}$ in Eq.\ref{nu23L} be
real,
\beq
|m^{\nu}_{33}|
\sin(\phi^{\nu}_{33}-\phi^{\nu}_{23}+\phi^{\nu_L}_2-\phi^{\nu_L}_3)
=
|m^{\nu}_{22}|
\sin(\phi^{\nu}_{22}-\phi^{\nu}_{23}+\phi^{\nu_L}_3-\phi^{\nu_L}_2)
\label{phase1}
\eeq
After the 23 rotation in Eq.\ref{step2},
the neutrino mass matrix in Eq.\ref{mnu2} becomes
\beq
\left( \begin{array}{ccc}
m^{\nu}_{11} & \tilde{m}^{\nu}_{12} & \tilde{m}^{\nu}_{13} \\
\tilde{m}^{\nu}_{12} &  \tilde{m}^{\nu}_{22} &  0 \\
\tilde{m}^{\nu}_{13} &  0 &  m_3'
\end{array}
\right)
\label{nustep2}
\eeq
The lower block elements are given by
\beq
\left( \begin{array}{cc}
\tilde{m}^{\nu}_{22} &  0\\
0 &  m_3'
\end{array}
\right)
\equiv
{R^{\nu_L}_{23}}^T
\left( \begin{array}{cc}
|m^{\nu}_{22}|e^{i(\phi^{\nu}_{22}-2\phi^{\nu_L}_2)}
& |m^{\nu}_{23}|e^{i(\phi^{\nu}_{23}-\phi^{\nu_L}_2-\phi^{\nu_L}_3)}  \\
|m^{\nu}_{23}|e^{i(\phi^{\nu}_{23}-\phi^{\nu_L}_2-\phi^{\nu_L}_3)}
& |m^{\nu}_{33}|e^{i(\phi^{\nu}_{33}-2\phi^{\nu_L}_3)}
\end{array}
\right)
R^{\nu_L}_{23}
\label{nustep222}
\eeq
which implies
\bea
\tilde{m}^{\nu}_{22}&=&
(c_{23}^{\nu_L})^2|m^{\nu}_{22}|e^{i(\phi^{\nu}_{22}-2\phi^{\nu_L}_2)}
-2s_{23}^{\nu_L}c_{23}^{\nu_L}|m^{\nu}_{23}|
e^{i(\phi^{\nu}_{23}-\phi^{\nu_L}_2-\phi^{\nu_L}_3)}
+(s_{23}^{\nu_L})^2|m^{\nu}_{33}|e^{i(\phi^{\nu}_{33}-2\phi^{\nu_L}_3)}
\nonumber \\
&&  \label{m22}\\
m_3'&=&
(s_{23}^{\nu_L})^2|m^{\nu}_{22}|e^{i(\phi^{\nu}_{22}-2\phi^{\nu_L}_2)}
+2s_{23}^{\nu_L}c_{23}^{\nu_L}|m^{\nu}_{23}|
e^{i(\phi^{\nu}_{23}-\phi^{\nu_L}_2-\phi^{\nu_L}_3)}
+(c_{23}^{\nu_L})^2|m^{\nu}_{33}|e^{i(\phi^{\nu}_{33}-2\phi^{\nu_L}_3)}
\nonumber \\
&&
\label{m3p}
\eea
and from Eq.\ref{step23}
\beq
\left( \begin{array}{c}
\tilde{m}^{\nu}_{12} \\
\tilde{m}^{\nu}_{13}
\end{array}
\right)
=
{R_{23}^{\nu_L}}^{T}
\left( \begin{array}{c}
|m^{\nu}_{12}|
e^{i(\phi^{\nu}_{12}-\phi^{\nu_L}_2)} \\
|m^{\nu}_{13}|
e^{i(\phi^{\nu}_{13}-\phi^{\nu_L}_3)}
\end{array}
\right)
\label{nustep23}
\eeq

We now perform a 13 rotation on the neutrino matrix in
Eq.\ref{nustep2} which diagonalises the outer 13 block of Eq.\ref{nustep2} and
determines the 13 neutrino mixing angle
$\theta_{13}^{\nu_L}$.
From Eq.\ref{13L} we find the 13 neutrino mixing angle
$\theta_{13}^{\nu_L}$ as
\beq
\theta_{13}^{\nu_L}\approx
\frac{\tilde{m}^{\nu}_{13}}{m_3'}
\label{nu13L}
\eeq
The absolute phases $\phi^{\nu_L}_2$, $\phi^{\nu_L}_3$ are fixed by the
requirement that the angle $\theta_{13}^{\nu_L}$ in Eq.\ref{nu13L} be
real,
\beq
s_{23}^{\nu_L}|m^{\nu}_{12}|
\sin(\phi^{\nu}_{12}-\phi^{\nu_L}_2-\phi_3')
+
c_{23}^{\nu_L}|m^{\nu}_{13}|
\sin(\phi^{\nu}_{13}-\phi^{\nu_L}_3-\phi_3')=0
\label{absphase}
\eeq
After the 13 rotation in Eq.\ref{step3},
Eq.\ref{nustep2} becomes
\beq
\left( \begin{array}{ccc}
\tilde{m}^{\nu}_{11} & \tilde{m}^{\nu}_{12} & 0 \\
\tilde{m}^{\nu}_{12} &  \tilde{m}^{\nu}_{22} &  0 \\
0 &  0 &  m_3'
\end{array}
\right)
\equiv
\left( \begin{array}{ccc}
|\tilde{m}^{\nu}_{11}|e^{i\tilde{\phi}^{\nu}_{11}}
& |\tilde{m}^{\nu}_{12}|e^{i\tilde{\phi}^{\nu}_{12}} & 0 \\
|\tilde{m}^{\nu}_{12}|e^{i\tilde{\phi}^{\nu}_{12}}
& |\tilde{m}^{\nu}_{22}|e^{i\tilde{\phi}^{\nu}_{22}} &  0 \\
0 &  0 &  |m_3'|e^{i\phi_3'}
\end{array}
\right)
\label{nustep3}
\eeq
To leading order in $\theta_{13}^{\nu_L}$
the only new element in Eq.\ref{nustep3} is
\beq
\tilde{m}^{\nu_L}_{11}\approx m^{\nu_L}_{11}-
\frac{(\tilde{m}^{\nu_L}_{13})^2}{m_3'}
\label{num11}
\eeq

It only remains to determine the 12 neutrino mixing angle
$\theta_{12}^{\nu_L}$ by diagonalising the upper 12 block
of Eq.\ref{nustep3}.
From Eq.\ref{12L} we find the 12 neutrino mixing angle
$\theta_{12}^{\nu_L}$ as
\beq
\tan 2\theta_{12}^{\nu_L}=
\frac{2\left[|\tilde{m}^{\nu}_{12}|
e^{i(\tilde{\phi}^{\nu}_{12}-\chi^{\nu_L})}
\right]}
{\left[|\tilde{m}^{\nu}_{22}|
e^{i(\tilde{\phi}^{\nu}_{22}-2\chi^{\nu_L})}
-|\tilde{m}^{\nu}_{11}|
e^{i\tilde{\phi}^{\nu}_{11}}
\right]}
\label{nu12L}
\eeq
The phase $\chi^{\nu_L}$ is fixed by the
requirement that the angle $\theta_{12}^{\nu_L}$ in Eq.\ref{nu12L} be
real,
\beq
|\tilde{m}^{\nu}_{22}|
\sin(\tilde{\phi}^{\nu}_{22}-\tilde{\phi}^{\nu}_{12}-\chi^{\nu_L})
=
|\tilde{m}^{\nu}_{11}|
\sin(\tilde{\phi}^{\nu}_{11}-\tilde{\phi}^{\nu}_{12}+\chi^{\nu_L})
\label{chi}
\eeq
After the 12 rotation the upper block of the matrix in
Eq.\ref{nustep3} is diagonal and the resulting matrix is
\beq
\left( \begin{array}{ccc}
m_1' & 0 & 0 \\
0 & m_2' &  0 \\
0 &  0 & m_3'
\end{array}
\right)
\equiv
\left( \begin{array}{ccc}
m_1e^{i{\phi}_1'} & 0 & 0 \\
0 & m_2e^{i{\phi}_2'} & 0 \\
0 &  0 &  m_3e^{i\phi_3'}
\end{array}
\right)
\label{nustep5}
\eeq
where from Eq.\ref{step51}
\bea
m_1'&=&
(c_{12}^{\nu_L})^2|\tilde{m}^{\nu}_{11}|e^{i\tilde{\phi}^{\nu}_{11}}
-2s_{12}^{\nu_L}c_{12}^{\nu_L}|\tilde{m}^{\nu}_{12}|
e^{i(\tilde{\phi}^{\nu}_{12}-\chi^{\nu_L})}
+(s_{12}^{\nu_L})^2|\tilde{m}^{\nu}_{22}|
e^{i(\tilde{\phi}^{\nu}_{22}-2\chi^{\nu_L})}
\nonumber \\
&& \\
m_2'&=&
(s_{12}^{\nu_L})^2|\tilde{m}^{\nu}_{11}|e^{i\tilde{\phi}^{\nu}_{11}}
+2s_{12}^{\nu_L}c_{12}^{\nu_L}|\tilde{m}^{\nu}_{12}|
e^{i(\tilde{\phi}^{\nu}_{12}-\chi^{\nu_L})}
+(c_{12}^{\nu_L})^2|\tilde{m}^{\nu}_{22}|
e^{i(\tilde{\phi}^{\nu}_{22}-2\chi^{\nu_L})}
\nonumber \\
&&
\label{m2p}
\eea

It is a simple matter to adjust the
phases $\omega^{\nu_L}_i$ in $P^{\nu_L}_3$
to remove the phases in Eq.\ref{nustep5} and make
the neutrino masses real,
as in Eq.\ref{step6},
\beq
\omega^{\nu_L}_i=\frac{{\phi}_i'}{2}
\eeq
This completes the diagonalisation.
In the case of neutrino masses, unlike the case of the charged
fermions, there is no left over phase freedom. This is the reason
why the lepton mixing matrix has three more physical phases than the
CKM matrix.

\subsection{Diagonalising the inverted hierarchical neutrino mass matrix}
\label{inverted}
In this appendix we shall consider the case of the
complex symmetric inverted hierarchical neutrino mass matrix of the
leading order form of Type IB in Table 1.
In this case the proceedure is as follows.

\begin{enumerate}
\item Re-phase $m^{\nu}_{LL}$ using the $P^{\nu_L}_2$.
\item Put zeroes in the 13=31 positions using $R^{\nu_L}_{23}$.
\item Put zeroes in the 23=32 positions using $R^{\nu_L}_{13}$.
\item Re-phase the mass matrix using $P^{\nu_L}_1$.
\item Put zeroes in the 12=21 positions using $R^{\nu_L}_{12}$.
\item Make the diagonal elements real using the $P^{\nu_L}_3$.
\end{enumerate}

We continue to write the neutrino mass matrix as
in Eq.\ref{mnu1}, but now it should be remembered that
for a Type IB matrix the 12,13 elements
are now larger than the other elements.
This is the reason why the above
proceedure differs from that for the case of the hierarchical neutrino
mass matrix.

We first perform the re-phasing as in Eq.\ref{mnu2}.
Then we determine the 23 neutrino mixing angle
$\theta_{23}^{\nu_L}$ by performing a 23 rotation such that
\beq
\left( \begin{array}{ccc}
m^{\nu}_{11} & \tilde{m}^{\nu}_{12} & 0 \\
\tilde{m}^{\nu}_{12} &  \tilde{m}^{\nu}_{22} &  \tilde{m}^{\nu}_{23}\\
0 &  \tilde{m}^{\nu}_{23} &  m_3'
\end{array}
\right)
\equiv
{R^{\nu_L}_{23}}^T
\left( \begin{array}{ccc}
|m^{\nu}_{11}|e^{i\phi^{\nu}_{11}}
& |m^{\nu}_{12}|e^{i(\phi^{\nu}_{12}-\phi^{\nu_L}_2)}
& |m^{\nu}_{13}|e^{i(\phi^{\nu}_{13}-\phi^{\nu_L}_3)}  \\
|m^{\nu}_{12}|e^{i(\phi^{\nu}_{12}-\phi^{\nu_L}_2)}
& |m^{\nu}_{22}|e^{i(\phi^{\nu}_{22}-2\phi^{\nu_L}_2)}
& |m^{\nu}_{23}|e^{i(\phi^{\nu}_{23}-\phi^{\nu_L}_2-\phi^{\nu_L}_3)} \\
|m^{\nu}_{13}|e^{i(\phi^{\nu}_{13}-\phi^{\nu_L}_3)}
& |m^{\nu}_{23}|e^{i(\phi^{\nu}_{23}-\phi^{\nu_L}_2-\phi^{\nu_L}_3)}
& |m^{\nu}_{33}|e^{i(\phi^{\nu}_{33}-2\phi^{\nu_L}_3)}
\end{array}
\right)
R^{\nu_L}_{23}
\label{invnustep2}
\eeq
where
\beq
\left( \begin{array}{c}
\tilde{m}^{\nu}_{12} \\
0
\end{array}
\right)
=
{R_{23}^{\nu_L}}^{T}
\left( \begin{array}{c}
|m^{\nu}_{12}|
e^{i(\phi^{\nu}_{12}-\phi^{\nu_L}_2)} \\
|m^{\nu}_{13}|
e^{i(\phi^{\nu}_{13}-\phi^{\nu_L}_3)}
\end{array}
\right)
\label{invnustep21}
\eeq
which gives the 23 neutrino mixing angle
$\theta_{23}^{\nu_L}$ in this case to be
\beq
\tan \theta_{23}^{\nu_L}=
\frac{-|m^{\nu}_{13}|
e^{i(\phi^{\nu}_{13}-\phi^{\nu_L}_3)}}
{|m^{\nu}_{12}|e^{i(\phi^{\nu}_{12}-\phi^{\nu_L}_2)}}
\label{invnu23L}
\eeq
Since the Euler angles are constrained to satisfy
$\theta_{ij}\leq \pi/2$, we must have
$\tan \theta_{23}^{\nu_L} \approx +1$,
and this then fixes
\beq
\phi^{\nu}_{13}-\phi^{\nu}_{12}+\phi^{\nu_L}_2-\phi^{\nu_L}_3=\pi
\eeq
This fixes $\phi^{\nu_L}_2-\phi^{\nu_L}_3$ and gives
\beq
\tan \theta_{23}^{\nu_L}=
\frac{|m^{\nu}_{13}|}
{|m^{\nu}_{12}|}
\label{invnu223L}
\eeq
and
\beq
\tilde{m}^{\nu}_{12}=c_{23}^{\nu_L}|m^{\nu}_{12}|
e^{i(\phi^{\nu}_{12}-\phi^{\nu_L}_2)}-
s_{23}^{\nu_L}|m^{\nu}_{13}|
e^{i(\phi^{\nu}_{13}-\phi^{\nu_L}_3)}
\eeq
The lower block elements are given by
\beq
\left( \begin{array}{cc}
\tilde{m}^{\nu}_{22} &  \tilde{m}^{\nu}_{23}\\
\tilde{m}^{\nu}_{23} &  m_3'
\end{array}
\right)
\equiv
{R^{\nu_L}_{23}}^T
\left( \begin{array}{cc}
|m^{\nu}_{22}|e^{i(\phi^{\nu}_{22}-2\phi^{\nu_L}_2)}
& |m^{\nu}_{23}|e^{i(\phi^{\nu}_{23}-\phi^{\nu_L}_2-\phi^{\nu_L}_3)}  \\
|m^{\nu}_{23}|e^{i(\phi^{\nu}_{23}-\phi^{\nu_L}_2-\phi^{\nu_L}_3)}
& |m^{\nu}_{33}|e^{i(\phi^{\nu}_{33}-2\phi^{\nu_L}_3)}
\end{array}
\right)
R^{\nu_L}_{23}
\label{invnustep22}
\eeq
which implies
\beq
\tilde{m}^{\nu}_{23}=
s_{23}^{\nu_L}c_{23}^{\nu_L}
(|m^{\nu}_{22}|e^{i(\phi^{\nu}_{22}-2\phi^{\nu_L}_2)}-
|m^{\nu}_{33}|e^{i(\phi^{\nu}_{33}-2\phi^{\nu_L}_3)})
+((c_{23}^{\nu_L})^2-(s_{23}^{\nu_L})^2)
|m^{\nu}_{23}|
e^{i(\phi^{\nu}_{23}-\phi^{\nu_L}_2-\phi^{\nu_L}_3)}
\label{mt23}
\eeq
and the remaining diagonal elements are given as before in
Eqs.\ref{m22},\ref{m3p}.

We next perform a small angle 13 rotation such that
\beq
\left( \begin{array}{ccc}
m^{\nu}_{11} & \tilde{m}^{\nu}_{12} & 0 \\
\tilde{m}^{\nu}_{12} &  \tilde{m}^{\nu}_{22} &  0\\
0 &  0 &  m_3'
\end{array}
\right)
\approx
{R^{\nu_L}_{13}}^T
\left( \begin{array}{ccc}
m^{\nu}_{11} & \tilde{m}^{\nu}_{12} & 0 \\
\tilde{m}^{\nu}_{12} &  \tilde{m}^{\nu}_{22} &  \tilde{m}^{\nu}_{23}\\
0 &  \tilde{m}^{\nu}_{23} &  m_3'
\end{array}
\right)
R^{\nu_L}_{13}
\label{invnustep3}
\eeq
where
\beq
\left( \begin{array}{c}
\tilde{m}^{\nu}_{12} \\
0
\end{array}
\right)
\approx
{R_{13}^{\nu_L}}^{T}
\left( \begin{array}{c}
\tilde{m}^{\nu}_{12}   \\
\tilde{m}^{\nu}_{23}
\end{array}
\right)
\label{invnustep31}
\eeq
Note that to leading order in $\theta_{13}^{\nu_L}$ the
large element $\tilde{m}^{\nu}_{12}$ is unchanged.
The remaining elements in Eq.\ref{invnustep3} are also unchanged
to leading order in $\theta_{13}^{\nu_L}$.
The 13=31 element in Eq.\ref{invnustep3} gets filled in
by a term $\theta_{13}^{\nu_L}(m^{\nu}_{11}-m_3')$ which is
of order $(\theta_{13}^{\nu_L})^2$ compared to $\tilde{m}^{\nu}_{12}$
and does not appear to leading order in $\theta_{13}^{\nu_L}$.
From Eq.\ref{invnustep31} the 13 neutrino mixing angle
$\theta_{13}^{\nu_L}$ is
\beq
\theta_{13}^{\nu_L}\approx
\frac{-\tilde{m}^{\nu}_{23}}{\tilde{m}^{\nu}_{12}}
\label{invnu13L}
\eeq
The requirement that $\theta_{13}^{\nu_L}$ is real fixes the absolute
value of the phases $\phi^{\nu_L}_2$, $\phi^{\nu_L}_3$.

The left hand side of Eq.\ref{invnustep3} now resembles the left hand
side of Eq.\ref{nustep3}, except that here $m^{\nu}_{11}$ is unchanged
due to the zero 13=31 element after the 23 rotation.
Therefore the rest of the diagonalisation process follows that of
the previous hierarchical case from Eq.\ref{nu12L} onwards,
where now
\beq
\tan 2\theta_{12}^{\nu_L}=
\frac{2\left[|\tilde{m}^{\nu}_{12}|
e^{i(\tilde{\phi}^{\nu}_{12}-\chi^{\nu_L})}
\right]}
{\left[|\tilde{m}^{\nu}_{22}|
e^{i(\tilde{\phi}^{\nu}_{22}-2\chi^{\nu_L})}
-|{m}^{\nu}_{11}|
e^{i{\phi}^{\nu}_{11}}
\right]}
\label{invnu12L}
\eeq
Note that in the inverted hierarchy case here we have
\beq
|\tilde{m}^{\nu}_{12}| \gg |\tilde{m}^{\nu}_{22}|,|{m}^{\nu}_{11}|
\eeq
which implies an almost degenerate pair of
pseudo-Dirac neutrino masses (with opposite sign eigenvalues) and an almost
maximal 12 mixing angle from Eq.\ref{invnu12L}.

\end{document}